\documentclass[a4paper,11pt]{article}
\pdfoutput=1 
\usepackage{graphicx}  
\usepackage{dcolumn}   
\usepackage{bm,relsize}        
\usepackage{amssymb, amsmath}
\usepackage{textcomp}
\usepackage{wasysym}
\usepackage{slashed}
\usepackage{caption, subcaption}
\usepackage{multirow}
\usepackage{gensymb}
\usepackage{ascmac}
\usepackage{comment}
\usepackage{amsmath,amsthm,amssymb,slashed,cancel}
\usepackage{cite,tabularx}
\usepackage{subcaption}
\usepackage{graphicx,color}
\usepackage[normalem]{ulem}
\usepackage{mathtools}
\usepackage{enumerate}

\usepackage{tikz}
\usetikzlibrary{shapes.geometric, arrows}

\usepackage{lipsum, color}
\usepackage[usenames,dvipsnames,svgnames]{xcolor}
\usepackage{braket}
\usepackage{jheppub} 
\usepackage{booktabs}
\usepackage{pdflscape}
\usepackage[utf8]{inputenc} 
\usepackage{}
\usetikzlibrary{decorations.pathreplacing, arrows.meta}

\usetikzlibrary{decorations.markings, decorations.pathreplacing}

\def\beq{\begin{equation}}
\def\eeq{\end{equation}}
 \def\be{\begin{equation}} \def\ee{\end{equation}}
\def\bea{\begin{eqnarray}} \def\eea{\end{eqnarray}}

\def\beal{\begin{align}} \def\eeal{\end{align}}

\newcommand{\bi}{\begin{itemize} }
\newcommand{\ei}{\end{itemize} }

\usepackage{caption}
\usepackage{subcaption}
\usepackage{multirow}
\usepackage[normalem]{ulem}
\allowdisplaybreaks

\usepackage[capitalise, english]{cleveref}

\begin{document}

\title{Bounds on nonlinear electrodynamics via
resummed relative entropy}

\author[a,b]{Pietro Conzinu,}
\author[c]{Daiki Ueda}

\emailAdd{p.conzinu@ssmeridionale.it}
\emailAdd{daiki.ueda@campus.technion.ac.il}

\affiliation[a]{Scuola Superiore Meridionale, Via Mezzocannone 4, 80138 Napoli, Italy}
\affiliation[b]{INFN-Napoli, Complesso Universitario di Monte S. Angelo, Via Cinthia Edificio 6, 80126 Napoli, Italy
}
\affiliation[c]{Physics Department, Technion -- Israel Institute of Technology, Haifa 3200003, Israel}

\date{\today}

\preprint{}

\abstract{We investigate nonlinear electrodynamic effective field theories (EFTs) through the relative entropy evaluated in suitable background electromagnetic fields. 
In this setup, the relative entropy encodes information about the infinite tower of higher-dimensional operators and provides a systematic probe of nonlinear EFT effects.
We study these features in fermionic QED, scalar QED, and Dirac-Born-Infeld theory using perturbative analyses, resummation techniques such as Borel--Laplace resummation, and non-perturbative approaches including the Schwinger proper-time method.
In the weak-coupling regime, we show that the non-negativity of the perturbative relative entropy imposes sign constraints on finite truncations of higher-dimensional operators, generalizing familiar positivity bounds on leading EFT coefficients.
We further show that violations of non-negativity in the strong-coupling regime admit qualitatively different interpretations depending on the framework: perturbatively analyzed violations diagnose the breakdown of the truncated EFT expansion, whereas violations in resummed or genuinely non-perturbative relative entropy signal physical instabilities of the system, such as the Schwinger effect.
Extending the analysis to broader classes of UV completions, including theories with factorial or power-law growth of EFT coefficients, we derive general constraints on nonlinear electrodynamic EFT effects from the non-negativity of the resummed relative entropy.
Our results suggest that relative entropy provides a unified diagnostic of perturbative consistency and non-perturbative stability in nonlinear EFTs.}

\maketitle

\section{Introduction}\label{sec:intro}

Nonlinear extensions of electrodynamics provide a paradigmatic arena for exploring the consistency of effective field theories (EFTs).
From the Euler--Heisenberg effective action in QED~\cite{Heisenberg:1936nmg,Weisskopf:1936hya,Schwinger:1951nm,Dunne:2012vv,Quevillon:2018mfl} to Born--Infeld and DBI-type theories~\cite{Born:1934gh,Dirac:1962iy,Fradkin:1985qd,Leigh:1989jq,Tseytlin:1999dj}, the low-energy dynamic of photon is generically described by nonlinear corrections organized as an infinite tower of higher-dimensional operators obtained after integrating out heavy degrees of freedom.
Understanding how such infinite EFT expansions remain compatible with unitarity, causality, and a consistent UV completion is therefore a central problem in quantum field theory~\cite{Adams:2006sv,deRham:2017avq,Henriksson:2021ymi,Henriksson:2022oeu,Bertucci:2024qzt,CarrilloGonzalez:2023cbf,Buoninfante:2024ibt,Abe:2023anf,Abe:2025vdj,Cheung:2018cwt,Cheung:2019cwi,Cheung:2025nhw,Fernandez-Sarmiento:2025tpy,Cheung:2026dng}.

Nonlinear EFTs are characterized by an infinite tower of higher-dimensional operators whose coefficients may exhibit nontrivial large-order behavior~\cite{Georgi:1993mps,Burgess:2007pt}, including factorial growth or more general power-law asymptotics.
Such behavior can encode information about the underlying UV dynamics~\cite{Dunne:1999uy,Dunne:1999vd,Dunne:2021acr,Aniceto:2018bis} and non-perturbative physics.
This raises a fundamental question: what constraints govern the full infinite tower of higher-dimensional operators beyond conventional perturbative positivity bounds, and what do they imply for the asymptotic large-order structure of the EFT expansion?

Several approaches may be conceivable for addressing this question.
Among them, relative entropy has recently emerged as a novel information-theoretic probe of EFT consistency~\cite{Cao:2022iqh,Cao:2022ajt}.
At leading order in perturbation theory, its non-negativity yields nontrivial constraints on EFT coefficients, reproducing or closely paralleling conventional positivity bounds derived from analyticity, unitarity and causality.
However, existing analyses remain restricted to perturbative regimes and do not address genuinely nonlinear EFT expansions involving infinitely many higher-dimensional operators.

Motivated by these developments, Ref.~\cite{Ueda:2024cyf} explored constraints on EFT coefficients beyond leading order and their implications for the validity of perturbative EFT expansions, focusing on scalar field theories.
However, the analysis in Ref.~\cite{Ueda:2024cyf} remained perturbative, relying on a truncation of the EFT expansion at finite order.
As a result, the connection between EFT consistency bounds and genuinely non-perturbative effects encoded in the large-order structure of the expansion remained unclear.

Building on these observations, in Ref.~\cite{Conzinu:2026cuf}, we investigated nonlinear EFTs with factorially growing perturbative expansions, focusing on a class of theories in which the relative entropy captures an infinite tower of higher-dimensional operators.
By performing a resummation of the relative entropy, we derived constraints on the large-order structure of EFT coefficients: the non-negativity of the resummed relative entropy fixes their asymptotic sign structure, while a violation of this non-negativity signals non-perturbative effects such as vacuum instabilities.
Fermionic QED provides a concrete example of this mechanism.
There, analytic continuation from Euclidean to Minkowski spacetime connects the violation of the non-negativity of the resummed relative entropy to the Schwinger effect, which appears as a non-perturbative vacuum instability.

In this paper, we extend Ref.~\cite{Conzinu:2026cuf} in three main directions within electrodynamics.
First, we provide a detailed analysis of fermionic QED using three complementary approaches: perturbative EFT analysis, Borel--Laplace resummation, and a non-perturbative analysis based on the Schwinger proper-time method.
Second, we study scalar QED~\cite{Weisskopf:1936hya,Gupta:2023gsw,Ye:2025zhs} and DBI-type~\cite{Born:1934gh,Dirac:1962iy,Fradkin:1985qd,Leigh:1989jq,Tseytlin:1999dj} theories to further test the general bounds derived in Ref.~\cite{Conzinu:2026cuf} and generalized in this work.
Third, we extend the framework to EFTs with power-law growth, going beyond the factorial-growth scenario considered in Ref.~\cite{Conzinu:2026cuf}.

This paper is structured as follows.
In Section~\ref{sec:relative entropy}, we provide a general framework for deriving the bounds in nonlinear EFTs using relative entropy.
In Section~\ref{sec:qed}, we present a detailed analysis of nonlinear electrodynamics in fermionic QED. 
In Sections~\ref{sec:scalar qed} and~\ref{sec:DBI}, we extend the analysis of fermionic QED to scalar QED and the DBI model, respectively.
In Section~\ref{sec:gen}, we derive general consistency bounds for nonlinear electrodynamics EFTs with both factorial-growth and power-law large-order behavior.
Section~\ref{sec:summary} summarizes our results and discusses future directions.

\section{General framework}\label{sec:relative entropy}
In this section, we provide a general framework for deriving the bounds in nonlinear electrodynamics using relative entropy.
In Section~\ref{sec:setup}, we explain our motivation, outline the basic idea, and introduce a formal setup for mathematically quantifying it.
In Sections~\ref{sec:pert} and \ref{sec:inst}, we present two computational methods for evaluating the relative entropy and show that apparent violations of its non-negativity can arise as signals of non-perturbative effects in the theory.
In Section~\ref{sec:imp}, we explore the use of relative entropy to quantify both non-perturbative effects in the theory and the system’s unitarity.

\subsection{Formal setup}
\label{sec:setup}
Although the general framework of this work largely follows Refs.~\cite{Cao:2022iqh,Cao:2022ajt,Ueda:2024cyf,Conzinu:2026cuf}, we begin by briefly outlining our motivation and basic ideas.
At low energies, information about UV physics associated with heavy degrees of freedom can be encoded in an effective field theory (EFT) of light fields.
Interactions between heavy and light fields can imprint this UV information into higher-dimensional operators of the EFT.
This simple observation motivates us to quantify the UV information carried by such interactions and to study its universal properties.
In particular, the difference between theories with and without these interactions isolates the transferred UV information.
Quantifying this difference provides a direct way to characterize how UV physics manifests itself in the EFT.
These considerations form the motivation and conceptual foundation of our analysis.

Throughout this paper, following Refs.~\cite{Cao:2022iqh,Cao:2022ajt,Ueda:2024cyf,Conzinu:2026cuf}, we quantify the difference between theories with and without interactions between heavy and light fields using the information measure known as relative entropy (the Kullback–Leibler divergence~\cite{Kullback:1951zyt}).
The relative entropy is defined for two density operators as
\begin{align}
S(\rho_{\rm R}\|\rho_{\rm T}) \equiv {\rm Tr}\,\left[ \rho_{\rm R} \ln \rho_{\rm R} - \rho_{\rm R} \ln \rho_{\rm T} \right]\,,\label{eq:rel0}
\end{align}
where $\rho_{\rm R}$ and $\rho_{\rm T}$ denote positive semi-definite density operators satisfying
$\rho_{\rm R} = \rho_{\rm R}^{\dagger}$, $\rho_{\rm T} = \rho_{\rm T}^{\dagger}$, and ${\rm Tr}\,[\rho_{\rm R}] = {\rm Tr}\,[\rho_{\rm T}] = 1$. 
Similar to Refs.~\cite{Cao:2022iqh,Cao:2022ajt,Ueda:2024cyf,Conzinu:2026cuf}, we assume that $\rho_{\rm T}$ represents the density operator of the theory including the interaction term ({\it i.e.}, the target theory from which we want to extract information), and $\rho_{\rm R}$ represents the density operator of the theory without the interaction term ({\it i.e.}, the reference theory used to extract information from the target theory).
One of the essential properties of the relative entropy is its non-negativity:
\begin{align}
S(\rho_{\rm R}\|\rho_{\rm T}) \geq 0\,, \label{eq:non}
\end{align}
with equality holding if and only if $\rho_{\rm R} = \rho_{\rm T}$.
From Eq.~\eqref{eq:non}, the relative entropy can be interpreted as a measure of the information that distinguishes the two states $\rho_{\rm R}$ and $\rho_{\rm T}$.
It is worth emphasizing that the derivation of the non-negativity in Eq.~\eqref{eq:non} crucially depends on the following properties of the positive semi-definite density operators: $\rho_{\rm R} = \rho_{\rm R}^{\dagger}$, $\rho_{\rm T} = \rho_{\rm T}^{\dagger}$, and ${\rm Tr}\,[\rho_{\rm R}] = {\rm Tr}\,[\rho_{\rm T}] = 1$.

In this paper, we mainly consider UV theories related to electrodynamics, and thus focus on a scenario in which the only light degree of freedom is the photon field, while the other degrees of freedom are heavy.
Specifically, we assume that this class of theories can be described by the following Hamiltonian:
\begin{align}
    H= H_0 + H_{\rm I}\,,\label{eq:H}
\end{align}
where $H_0$ denotes the operator describing the system without interactions between the heavy and light degrees of freedom, and $H_{\rm I}$ denotes these interactions.
While $H_0$ contains interactions within the heavy sector and within the light sector, it excludes any interactions between the heavy and light sectors.
We assume that $H_0$ has a ground state, ensuring the stability of both heavy and light particles in the absence of heavy-light interactions.
To facilitate subsequent discussions, we introduce a parameter $\lambda$ that characterizes the interaction as follows:
\begin{align}
    H_\lambda\equiv H_0 +\lambda\, H_{\rm I}\,.\label{eq:H_lam}
\end{align}
The parameter $\lambda$ is introduced to track the order of perturbation in $H_{\rm I}$ and can be set to 1 after computing the relative entropy.
The information from the UV theory is transmitted to the EFT through the interaction $H_{\rm I}$ by integrating out the heavy degrees of freedom, and the difference between the theories with and without $H_{\rm I}$ represents the transferred information.
In Refs.~\cite{Cao:2022ajt,Cao:2022iqh,Ueda:2024cyf,Conzinu:2026cuf}, the relative entropy between two theories is computed using two density operators defined as follows:
\begin{align}
    \rho_{\rm R}\equiv \frac{e^{-\beta H_0}}{Z_0}\,,\qquad \rho_{\rm T}\equiv \frac{e^{-\beta H_\lambda}}{Z_\lambda}\,,\label{eq:Prob}
\end{align}
where the partition functions are defined as $Z_0\equiv {\rm Tr}\,[e^{-\beta H_0}]$ and $Z_\lambda\equiv {\rm Tr}\,[e^{-\beta H_\lambda}]$.
The parameter $\beta$ corresponds to the inverse temperature\footnote{However, there is freedom in the choice of the density operator corresponding to the theory when computing the relative entropy.
Depending on the information we wish to extract, the density operator can be adjusted accordingly.
Therefore, it is not necessary to interpret $\beta$ strictly as the inverse temperature.
} of the system if Eq.~\eqref{eq:Prob} is assumed to represent the canonical distribution of a thermodynamic system.
Under this standard definition, assuming that the theory satisfies unitarity, {\it i.e.}, that the Hamiltonians $H_0$ and $H_{\rm I}$ are Hermitian and that the corresponding density operators are positive semi-definite, the conditions $\rho_{\rm R} = \rho_{\rm R}^{\dagger}$ and $\rho_{\rm T} = \rho_{\rm T}^{\dagger}$ are in principle satisfied according to Eq.~\eqref{eq:Prob}.
The Hermiticity of these density operators ensures the non-negativity of the relative entropy~\eqref{eq:non}.
Importantly, the non-negativity of the relative entropy, when evaluated with the probability distribution functions in Eq.~\eqref{eq:Prob}, can be seen as a direct manifestation of unitarity.
Conversely, if this non-negativity is lost, it signals a breakdown of the Hermiticity of the density operators~\eqref{eq:Prob}, {\it i.e.}, a violation of the theory’s unitarity.
This correspondence between the non-negativity of relative entropy and unitarity plays a central role throughout this paper.

According to Eqs.~\eqref{eq:rel0} and \eqref{eq:Prob}, the relative entropy between $\rho_{\rm R}$ and $\rho_{\rm T}$ (representing the information transferred from the UV theory~\eqref{eq:H_lam} to the EFT) is calculated as follows:
\begin{align}
    S(\rho_{\rm R}\|\rho_{\rm T})&={\rm Tr}\,\left[
    \rho_{\rm R}\ln \rho_{\rm R}- \rho_{\rm R}\ln \rho_{\rm T}
    \right]\notag
    \\
    &=-\ln Z_0 +\ln Z_\lambda +\lambda\, {\rm Tr}\,\left[\rho_{\rm R}\beta H_{\rm I}\right]\notag
    \\
    &=W_0 -W_\lambda +\lambda \,\left(\frac{d W_\lambda}{d \lambda}\right)_{\lambda=0}\geq 0\,,\label{eq:quanrel}
\end{align}
where  $W_0\equiv -\ln Z_0$ and $W_\lambda\equiv -\ln Z_\lambda$ represent effective actions, and we performed several straightforward calculations, $\ln \rho_{\rm R}=-\beta H_0-\ln Z_0$, $\ln \rho_{\rm T}=-\beta H_\lambda-\ln Z_\lambda$, and used the non-negativity of relative entropy~\eqref{eq:non}.
Moreover, in the last line of Eq.~\eqref{eq:quanrel}, we have used the following relation:
\begin{align}
    \frac{d W_\lambda}{d\lambda}&=-\frac{1}{Z_\lambda}\frac{d Z_\lambda}{d\lambda}=-\frac{1}{Z_\lambda}\frac{d }{d\lambda}{\rm Tr}\, [e^{-\beta H_{\lambda}}]=-\frac{1}{Z_\lambda} {\rm Tr}\,\left[
    -\beta \left(\frac{d}{d\lambda} H_{\lambda}\right)e^{-\beta H_{\lambda}}
    \right]={\rm Tr}\,\left[\rho_{\rm T} \beta H_{\rm I}\right]\,,
\end{align}
and
\begin{align}
    \left(\frac{d W_\lambda}{d\lambda}\right)_{\lambda=0}&=\left({\rm Tr}\,\left[\rho_{\rm T} \beta H_{\rm I}\right]\right)_{\lambda=0}={\rm Tr}\,\left[\rho_{\rm R} \beta H_{\rm I}\right]\,,
\end{align}
where $\lim_{\lambda\to 0}\rho_{\rm T}=\rho_{\rm R}$.
In particular, throughout this paper, we focus on the limit $\beta \to \infty$ in the computations of the relative entropy~\eqref{eq:quanrel}, in order to clarify its connection with the effective action at zero temperature.
From Eq.~\eqref{eq:quanrel}, the relative entropy can be expressed in terms of the effective actions, so its calculation reduces to that of the effective action.
In the present work, we restrict our attention to low-energy regions where the heavy degrees of freedom are non-dynamical.
Accordingly, the effective action $W_\lambda$, defined for the background fields of the light sector, represents the EFT corresponding to $H_\lambda$ within this low-energy domain.
In this context, the inequality~\eqref{eq:quanrel} implies that the low-energy EFT of the theory $H_\lambda$ must satisfy a specific relation.
In our applications, this relation corresponds to the bounds on the electromagnetic EFT, as discussed in later sections.

As emphasized above, the non-negativity of the relative entropy follows from the unitarity of the system ({\it i.e.}, the Hermiticity of the density operators) when the density operators are defined as in Eq.~\eqref{eq:Prob}.
Formally, if the system’s Hamiltonian is Hermitian, this non-negativity holds under the standard definition of the density operators.
However, in practice, this non-negativity can be violated due to the computational techniques used to evaluate the relative entropy or a physical instability of the systems, as explained in the following subsections.
In this paper, we focus on two widely used methods: perturbative analysis, which expands in powers of the interaction and is typically truncated at finite order, and analytic-continuation-based computations, which are necessary when dealing with unstable systems.
Consequently, assuming that the system’s Hamiltonian is Hermitian, the observed violations of non-negativity can be attributed to the specific treatment adopted in these computational techniques.
As will be seen, within these frameworks, such violations signal the presence of non-perturbative effects in the theory.

    \subsection{Perturbative analysis} 
    \label{sec:pert}
    We begin with a perturbative analysis, which is one of the two computational techniques mentioned above.
    The effective action $W_{\lambda}$ appearing in the relative entropy~\eqref{eq:quanrel} is often evaluated perturbatively, {\it i.e.}, by expanding in powers of the interaction and typically truncating the series at finite order.
    The perturbative analysis is applicable, and the relative entropy can be reliably computed in a weakly coupled theory, {\it i.e.}, one in which both the loop expansion and the operator expansion are valid (see Ref.~\cite{Ueda:2024cyf} for further details).
    However, when the interaction becomes strong, this truncation can lead to inaccurate results for the relative entropy, and its non-negativity may be violated even if the Hamiltonian of the system is Hermitian.
    This situation has been extensively examined in Ref.~\cite{Ueda:2024cyf}, with a focus on scalar field theories at tree level.

    To illustrate that perturbative analyses may produce spurious non-Hermitian density operators, we present a perturbative analysis of the relative entropy in this subsection.
Let us consider the density operator $\rho_{\rm T}$ in Eq.~\eqref{eq:Prob}.
By expanding in powers of the interaction, we can formally rewrite
\begin{align}
    e^{-\beta H_{\lambda}}= \sum_{n=0}^\infty \lambda^n h^{(n)}\quad \Rightarrow \quad \beta H_{\lambda}= \ln \left(\sum_{n=0}^\infty \lambda^n h^{(n)}\right)^{-1}\,.
\end{align}
By truncating at a finite order $N$, we define the perturbative Hamiltonian as
\begin{align}
    \widetilde{H}_{\lambda}\equiv  \ln \left(\sum_{n=0}^N \lambda^n h^{(n)}\right)^{-1}\,,\qquad e^{-\widetilde{H}_{\lambda}}=\sum_{n=0}^N \lambda^n h^{(n)}\,,\label{eq:Htilde}
\end{align}
with $\widetilde{H}_0 =\beta H_0$.
For later convenience, we also provide a useful relation: 
the interaction term in the Hamiltonian can be rewritten as
\begin{align}
    \lambda H_{\rm I}= H_{\lambda}-H_0=\beta^{-1}\left[\ln \left(e^{-\beta H_{0}}\right)-\ln \left(e^{-\beta H_{\lambda}}\right)\right]\,.\label{eq:form_per}
\end{align}
Since the left-hand side is of first order in $\lambda$, the terms of order higher than the first in $\lambda$ appearing on the right-hand side of Eq.~\eqref{eq:form_per} vanish.
Thus, Eq.~\eqref{eq:form_per} also holds even when $e^{-\beta H_{\lambda}}$ is expanded in powers of the interaction and truncated at a finite order $N$.  By the truncation at a finite order $N$, we therefore find
\begin{align}
    \lambda H_{\rm I}=\beta^{-1}\left[\ln \left(e^{-\widetilde{H}_{0}}\right)-\ln \left(e^{-\widetilde{H}_{\lambda}}\right)\right]+\mathcal{O}\left(\lambda^{N+1}\right)\,.\label{eq:fom1}
\end{align}
From Eq.~\eqref{eq:Htilde}, one can perturbatively define the density operator corresponding to the target theory as
\begin{align}
    \widetilde{\rho}_{\rm T}\equiv \frac{e^{-\widetilde{H}_{\lambda}}}{\widetilde{Z}_{\lambda}}\,,\qquad  \widetilde{Z}_{\lambda}\equiv {\rm Tr}\,\left[e^{-\widetilde{H}_{\lambda}}\right]\,.
\end{align}
From these constructions, it follows
\begin{align}
    \left(\widetilde{\rho}_{\rm T}\right)_{\lambda=0}= \rho_{\rm R}\,.
\end{align}
Similar to Eq.~\eqref{eq:quanrel}, the relative entropy for the density operators $\widetilde{\rho}_{\rm T}$ and $\widetilde{\rho}_{\rm R}\equiv {\rho}_{\rm R}$ is calculated as follows:
\begin{align}
    S\left(\widetilde{\rho}_{\rm R}\|\widetilde{\rho}_{\rm T}\right)&= {\rm Tr}\,\left[
    \widetilde{\rho}_{\rm R}\ln\widetilde{\rho}_{\rm R}-\widetilde{\rho}_{\rm R} \ln\widetilde{\rho}_{\rm T}
    \right]\notag
    \\
    &=-\ln \widetilde{Z}_0+ \ln \widetilde{Z}_\lambda + {\rm Tr}\,\left[
    \widetilde{\rho}_{\rm R} \left(
    \ln \left(e^{-\widetilde{H}_0}\right)
    -
    \ln \left(e^{-\widetilde{H}_\lambda}\right)
    \right)
    \right]\notag
    \\
    &=\widetilde{W}_{0}-\widetilde{W}_{\lambda}+{\rm Tr}\,\left[\rho_{\rm R} \beta\lambda H_{\rm I}\right]+\mathcal{O}\left(\lambda^{N+1}\right)\notag
    \\
    &=\widetilde{W}_{0}-\widetilde{W}_{\lambda}+\lambda\, \left(\frac{d\widetilde{W}_{\lambda}}{d\lambda}\right)_{\lambda=0}+\mathcal{O}\left(\lambda^{N+1}\right)\,,\label{eq:rel_pert}
\end{align}
where we have used $\widetilde{W}_{\lambda}\equiv -\ln \widetilde{Z}_\lambda$, Eq.~\eqref{eq:fom1}, and $({d\widetilde{W}_{\lambda}}/{d\lambda})_{\lambda=0}=\left({d{W}_{\lambda}}/{d\lambda}\right)_{\lambda=0}={\rm Tr}\,\left[ \rho_{\rm R}\beta H_{\rm I}\right]$.
According to the definition of the density operators, the relative entropy in Eq.~\eqref{eq:rel_pert} coincides with Eq.~\eqref{eq:quanrel} to finite order $N$.
From Eq.~\eqref{eq:Htilde}, the Hermiticity of $e^{-\tilde{H}_{\lambda}}$ may be violated when $\sum_{n=0}^N \lambda^n h^{(n)}$ develops negative eigenvalues, signaling the breakdown of the perturbative expansion.
Such breakdowns indicate a loss of non-negativity in the relative entropy and can be interpreted as the onset of non-perturbative effects beyond the regime of validity of perturbation theory.
As discussed in Section~\ref{sec:DBI}, a similar violation can occur when $H_{\rm I}$ depends on a physical parameter characterizing the interaction, such as $g_e^{-2}$, rather than on an artificial expansion parameter like $\lambda$, and the corresponding series is truncated at finite order in $g_e^{-2}$.
These observations imply that, in a unitary theory, the non-negativity of the relative entropy is recovered once the perturbative expansion is properly resummed to all orders.
However, as we explain below, resummation can incorporate information beyond the perturbative expansion, such as dynamical instabilities of the system, potentially leading to a violation of the non-negativity of the relative entropy.
The resummation will be examined in later sections using, for example, Borel–Laplace techniques.

    \subsection{System instabilities}
    \label{sec:inst}
    Next, we consider the computation using analytic continuation in an unstable system, which is the second computational technique employed in this paper.
    An unstable system typically features a Hamiltonian that is unbounded from below or lacks a well-defined ground state. 
    As a result, the partition function may diverge, requiring special techniques for its evaluation.
In such cases, analytic continuation can serve as a powerful tool for evaluating the partition function.
Suppose we want to compute the partition function $Z_{\rm unstable}$ of an unstable system described by the Hamiltonian $H_{\rm unstable}$.
If $H_{\rm unstable}$ can be obtained by analytically continuing the Hamiltonian of a related stable system, $H_{\rm stable}$, which possesses a ground state, we proceed as follows. 
First, we compute the partition function $Z_{\rm stable}$.
Then, through analytic continuation, we can obtain the desired partition function $Z_{\rm unstable}$.
However, this procedure may lead to an effective non-Hermitian description of the density operator in the analytically continued theory, thereby potentially violating the assumptions underlying the positivity of relative entropy.
Since the energy spectrum obtained through analytic continuation generally contains an imaginary part, the resulting unstable system can be viewed as admitting an effective non-Hermitian description.

   For clarity, we present a simple example illustrating how the system’s instability manifests via the spurious non-Hermitian density operator that arises through analytic continuation.
Let us consider a one-dimensional stable harmonic oscillator, defined by the following Hamiltonian:
\begin{align}
    H_{\text{stable}}\equiv\frac{p^2}{2m}+\frac{1}{2}m\omega^2 x^2\,.
\end{align}
The eigenvalues of this system are $E_n = \omega \left(n + 1/2\right)$ (with $\hbar = 1$ assumed throughout this paper), and the partition function is then computed as follows:
\begin{align}
    Z_{\rm stable}\equiv {\rm Tr}\, \left[e^{-\beta H_{\text{stable}}}\right]=\sum_{n=0}^{\infty} e^{-\beta \omega (n+1/2)}=\frac{e^{-\beta \omega/2}}{1-e^{-\beta \omega}}=\frac{1}{2\sinh\left(\beta \omega/2\right)}\,.\label{eq:sta_Z}
\end{align}
Building on these results, we consider an inverted harmonic oscillator, which serves as an example of an unstable system:
\begin{align}
    H_{\text{unstable}}\equiv\frac{p^2}{2m}-\frac{1}{2}m\Omega^2 x^2\,.
\end{align}
Since this system lacks a ground state, the partition function $Z_{\rm unstable} \equiv {\rm Tr}\, \left[e^{-\beta H_{\text{unstable}}}\right]$ diverges and is ill-defined.
Nevertheless, by performing the analytic continuation $\omega \to i\Omega$, following Eq.~\eqref{eq:sta_Z}, we can obtain the partition function of this unstable system as
\begin{align}
    Z_{\rm stable}\to Z_{\rm unstable}=\frac{1}{2\sinh\left(\beta i\Omega/2\right)}=-\frac{i}{2 \sin \left(\beta \Omega/2\right)}\,.
\end{align}
This imaginary part arises from the imaginary energy eigenvalues and thus clearly characterizes the system’s instability.
It should be emphasized that in analytic-continuation-based computations, necessitated by the instability, the partition function may acquire an imaginary part even when the Hamiltonian $H_{\text{unstable}}$ is Hermitian.
In such cases, the Hermiticity of the density operator is violated.
Refs.~\cite{Cao:2022iqh,Cao:2022ajt} also showed that the non-negativity of the relative entropy can be violated in unstable systems.
In QED, the physical partition function obtained by analytically continuing the Euclidean path integral, which describes a stable system, develops an imaginary part associated with the Schwinger effect, even though the underlying QED Hamiltonian is Hermitian.

In the cases considered throughout this paper ({\it i.e.}, where $H_0$ in Eq.~\eqref{eq:H} is assumed to have a ground state\footnote{For instance, when $H_0$ in Eq.~\eqref{eq:H} represents the Hamiltonian of {\it particles}, the existence of a ground state is automatically assumed. This is because, by definition, the particle picture presupposes a vacuum (ground) state.
A potential exception is, for example, the instability arising from self-interactions in $H_0$.
}), as in the first argument of Section~\ref{sec:pert}, the second argument for the violation of non-negativity discussed above ultimately stems from the breakdown of perturbation theory.
When $H_0$ possesses such a ground state, the perturbation due to the interaction $H_{\rm I}$ is stable, meaning that the ground state remains well defined.
In other words, since the perturbative corrections remain finite and $H_0$ has a ground state, the ground state of $H_\lambda$ is guaranteed within perturbative analysis.
The key point is that, in our setup ({\it i.e.}, $H_0$ in Eq.~\eqref{eq:H} has the ground state), the spurious loss of Hermiticity arising from analytic continuation is related to a genuinely non-perturbative physical effect.
Consequently, the second type of non-negativity violation is linked to the breakdown of perturbative analysis.
As will be explained later, in QED the spurious violation of Hermiticity due to analytic continuation originates from a non-perturbative effect, namely the Schwinger effect.
Conversely, once we perform the perturbative expansion for such an unstable system, purely non-perturbative effects do not appear.
Therefore, it is naively expected that the perturbative expansion, when properly resummed using techniques such as Borel-Laplace resummation, remains stable.
However, as will be shown, the resummed perturbative expansion can fail to respect the non-negativity of the relative entropy in the strong-coupling regime and signals non-perturbative effects such as the instability of the system.
In other words, the sign of the relative entropy may probe the non-perturbative physical instability solely from the perturbative expansion.
This will be illustrated, for instance, in Section~\ref{sec:Borel_fer}.

\subsection{Interpretation of relative entropy: preservation and violation of non-negativity}
\label{sec:imp}
As explained above, in this paper we employ two widely used methods to compute the relative entropy.
Although the non-negativity of relative entropy is ensured under the standard definition of the density operator for a Hermitian Hamiltonian, the computational methods employed here may still yield apparent violations.
Furthermore, if $H_0$ in Eq.~\eqref{eq:H} possesses a ground state, these violations of non-negativity signal the presence of non-perturbative effects, either purely non-perturbative effects or those arising from a breakdown of the perturbative analysis.
Motivated by this observation, we propose to use the violation or preservation of the non-negativity of the relative entropy, computed using the two methods, as an indicator of non-perturbative effects in the theory.

Owing to the two computational methods presented in Sections~\ref{sec:pert} and \ref{sec:inst}, two distinct types of spurious non-Hermiticity of $\rho_{\rm T}$ may arise, as described below:

\begin{enumerate}[(1)]
    \item {\bf Hermiticity of $\rho_{\rm T}$ is violated in the perturbative analysis} --- As discussed in Section~\ref{sec:pert}, the perturbative analysis, including both the loop and operator expansions in field-theoretical calculations, breaks down, signaling that the interaction $H_{\rm I}$ resides in the non-perturbative regime.
    In this case, the non-negativity of the relative entropy may be violated, and the resulting violation can serve as a signal of non-perturbative effects ({\it i.e.}, a breakdown of the perturbative analysis).
    Conversely, if $\rho_{\rm T}$ remains Hermitian in the perturbative analysis, the relative entropy remains non-negative, reflecting the unitarity inherent in the perturbative framework. 

    \item {\bf Hermiticity of $\rho_{\rm T}$ is violated due to the instability of the system} --- As discussed in Section~\ref{sec:inst}, for unstable theories, analytic continuation from a related stable system provides a powerful computational technique, at the cost of potentially violating the Hermiticity of the density operator due to the underlying instability.
    Since the instability under the condition that $H_0$ is stable originates from non-perturbative effects of $H_{\rm I}$, the violation of the non-negativity of the relative entropy serves as an indicator of such effects.
    Conversely, if the system is stable ($\rho_{\rm T}$ remains Hermitian), the relative entropy remains non-negative, reflecting the system’s unitarity.
    
\end{enumerate}

    In both cases (1) and (2), corresponding to the two computational techniques adopted in this paper, the violation of non-negativity of the relative entropy is associated with {\it non-perturbative} effects in the theory.
    It should be emphasized that the apparent violation of non-negativity due to (1) is resolved once the perturbative expansion is properly resummed to all orders.
    However, the apparent violation in (2) may become more pronounced after a proper all-order resummation of the perturbative expansion.
    For further details, see, {\it e.g.}, Sections~\ref{sec:Borel_fer} and \ref{sec:nonpe_fer}.
    In the following sections, we consider fermionic and scalar QED as illustrative examples of the formal results discussed above.

\section{Fermionic QED}\label{sec:qed}
As a first case study, we examine a charged fermion in electromagnetic fields.
By analyzing the system both perturbatively and non-perturbatively, we find that the positivity of the relative entropy corresponds to the positivity bounds on nonlinear corrections to Maxwell theory, whereas a violation of its non-negativity signals non-perturbative effects such as the Schwinger effect.

\subsection{Setup and conventions}

We consider the Lagrangian in Minkowski spacetime:
\begin{align}
    \mathcal{L}_{\lambda}=-\frac{1}{4}F_{\mu\nu}F^{\mu\nu} +\overline{\psi} \left(i\slashed{D}-m\right)\psi\,, \label{eq:qed_fem}
\end{align}
with
\begin{align}
    \slashed{D} = \gamma^{\mu} D_{\mu}\,,\quad 
D_{\mu} = \partial_{\mu} + i \lambda e A_{\mu}\,,\quad \overline{\psi}=\psi^\dagger \gamma^0\,.
\end{align}
Our conventions are 
$g_{\mu\nu} = (+,-,-,-)$; 
Greek (Latin) indices run over $0\text{--}3$ ($1\text{--}3$); 
and $\{\gamma^{\mu}, \gamma^{\nu}\} = 2 g^{\mu\nu}$.
We denote Minkowski and Euclidean gauge fields by $A_\mu$ and $A^{\rm E}_I$, respectively.
As explained in Section~\ref{sec:setup}, the parameter $\lambda$ keeps track of the perturbative order, and we set $\lambda = 1$ after computing the relative entropy.
The density operator associated with the Hamiltonian $H_{\lambda}[A^{\rm cl}]$ corresponding to the Lagrangian~\eqref{eq:qed_fem} is given by
\begin{align}
    \rho_{\rm T} =  \frac{1}{{Z}_{\lambda}[A^{\rm cl}]}\,\lim_{\beta\to \infty}e^{-\beta H_{\lambda}[A^{\rm cl}]}\,,\qquad Z_{\lambda}[A^{\rm cl}]=\lim_{\beta\to \infty}{\rm Tr}\,\left[e^{-\beta H_{\lambda}[A^{\rm cl}]}\right]\,,\label{eq:part_0}
\end{align}
where the trace is taken over the full Hilbert space.
The background gauge field $A_\mu^{\rm cl}$ denotes a stationary Minkowski configuration.

As is commonly done, we evaluate the partition function $Z_{\lambda}[A^{\rm cl}]$ through the following steps:
(i) We introduce the Euclidean counterpart $Z_{\lambda}^{\rm E}[A^{\rm E, cl}]\equiv \lim_{\beta\to \infty}{\rm Tr}\left[e^{-\beta H_{\lambda}[A^{\rm E,cl}]}\right]$ of $Z_{\lambda}[A^{\rm cl}]$ and perform the Euclidean path integral\footnote{In Euclidean spacetime, the Hamiltonian of fermionic QED is generally not Hermitian due to $A_4^{\rm E}\in\mathbb{R}$.
A key point is that, in the purely magnetic case, the Hamiltonian is Hermitian in both Minkowski and Euclidean spacetimes, and the instability of the system is inferred through analytic continuation from this magnetic background.
}, where all fields in Eq.~\eqref{eq:qed_fem} are defined in Euclidean spacetime ($A_I^{\rm E, cl}$ is real). 
As will be seen, this Euclidean path integral corresponds to the stable system discussed in Section~\ref{sec:inst}.
(ii) We then obtain $Z_{\lambda}[A^{\rm cl}]$ by analytically continuing the Euclidean field $A_I^{\rm E, cl}$ to the Minkowski-space field $A_{\mu}^{\rm cl}$, so that the real field $A^{\rm E, cl}$ generally becomes complex.
It should be noted that the Hamiltonian $H_{\lambda}[A^{\rm cl}]$ is expressed in terms of the physical electromagnetic fields $A_\mu^{\rm cl}$ in Minkowski spacetime.
Following steps (i) and (ii), we evaluate the partition function $Z_{\lambda}[A^{\rm cl}]$ (see Appendix~\ref{sec:QED in E-space} for further details). 

\begin{enumerate}[{\bf (i)}]
    \item {\bf Euclidean path integral} --- Using the gauge field $A^{\rm E}$ in Euclidean spacetime, we express the partition function as
\begin{align}
    Z^{\rm E}_\lambda[A^{\rm E, cl}]=\int \mathcal{D}A^{\rm E}\, z^{\rm E}_\lambda[A^{\rm E}]\,,\qquad z^{\rm E}_\lambda[A^{\rm E}]\equiv \lim_{\beta\to \infty}{\rm Tr}_{\text{partial}}\left[e^{-\beta H_{\lambda}[A^{\rm E,cl}]}\right]\,,\label{eq:def_z_fem}
\end{align}
where the partial trace ${\rm Tr}_{\text{partial}}$ is taken over all degrees of freedom except the gauge field $A_I^{\rm E}$.
In what follows, $A^{\rm E, cl}$ denotes the stationary configurations, whereas $A^{\rm E}$ denotes the integration variables.
In the one-loop calculations considered in this paper, where photon loops are neglected, the path integral over $A_I^{\rm E}$ in Eq.~\eqref{eq:def_z_fem} contributes only via the stationary configurations\footnote{Consequently, the Euclidean path integral~\eqref{eq:def_z_fem} is evaluated under the background field $A^{\rm E, cl}$.
In this procedure, integrating over $A^{\rm E}$ automatically accounts for the wave function renormalization of the gauge fields via the stationary configurations.
} $A_I^{\rm E, cl}$ to the partition function $Z^{\rm E}_{\lambda}[A^{\rm E, cl}]$.

In Euclidean spacetime, the partition function can be written as
\begin{align}
    Z_{\lambda}^{\rm E}[A^{\rm E,cl}]&=e^{-\int d^4x_{\rm E}(F_{IJ})^2/4}\,{\rm det}\,{\rm M}_{\lambda}[A^{\rm E,cl}]\,,\label{eq:ZA_fm}
\end{align}
where $F_{IJ}=\partial_I A_J^{\rm E,cl}-\partial_J A_I^{\rm E,cl}$, and 
\begin{align}
{\rm M}_\lambda [A]\equiv \gamma^I_{\rm E} \left(\partial_I+i\lambda e A_I\right) +m\,.
\end{align}
We adopt the following conventions for the Euclidean gamma matrices $\gamma^I_{\rm E}$: $\gamma^4_{\rm E}\equiv\gamma^0$, and $\gamma^i_{\rm E}=-i\gamma^i$ for $i=1, 2, 3$, using the Weyl representation of the gamma matrices (see Appendix~\ref{sec:QED in E-space}).
From these definitions, it follows that $\{\gamma^I_{\rm E}, \gamma^J_{\rm E}\}=2\delta^{IJ}$ and $(\gamma^I_{\rm E})^\dagger=\gamma^I_{\rm E}$, for $I,J=1,2,3,4$.
From Eq.~\eqref{eq:ZA_fm}, the Euclidean effective action is calculated as
\begin{align}
    W_{\lambda}^{\rm E}[A^{\rm E,cl}]&\equiv-\ln Z^{\rm E}_{\lambda}[A^{\rm E,cl}]=\int d^4x_{\rm E}\,\left[\frac{1}{4}(F_{IJ})^2\right]-{\rm Tr}\,\left[\ln \left(\slashed{D}_{\rm E}+m\right)\right]\,,\label{eq:WA}
\end{align}
where ${\rm Tr}$ denotes the trace, including both the spacetime integration $\int d^4x_{\rm E}$ and the trace over Dirac indices, while ${\rm tr}_{\rm D}$ denotes the trace over Dirac indices only, and $\slashed{D}_{\rm E}\equiv\gamma^I_{\rm E}D_I$ with $D_I\equiv\partial_I+i \lambda e A_I^{\rm E,cl}$.
Using the relation
\begin{align}
    \frac{d W_\lambda^{\rm E}[A^{\rm E,cl}]}{dm^2}=-\frac{1}{2}\int d^4 x_{\rm E}\, {\rm tr}_{\rm D}\left[{\langle x|\frac{1}{\hat{H}_\lambda+m^2}|x\rangle}\right]\,,\label{eq:dWdm2} 
\end{align}
with the positive semidefinite operator $\hat{H}_\lambda\equiv\left((\hat{p}+\lambda e A^{\rm E,cl})_I \gamma^I_{\rm E}\right)^2$, we introduce the Schwinger proper time $s$ via the formula $1/\mathcal{A}=\int_0^\infty ds\, e^{-s\mathcal{A}}$ for real $\mathcal{A}>0$.
It should be emphasized that from Eq.~\eqref{eq:dWdm2}, the eigenvalues of $\hat{H}_\lambda+m^2$ are positive and no poles appear; {\it i.e.}, fermionic QED in Euclidean spacetime is a {\it stable} system\footnote{
In Euclidean spacetime, all external fields behave effectively as magnetic fields.
Therefore, for the electric model defined by an electric field $\vec{E}$, the analytic continuation from Euclidean to Minkowski spacetime can be viewed as the continuation from a magnetic model with magnetic field $\vec{B}$ to an imaginary magnetic field $i\vec{E}$, where $\vec{E}$ is identified with the physical electric field in Minkowski spacetime.
As a consequence, the analytically continued theory is described by a non-Hermitian Hamiltonian and Lagrangian due to the presence of the imaginary magnetic field.
}.
This stability relies on the Euclidean gauge field $A^{\rm E}_I$ taking real values.
Therefore, the Euclidean QED system serves as the stable system mentioned in Section~\ref{sec:inst}.
Using the Schwinger proper time method, we obtain
\begin{align}
W^{\rm E}_{\lambda}[A^{\rm E,cl}]&=\int d^4x_{\rm E}\,\left[\mathcal{F}+\frac{(\lambda e)^2}{8\pi^2}
    \int_0^{i\infty}ds\, \frac{e^{-sm^2}}{s} \frac{\cosh\left(\lambda es\, x_+\right)+\cosh\left(\lambda es\, x_-\right)}{\cosh(\lambda es\,x_+)-\cosh(\lambda es\,x_-)} \,\mathcal{G} \right]\,,\label{eq:WAe_fer}
\end{align}
where $x_\pm\equiv\sqrt{2\left(\mathcal{F}\pm \mathcal{G}\right)}$ with $\mathcal{F}\equiv(F_{IJ})^2/4$, $\mathcal{G}\equiv-F_{IJ}\widetilde{F}^{IJ}/4$, and $\widetilde{F}^{IJ}\equiv\epsilon^{IJKL}F_{KL}/2$. 

In Euclidean spacetime, the operator $\hat{H}_\lambda$ is positive semidefinite, and there are no poles in the complex $s$-plane.
Therefore, the integration contour can be deformed from $[0,+\infty)$ to $[0,+i\infty)$ by Cauchy's theorem.
However, as discussed below, after analytically continuing the electromagnetic fields, poles appear due to the electric field.
To remove the ambiguity in the choice of the $s$-integration contour, we introduce the standard $i\epsilon$ prescription,
$\hat{H}_\lambda\to \hat{H}_\lambda-i\epsilon^+$ with $\epsilon^+>0$,
which defines the forward-time evolution.

    \item {\bf Analytic continuation} --- By analytically continuing the Euclidean gauge fields in Eq.~\eqref{eq:WAe_fer} to the physical electromagnetic fields defined in Minkowski spacetime, we compute the effective action and the partition functions appearing in the relative entropy~\eqref{eq:quanrel}.
    This is formally achieved by the following analytic continuation:
    \begin{align}
    W_{\lambda}[A^{\rm cl}]=W^{\rm E}_{\lambda}[A^{\rm E,cl}]\bigg|_{F_{4k} \to -i F_{0k}=-i E^k}\,,\label{eq:W_an}
    \end{align}
    where $E^k\in  \mathbb{R}$ for $k=1,2,3$.
    Under this continuation, the Euclidean antisymmetric tensor $F_{IJ} = \partial_I A_J^{\rm E} - \partial_J A_I^{\rm E}$ can be expressed in terms of the physical electric and magnetic fields $\vec{E}$ and $\vec{B}$ in Minkowski spacetime as
\begin{align}
    F_{23}=-B_x\,,\quad F_{31}=-B_y\,,\quad F_{12}=-B_z\,,\quad F_{4k}=-i E^k\,,
\end{align}
with $E^1=E_x$, $E^2=E_y$, and $E^3=E_z$.  
Consequently, $\mathcal{F}$ and $\mathcal{G}$ in Eq.~\eqref{eq:WAe_fer} can be expressed in terms of the physical electromagnetic fields as
\begin{align}
    \mathcal{F}=\frac{1}{2}\left(\vec{B}^2-\vec{E}^2\right)\, (1+(\lambda e)^2\,\delta_{\rm F})^{-1}\,,\qquad \mathcal{G}=-i \left(\vec{E}\cdot \vec{B}\right)\, (1+(\lambda e)^2\,\delta_{\rm F})^{-1}\,,\label{eq:ferm_GF}
\end{align}
where $\delta_{\rm F} \equiv \tfrac{1}{12\pi^2}\int_0^{\infty} ds\,\tfrac{1}{s} e^{-sm^2}$ represents the wavefunction renormalization (WFR) effect on the gauge fields and $\vec E$ and $\vec B$ denote the physical fields after the WFR has been implemented.
This WFR is implemented through the choice of stationary configurations of the gauge fields and removes the divergences appearing in the effective action and, consequently, in the relative entropy.
Note that the electric field is a potential source of an imaginary part and the resulting instability of the system, whereas the magnetic field is not.
This is because the positive semidefiniteness of $\hat{H}_\lambda+m^2$ in Eq.~\eqref{eq:dWdm2} can be violated by the imaginary part generated by the analytic continuation of the electric field.

\end{enumerate}

From steps (i) and (ii), we obtain the effective action~\eqref{eq:W_an}, which enters into the relative entropy~\eqref{eq:quanrel}:
\begin{align}
    W_\lambda [A^{\rm cl}]&=\int d^4x_{\rm E}\,\left[\epsilon_{\rm F} +\frac{1}{2}\left(\vec{B}^2-\vec{E}^2\right)- \mathcal{L}_{\rm F}\right]\,,\label{eq:W_nonlin_ferm}
\end{align}
where $\epsilon_{\rm F} \equiv \tfrac{1}{8\pi^2}\int_0^{\infty} ds\,\tfrac{1}{s^3} e^{-sm^2}$ denotes the vacuum energy, and the one-loop nonlinear effects in Maxwell theory are given by
\begin{align}
    \mathcal{L}_{\rm F}&=
    \frac{(\lambda e)^2}{32\pi^2}\int_0^{i\infty} ds\, \frac{e^{-sm^2}}{s}\left[\frac{\cosh\left(\lambda es\, x_+\right)+\cosh\left(\lambda es\, x_-\right)}{\cosh(\lambda es\,x_+)-\cosh(\lambda es\,x_-)}  \left(i4 \vec{E}\cdot \vec{B}\right)\right]
    \notag\\
    &\quad
    +\frac{(\lambda e)^2\,\delta_{\rm F}}{2}\left(\vec{B}^2-\vec{E}^2\right)\,,
\end{align}
with $x_\pm\equiv\sqrt{\vec{B}^2-\vec{E}^2\mp i 2\,\vec{E}\cdot \vec{B}}$.
In what follows, we combine Eq.~\eqref{eq:W_nonlin_ferm} with Eq.~\eqref{eq:quanrel} and evaluate the relative entropy using three different approaches, namely perturbative analysis in Section~\ref{sec:pe_fer}, Borel--Laplace resummation in Section~\ref{sec:Borel_fer}, and non-perturbative analysis in Section~\ref{sec:nonpe_fer}.

\subsection{Perturbative analysis}
\label{sec:pe_fer}
First, we compute the relative entropy using perturbation theory, expanding in powers of the interaction and truncating at a finite order.
As shown in Eq.~\eqref{eq:rel_pert}, the perturbative analysis of the relative entropy reduces to the perturbative evaluation of the effective action $\widetilde{W}_\lambda$, obtained by applying the same truncation procedure to $W_\lambda$ in Eq.~\eqref{eq:quanrel}.
We begin by expanding Eq.~\eqref{eq:W_an} in powers of $\lambda$.
This leads to:
\begin{align}
    W_{\lambda}[A^{\rm cl}]&=\int d^4x_{\rm E}\,\left[
   \epsilon_{\rm F}
    +\left(1+(\lambda e)^2\,\delta_{\rm F}\right)\mathcal{F}-\frac{(\lambda e)^4}{32\pi^2} \int_0^{\infty}ds\,\frac{e^{-sm^2}}{s}
    \left(
    \frac{4}{45}s^2 \left(4\mathcal{F}^2 -7 \mathcal{G}^2\right)
    +\cdots
    \right)
    \right]\,,
\end{align}
where $\mathcal{F}$ and $\mathcal{G}$ are defined in Eq.~\eqref{eq:ferm_GF}, and $\epsilon_{\rm F}$ denotes the vacuum energy.
Consequently, at the one-loop level, we obtain:
\begin{align}
    W_{\lambda}[A^{\rm cl}]=\int d^4x_{\rm E}\,\bigg[
    \epsilon_{\rm F}
    +
    \frac{1}{2}\left(\vec{B}^2-\vec{E}^2\right)
    -
    \underbrace{\frac{(\lambda e)^4}{360\pi^2 m^4}
    \left\{
    \left(\vec{B}^2-\vec{E}^2\right)^2
    +
    7\left(\vec{E}\cdot \vec{B}\right)^2
    \right\}
    +\cdots}_{= \mathcal{L}_{\rm F}~\text{in Eq.~\eqref{eq:W_nonlin_ferm}}}
    \bigg]\,,\label{eq:frm_Wlam}
\end{align}
where the WFR effect has been absorbed into the stationary configuration~\eqref{eq:ferm_GF}.
We also obtain from Eq.~\eqref{eq:frm_Wlam}:
\begin{align}
    \left(\frac{d W_{\lambda}[A^{\rm cl}]}{d \lambda}\right)_{\lambda=0}=0\,,\qquad W_{\lambda=0}[A^{\rm cl}]=\int d^4x_{\rm E}\,\left[
    \epsilon_{\rm F}
    +
    \frac{1}{2}\left(\vec{B}^2-\vec{E}^2\right)
    \right]\,. \label{eq:frm_dWlam}
\end{align}
By substituting Eqs.~\eqref{eq:frm_Wlam} and \eqref{eq:frm_dWlam} into Eq.~\eqref{eq:quanrel} and setting $\lambda=1$, we calculate the relative entropy as 
\begin{align}
    S\left(\rho_{\rm R}\|\rho_{\rm T}\right)
    &=
    W_{\lambda=0}[A^{\rm cl}]
    -
    W_{\lambda=1}[A^{\rm cl}]
    +
    \left(\frac{d W_{\lambda}[A^{\rm cl}]}{d \lambda}\right)_{\lambda=0}
    \notag\\
    &=
    \int d^4 x_{\rm E}\, \mathcal{L}_{\rm F}
    \notag\\
    &=
    \frac{e^4}{360\pi^2 m^4}
    \int d^4x_{\rm E}\,
    \left[
    \left(\vec{B}^2-\vec{E}^2\right)^2
    +
    7\left(\vec{E}\cdot \vec{B}\right)^2
    \right]
    +\cdots
    \geq 0\,.
    \label{eq:per_ferm_EB}
\end{align}
At leading order, the integrand is positive semidefinite for real physical electromagnetic fields. Therefore, the non-negativity of the relative entropy is explicitly preserved at this order.
It should be emphasized that the relative entropy is equivalent to the nonlinear effects $ \mathcal{L}_{\rm F}$ in Maxwell theory~\eqref{eq:W_nonlin_ferm}.
This follows naturally, since the relative entropy quantifies the difference between two theories, and in the present case, the difference between theories with and without interactions between heavy and light particles corresponds to the higher-dimensional operators at low energies.
Thus, the positivity of the relative entropy reflects the positivity of the sum of all higher-dimensional operators in the EFT.
This was shown in Refs.~\cite{Cao:2022ajt,Cao:2022iqh}, where the leading-order perturbative analysis was studied.
To further examine the consequences of the non-negativity of the relative entropy in the nonlinear corrections, we focus on two scenarios, namely the magnetic model and the electric model, as described below:
\begin{itemize}
   
\item {\bf Magnetic model} ---  
We first consider the case where the electric field vanishes ($\vec{E}=0$), and examine higher-order perturbative effects on the relative entropy.
We refer to this as the magnetic model.
From Eq.~\eqref{eq:ferm_GF}, we find
\begin{align}
    \mathcal{F}=\frac{1}{2}\vec{B}^2\, \left(1+(\lambda e)^2\, \delta_{\rm F}\right)^{-1}\,,\quad \mathcal{G}=0\,.\label{eq:mag_BG}
\end{align}
From Eqs.~\eqref{eq:W_nonlin_ferm} and~\eqref{eq:mag_BG}, we obtain
\begin{align}
    W_\lambda[A^{\rm cl}]=\int d^4x_{\rm E}\,\left[
    \epsilon_{\rm F}
    +
    \frac{1}{2}\vec{B}^2
    -
    \mathcal{L}_{\rm F}\left((\lambda e\hat{B})^2\right)
    \right]\,,\label{eq:W_ferm_B}
\end{align}
where the nonlinear EFT effect is given by
\begin{align}
     \mathcal{L}_{\rm F}\left((\lambda e\hat{B})^2\right)
     =
     \frac{m^4}{8\pi^2}\int_0^\infty 
     e^{-t} t^{-3}\mathcal{K}\left(\lambda e \hat{B}t\right)\, dt\,,
     \quad
     \mathcal{K}(x)\equiv-ix \cot (ix)+1- \frac{(ix)^2}{3}\,,\label{eq:Lnon_B_F}
\end{align}
where $\hat{B} \equiv |\vec{B}|/m^2$ and $t \equiv m^2 s$ are dimensionless quantities.
Similar to Eq.~\eqref{eq:per_ferm_EB}, we obtain the relative entropy by substituting Eq.~\eqref{eq:W_ferm_B} into Eq.~\eqref{eq:quanrel} and setting $\lambda=1$:
\begin{align}
    {S}\left({\rho}_{\rm R}\|{\rho}_{\rm T}\right)&=
    \int d^4x_{\rm E}\,\mathcal{L}_{\rm F}\left(( e\hat{B})^2\right)
    =
    \frac{m^4}{8\pi^2}\int d^4x_{\rm E}\,\int_0^\infty 
     e^{-t} t^{-3}\mathcal{K}\left( e \hat{B}t\right)\, dt\,.\label{eq:rel_B_F}
\end{align} 

Now let us perturbatively analyze the relative entropy.
Accordingly, we consider the following series expansions of the nonlinear EFT effect in Eq.~\eqref{eq:Lnon_B_F}:
\begin{align}
    \mathcal{L}_{\rm F}^{(N)}\left((e\hat{B})^2\right)
    \equiv
    m^4\,\sum_{n=2}^N c_n\,\left( e\hat{B}\right)^{2n}\,,\quad c_n=\frac{1}{4\pi^2}\zeta (2n) \frac{1}{(i\pi)^{2n}}(2n-3)!\,,\label{eq:LN_B_F}
\end{align}
where $N$ is a finite integer.
From Eqs.~\eqref{eq:rel_B_F} and~\eqref{eq:LN_B_F}, we obtain the perturbative definition of the relative entropy~\eqref{eq:rel_pert} at the $N$-th order:
    \begin{align}
{S}^{(N)}\left(\widetilde{\rho}_{\rm R}\|\widetilde{\rho}_{\rm T}\right)&=\int d^4x_{\rm E}\, \mathcal{L}^{(N)}_{\rm F}\left((e\hat{B})^2\right)
=
m^4\,  \int d^4x_{\rm E}\, \sum_{n=2}^N\, c_n \left(e\hat{B}\right)^{2n}\,.\label{eq:perp_rel_B}
\end{align}
For convenience, we present the coefficient $c_n$ up to sixth order in $(e\hat{B})^2$ as follows:
\begin{align}
    c_2=\frac{1}{45}\frac{1}{8\pi^2}\,,~ c_3=-\frac{4}{315}\frac{1}{8\pi^2}\,,~ c_4=\frac{8}{315}\frac{1}{8\pi^2}\,,~ c_5= -\frac{32}{297}\frac{1}{8\pi^2}\,,~c_6=\frac{176896}{225225}\frac{1}{8\pi^2}\,.\label{eq:Slis}
\end{align}
From these expressions, it is evident that the second-order term in $(e\hat{B})^2$ is non-negative, and the leading contribution in the operator expansion respects the non-negativity of the relative entropy.
In contrast, as seen from Eq.~\eqref{eq:Slis}, the third and higher-order terms in $(e\hat{B})^2$ can become negative.
As explained in Section~\ref{sec:pert}, when the EFT operator expansion is truncated at a finite order, the non-negativity of the relative entropy may be violated, particularly for large $(e\hat{B})^2$.
Accordingly, the non-negativity of the relative entropy, which serves as an implicit indicator of unitarity, constrains the parameter region where the operator expansion, {\it i.e.}, a finite truncation of the $(e\hat{B})^2$ series, remains quantitatively reliable.

We now turn to a further assessment of the relative entropy to clarify the domain of validity of the perturbative analysis.
Let us normalize the relative entropy~\eqref{eq:perp_rel_B} by ${S}^{(2)}\left(\widetilde{\rho}_{\rm R}\|\widetilde{\rho}_{\rm T}\right)$ and define the normalized relative entropy up to the $N$-th order of
the perturbative expansion as follows:
\begin{align}
    s^{(N)}\equiv \frac{ {S}^{(N)}\left(\widetilde{\rho}_{\rm R}\|\widetilde{\rho}_{\rm T}\right)}{ {S}^{(2)}\left(\widetilde{\rho}_{\rm R}\|\widetilde{\rho}_{\rm T}\right)} =  \sum_{n= 2}^N\, (e\hat{B})^{2(n-2)}\,\frac{c_n}{c_2}\,.\label{eq:s(N)_pert}
\end{align}
Since the exact relative entropy is non-negative, a truncated perturbative approximation is expected to remain non-negative within its regime of validity.
We therefore use the non-negativity of $s^{(N)}$ as a consistency criterion for the perturbative expansion.
From Eq.~\eqref{eq:s(N)_pert}, this criterion leads to explicit inequalities.
For example, for $N=3$ and $N=5$ one obtains
\begin{align}
    s^{(3)}&=1 -\frac{4}{7}(e\hat{B})^2 \gtrsim 0\,,\label{eq:s3}
    \\
    s^{(5)}&=1 -\frac{4}{7}(e\hat{B})^2 +  \frac{8}{7}(e\hat{B})^4 -\frac{160}{33} (e\hat{B})^6 \gtrsim 0\,.\label{eq:s5}
\end{align}
Here, we use the symbol $\gtrsim$ to indicate that the relative entropy is evaluated perturbatively, so its non-negativity holds only within the perturbative approximation.
Equations~\eqref{eq:s3} and \eqref{eq:s5} show that, for $(e\hat{B})^2\sim 1$, non-negativity is violated, signaling the breakdown of the perturbative analysis.
Within the regime where the perturbative expansion is valid, lower-order terms provide the dominant contributions. 
Accordingly, the domain of validity of the perturbative analysis can be quantified from Eq.~\eqref{eq:s3} as follows:
\begin{align}
     s^{(3)} \gtrsim 0\quad \Rightarrow \quad  \frac{\sqrt{7}}{2}  \gtrsim |e\hat{B}|\,,
\end{align}
where $e\hat{B}= e |\vec{B}|/m^2$.
In the parameter region $|e\hat{B}| \gtrsim \tfrac{\sqrt{7}}{2}$, the perturbative analysis breaks down, indicating that the perturbative description of the relative entropy is no longer reliable, as implied by the unitarity of the theory encoded in the relative entropy.
As explained in detail in Section~\ref{sec:Borel_fer}, the above power series forms a divergent series with zero radius of convergence and should therefore be interpreted as an asymptotic series.
In this sense, the parameter region $|e\hat{B}| \gtrsim \tfrac{\sqrt{7}}{2}$ marks the point at which the truncated perturbative expansion ceases to provide a reliable approximation, suggesting that non-perturbative contributions become increasingly important.
As explained in Section~\ref{sec:Borel_fer}, once the perturbative expansion is properly resummed to all orders, the non-negativity of the relative entropy is restored, even in regions where the naive perturbative expansion would otherwise be invalid.

\item {\bf Electric model} --- 
Next, we consider the case where the magnetic field vanishes ($\vec{B}=0$), referred to as the electric model.
Equation~\eqref{eq:ferm_GF} shows that the electric model can be obtained from the magnetic model by replacing $\vec{B}$ with $i\vec{E}$. 
Using Eq.~\eqref{eq:perp_rel_B}, we can then write the perturbative definition of the relative entropy for the electric model as
\begin{align}
    S^{(N)}\left(\widetilde{\rho}_{\rm R}\|\widetilde{\rho}_{\rm T}\right)=\int d^4x_{\rm E}\, \mathcal{L}^{(N)}_{\rm F}\left((ie\hat{E})^2\right)=m^4\,  \int d^4x_{\rm E}\, \sum_{n=2}^N\, c_n \left(ie\hat{E}\right)^{2n}\,,\label{eq:tild_LN}
\end{align}
where $\hat{E}\equiv |\vec{E}|/m^2$, and
\begin{align}
    c_n\,i^{2n}= \frac{1}{4\pi^2}\zeta(2n) \frac{1}{\pi^{2n}} (2n-3)!>0\,.\label{eq:cn_def}
\end{align}
From Eq.~\eqref{eq:cn_def}, we find that each term in the perturbative expansion of the relative entropy is non-negative.
Note that the above perturbative analysis is valid only when higher-order contributions in the expansion are sufficiently smaller than the lower-order effects.
Within this regime, the positivity of the relative entropy reflects the positivity of the combined contributions from higher-dimensional operators in the EFT, $ \mathcal{L}^{(N)}_{\rm F}\left((ie\hat{E})^2\right)$.
If higher-order effects become comparable to lower-order ones, the limit $N \to \infty$ must be considered.
Based on these observations, we can infer that the resummation of the perturbative relative entropy should also remain non-negative in the weak-coupling regime, given that the relative entropy is non-negative at each order in perturbation theory.
As will be discussed in Section~\ref{sec:Borel_fer} using Borel--Laplace resummation, this expectation is indeed correct in that regime.
However, non-perturbative effects—becoming significant in the strong-coupling regime, where the perturbative expansion breaks down—may invalidate this behavior, leading to a violation of non-negativity and signaling an instability of the system (see Section~\ref{sec:inst}).

\end{itemize}

\subsection{Borel--Laplace resummation}
\label{sec:Borel_fer}
Thus far, we have explored the implications of the non-negativity of relative entropy within perturbative analyses.
We now evaluate the summed perturbative contributions derived in Section~\ref{sec:pe_fer} by applying Borel--Laplace resummation.
Within the framework of resurgence theory, the singularity structure of the Borel transform encodes information about non-perturbative effects.
In particular, it can reveal signatures of phenomena such as the Schwinger effect that are invisible at any finite order in perturbation theory.
As in Section~\ref{sec:pe_fer}, we consider two scenarios: the magnetic model and the electric model.

\begin{itemize}
   
\item {\bf Magnetic model} --- 
We evaluate the summed perturbative contributions to the relative entropy using the perturbative calculations presented in Section~\ref{sec:pe_fer}.
As shown in Eq.~\eqref{eq:perp_rel_B}, in the limit $N \to \infty$, the nonlinear EFT effect $\mathcal{L}^{(N)}_{\rm F}\left((e\hat{B})^2\right)$ forms a divergent series with zero radius of convergence, since $c_n \propto (2n-3)!$.
That is, the series $\mathcal{L}^{(\infty)}_{\rm F}\left((e\hat{B})^2\right)$  diverges for any finite $(e\hat{B})^2$ and therefore represents an unphysical quantity.
Consequently, an appropriate method is required to resum the perturbative expansion.
From Eq.~\eqref{eq:perp_rel_B}, we obtain a formal power series that does not converge and must be treated formally:
\begin{align}
S^{(\infty)}\left(\widetilde{\rho}_{\rm R}\|\widetilde{\rho}_{\rm T}\right)&=\int d^4x_{\rm E}\, \mathcal{L}^{(\infty)}_{\rm  F}\left((e\hat{B})^2\right)\,,\quad
 \mathcal{L}_{\rm F}^{(\infty)}\left((e\hat{B})^2\right)
    =
    m^4\,\sum_{n=2}^\infty c_n \left(e\hat{B}\right)^{2n}\,.\label{eq:inf_rel}
\end{align}
Based on these results, we define the Borel transform as
\begin{align}
    \mathcal{B}_{\rm F}(\tau)\equiv \sum_{n=2}^\infty c_n\, \frac{\tau^{2n-3}}{(2n-3)!}
    =
    \sum_{n=2}^\infty \bar{c}_n\, \tau^{2n-3}\,,\label{eq:Btau}
\end{align}
where we have used $c_n=\bar{c}_n\, (2n-3)!=(4\pi^2)^{-1}\zeta (2n)(i\pi)^{-2n} (2n-3)!$.
The essential feature of the Borel transformation above is that the factorial growth of the coefficients, $c_n\propto (2n-3)!$ completely cancels the $(2n-3)!$ in $\mathcal{B}_{\rm F}(\tau)$, thus rendering the series convergent.
In the limit $n\to\infty$, $\bar{c}_n$ is given by
\begin{align}
    \bar{c}_n=\frac{1}{4\pi^2}\zeta (2n)(i\pi)^{-2n}\to \frac{1}{4\pi^2}\,(i\pi)^{-2n}\,,\label{eq:an}
\end{align}
where we have used $\lim_{n\to\infty}\zeta (2n)=1$.
From Eqs.~\eqref{eq:Btau} and \eqref{eq:an}, we see that the Borel transformation converges uniformly for $|\tau|<\pi$, in contrast to $\mathcal{L}^{(\infty)}_{\rm F}$.
Consequently, $\mathcal{B}_{\rm F}(\tau)$ is analytic in a neighborhood of the origin and can be analytically continued to obtain meaningful values at larger $\tau$.
A key observation is that Eqs.~\eqref{eq:Btau} and \eqref{eq:an} allow us to infer the location of the nearest Borel singularity, \emph{i.e.}, the pole of $\tau^3\,\mathcal{B}_{\rm F}(\tau)$, from the large-order behavior of perturbation theory, yielding $\tau_{\rm pole} = i\pi$.
We thus infer the following form of the singular part of the Borel transformation near the singularity at $\tau_{\rm pole} = i\pi$:
\begin{align}
    \tau^3\,\mathcal{B}_{\rm F}(\tau)\bigg|_{\rm sing}= \frac{1}{4\pi^2}\frac{1}{1-(\tau/\tau_{\rm pole})^2}\,.\label{eq:pole_Btau}
\end{align}
Since the sign of the factorially growing coefficient $c_n$ changes depending on $n$, the Borel singularity does not lie on the real $\tau$ axis. 
Since Eq.~\eqref{eq:pole_Btau} captures only the singular part of the Borel transform, \emph{i.e.} the asymptotic behavior of the perturbative coefficients, the regular part governing the lower-order coefficients is not fully determined.
However, as explained below, the fact that the relative entropy is sensitive to nonlinear effects—corresponding to higher-derivative operators that do not include renormalizable terms and vanish in the zero-field limit—allows us to fix the leading two coefficients that potentially control the regular part of the Borel transformation.
The corresponding Borel transform is then inferred to be approximately
\begin{align}
    \tau^3\,\mathcal{B}_{\rm F}(\tau)\simeq \frac{1}{4\pi^2}\frac{1}{1-(\tau/\tau_{\rm pole})^2}
    -
    \frac{1}{4\pi^2}\left(1+(\tau/\tau_{\rm pole})^2\right)
    =
    \frac{1}{4\pi^2}\frac{(\tau/\tau_{\rm pole})^4}{1-(\tau/\tau_{\rm pole})^2}\,.\label{eq:pole_Btau2}
\end{align}

Now, to resum the perturbative expansion using the Borel transformation, we consider the Borel--Laplace resummation given by
\begin{align}
    \mathcal{L}_{\rm F}^{(\infty)}\left((e\hat{B})^2\right)
    =m^4\,\int_0^\infty e^{-t} \left(e\hat{B}\right)^3 \mathcal{B}_{\rm F}\left(e\hat{B}t\right)dt\,.\label{eq:Borel_resum}
\end{align}
By the analytic continuation of the Borel transformation $\mathcal{B}_{\rm F}\left(e\hat{B}t\right)$ to large $e\hat{B}t$ regions, the above resummation can reconstruct the physical quantity.
The definition of the Borel--Laplace resummation relies on the following relations:
\begin{align}
     \int_0^\infty e^{-t}\mathcal{B}_{\rm F}\left(o t\right)dt&\to \sum_{n=2}^\infty  \frac{c_n o^{2n-3}}{(2n-3)!}
     \int_0^\infty e^{-t}\, t^{2n-3}dt=
     \sum_{n=2}^\infty c_n o^{2n-3}
\,,\label{eq:Borel_just}
\end{align}
where we have used $ \int_0^\infty e^{-t}\, t^{2n-3}dt=(2n-3)!$.
From the above Borel transformation, the relative entropy (nonlinear EFT effect) evaluated by the Borel--Laplace resummation is given by 
\begin{align}
    S^{(\infty)}\left(\widetilde{\rho}_{\rm R}\|\widetilde{\rho}_{\rm T}\right)&=\int d^4x_{\rm E}\, \mathcal{L}^{(\infty)}_{\rm F}\left((e\hat{B})^2\right)\notag
    \\
    &\simeq   \frac{m^4}{4\pi^2}\int d^4 x_{\rm E}\, \int_0^{\infty}e^{-t}t^{-3} \frac{(e\hat{B}t/\tau_{\rm pole})^4}{1-(e\hat{B} t/\tau_{\rm pole})^2}dt\geq 0\,,\label{eq:resum_pert_B}
\end{align}
where $\tau_{\rm pole}=i\pi$.
This resummed perturbative result is manifestly positive.
Therefore, the apparent violation of non-negativity in the magnetic model, observed in the perturbative analysis (Section~\ref{sec:pe_fer}), can be resolved by resumming the perturbative expansion to all orders.
Since the magnetic model contains no instability, there is no fundamental reason for the non-negativity of the relative entropy to be violated in all-order calculations.
Consequently, it is not surprising that the non-negativity is restored once the resummation is performed.

 \item {\bf Electric model} --- Replacing $\vec{B}$ with $i\vec{E}$ at the appropriate steps allows one to derive the electric model from the magnetic model.
From Eq.~\eqref{eq:inf_rel}, the formal power series of the relative entropy in the electric model is given by
\begin{align}
S^{(\infty)}\left(\widetilde{\rho}_{\rm R}\|\widetilde{\rho}_{\rm T}\right)&=\int d^4x_{\rm E}\, \mathcal{L}^{(\infty)}_{\rm F}\left((ie\hat{E})^2\right)\,,\quad
 \mathcal{L}_{\rm F}^{(\infty)}\left((ie\hat{E})^2\right)
    =
    m^4\,\sum_{n=2}^\infty c_n \left(ie\hat{E}\right)^{2n}\,,\label{eq:infrel_E}
\end{align}
where $\hat{E}=|\vec{E}|/m^2$, and $c_n$ is defined in Eq.~\eqref{eq:LN_B_F}.
Then, the Borel--Laplace resummation is given by
\begin{align}
        \mathcal{L}_{\rm F}^{(\infty)}\left((ie\hat{E})^2\right)
    =m^4\,\int_0^\infty e^{-t} \left(ie\hat{E}\right)^3 \mathcal{B}_{\rm F}\left(ie\hat{E}t\right)dt\,.
\end{align}
In contrast to the magnetic model, the Borel singularity lies on the real axis of $t$.
By avoiding the pole of $\mathcal{B}_{\rm F}\left(ie\hat{E}t\right)$, the Borel--Laplace resummation can be rewritten as
\begin{align}
    \mathcal{L}_{\rm F}^{(\infty)}\left((ie\hat{E})^2\right)
    =m^4\,&\left(\mathcal{P}\,\int_0^\infty e^{-t} \left(ie\hat{E}\right)^3 \mathcal{B}_{\rm F}\left(ie\hat{E}t\right)dt \right.
    \notag\\&\left. 
    +
    \lim_{\epsilon\to 0^+} \int_{C_\epsilon} e^{-t} \left(ie\hat{E}\right)^3 \mathcal{B}_{\rm F}\left(ie\hat{E}t\right)dt
    \right)\,,\label{eq:BL_C}
\end{align}
where the first term on the right-hand side denotes the principal value integral defined as
\begin{align}
    \mathcal{P}\,\int_0^{\infty}e^{-t} \left(ie\hat{E}\right)^3 \mathcal{B}_{\rm F}\left(ie\hat{E} t\right)dt\equiv
    \lim_{\epsilon\to 0^+}&\left(\int_0^{\tau_{\rm pole}/i|e\hat{E}|-\epsilon}e^{-t} \left(ie\hat{E}\right)^3\mathcal{B}_{\rm F}\left(ie\hat{E}t\right)dt
    \right.\notag
    \\&+\left.
    \int_{\tau_{\rm pole}/i|e\hat{E}|+\epsilon}^{\infty}e^{-t}\left(ie\hat{E}\right)^3 \mathcal{B}_{\rm F}\left(ie\hat{E}t\right)dt
    \right)\,,\label{eq:P_B}
\end{align}
and the second term on the right-hand side is given by
\begin{align}
    \lim_{\epsilon\to 0^+}
    \int_{C_\epsilon}
    e^{-t}\left(ie\hat{E}\right)^3
    \mathcal{B}_{\rm F}\left(ie\hat{E}t\right)dt
    &=
    \pm i\pi\,
    {\rm Res}_{t=\pi/|e\hat{E}|}
    \left[
    e^{-t}\left(ie\hat{E}\right)^3
    \mathcal{B}_{\rm F}\left(ie\hat{E}t\right)
    \right]
    \notag\\
    &=
    \pm i\,
    \frac{1}{4\pi^2}
    \frac{\pi}{2}
    \frac{e^{-\pi/|e\hat{E}|}}
    {(\pi/e\hat{E})^2}\,,
    \label{eq:non_B}
\end{align}
where $C_\epsilon$: $t(\theta)=\pi/|e\hat{E}| +\epsilon\, e^{i\theta}$, $\theta\in [\pi, 0]$, or $[\pi, 2\pi]$, and $\mathcal{B}_{\rm F}(\tau)$ is given by Eq.~\eqref{eq:pole_Btau2}.
Combining Eqs.~\eqref{eq:BL_C}, \eqref{eq:P_B}, and \eqref{eq:non_B} with Eq.~\eqref{eq:infrel_E}, we thus obtain an approximate Borel--Laplace-resummed relative entropy (nonlinear EFT correction):
\begin{align}
    S^{(\infty)}\left(\widetilde{\rho}_{\rm R}\|\widetilde{\rho}_{\rm T}\right)&=\int d^4x_{\rm E}\, \mathcal{L}^{(\infty)}_{\rm F}\left((ie\hat{E})^2\right)\notag
    \\
    &\simeq  \frac{m^4}{4\pi^2}\int d^4 x_{\rm E}\, \left[\mathcal{P}\,\int_0^{\infty}e^{-t} t^{-3}\frac{(e\hat{E} t/\pi)^4}{1-(e\hat{E} t/\pi)^2}dt
    \pm i \frac{\pi}{2} \frac{e^{-\pi/|e\hat{E}|}}{(\pi/ e\hat{E})^2} 
    \right]\,.\label{eq:rel_resurgent}
\end{align}
In particular, the second term in Eq.~\eqref{eq:rel_resurgent} exhibits a sign ambiguity arising from the choice of integration contour, depending on whether the pole is avoided from above or below.
It thus reflects the contour dependence of the resummation.
Since perturbation theory alone does not, in general, fix this choice, an additional prescription is required to remove the resulting non-perturbative ambiguity\footnote{From the UV perspective based on the Hamiltonian of the UV theory, the contour choice is fixed by imposing forward-time evolution through the standard $i\epsilon$ prescription.}.
This imaginary part arises solely from the electric field, implying that the system’s instability is directly linked to the analytic continuation of the electric field in Eq.~\eqref{eq:W_an}. 
If this non-perturbative ambiguity in Eq.~\eqref{eq:rel_resurgent} is neglected, one expects to recover only the resummed perturbative expansion discussed in Section~\ref{sec:pe_fer}.
In the weak-coupling regime, this expectation is supported by the perturbative analysis in Section~\ref{sec:pe_fer}.
However, as will be explained below, the contributions from the principal-value integral in Eq.~\eqref{eq:rel_resurgent} are also affected by Borel singularities in the strong-coupling regime.
As a result, the sign of the relative entropy also captures the system’s instability.
We now turn to the principal-value integral appearing in Eq.~\eqref{eq:rel_resurgent}:
\begin{align}
    {\rm Re}\, S^{(\infty)}\left(\widetilde{\rho}_{\rm R}\|\widetilde{\rho}_{\rm T}\right)\simeq   \frac{m^4}{4\pi^2}\int d^4 x_{\rm E}\, \mathcal{P}\,\int_0^{\infty}e^{-t}t^{-3} \frac{(e\hat{E} t/\pi)^4}{1-(e\hat{E} t/\pi)^2}dt
    >0~{\rm for}~\frac{\pi}{|e\hat{E}|}\geq 1\,.\label{eq:resum_pert_ferm}
\end{align}
From this expression and Eq.~\eqref{eq:pole_Btau2}, it follows that the right-hand side starts at fourth order in $e\hat{E}$.
This shows that the relative entropy is sensitive only to nonlinear effects, corresponding to higher-derivative operators that exclude renormalizable terms and vanish in the zero-field limit.
This observation motivates our choice of Eq.~\eqref{eq:pole_Btau2} as the Borel transform; see also Ref.~\cite{Conzinu:2026cuf}.
As shown analytically in Appendix~\ref{app:pos_func} and Ref.~\cite{Conzinu:2026cuf}, the resummed relative entropy is positive in the perturbative regime where $\pi/|e\hat{E}| \ge 1$.
Furthermore, this approximately resummed expression is fully consistent with the contribution from the nearest pole in the exact non-perturbative analysis presented in Section~\ref{sec:nonpe_fer}.

Consequently, the resummed nonlinear effects in Maxwell theory, $ \mathcal{L}^{(\infty)}_{\rm F}$, become positive in the weak-coupling regime, owing to the non-negativity of the perturbatively defined relative entropy.
These results are consistent with the perturbative analysis in Section~\ref{sec:pe_fer}, since the relative entropy remains non-negative order by order in the perturbative expansion.
However, in the strong-coupling regime ($\pi/|e\hat{E}|\ll 1$), the non-negativity of the resummed relative entropy~\eqref{eq:resum_pert_ferm} is shown analytically in Appendix~\ref{app:pos_func} and Ref.~\cite{Conzinu:2026cuf} to be violated.
This indicates that the effect of the nearest pole—corresponding to the instability associated with the Schwinger effect—enters the real part of the resummed relative entropy, leading to a violation of the positivity bound on the nonlinear sector of Maxwell theory, $ \mathcal{L}^{(\infty)}_{\rm F}$.

\end{itemize}

\subsection{Non-perturbative analysis}
\label{sec:nonpe_fer}
Finally, we evaluate the relative entropy using a non-perturbative analysis at the one-loop level, including all-order operator effects, and verify the results of the Borel--Laplace resummation within a more robust framework.
As will be shown, the results of the Borel--Laplace resummation presented in Section~\ref{sec:Borel_fer} are consistent with a non-perturbative analysis based solely on the nearest singularity.
In what follows, we present the details of the two scenarios: the magnetic model and the electric model.

\begin{itemize}
\item{\bf Magnetic model} ---
Let us begin with expression~\eqref{eq:rel_B_F}, together with Eq.~\eqref{eq:Lnon_B_F}. 
From Eq.~\eqref{eq:Lnon_B_F}, it is clear that there is no pole in $\mathcal{K}\left(e\hat{B}t\right)$ on the real $t$ axis, and the Schwinger effect (instability of system) does not arises.
From Eqs.~\eqref{eq:rel_B_F} and \eqref{eq:Lnon_B_F} (see also Appendix~\ref{sec:Magnetic EH}), the non-perturbative relative entropy in the magnetic model is given by
\begin{align}
    S\left(\rho_{\rm R}\|\rho_{\rm T}\right)&=\int d^4x_{\rm E}\, \mathcal{L}_{\rm F}\left((e\hat{B})^2\right)= \frac{m^4}{8\pi^2}\int d^4x_{\rm E}\,\int_0^\infty e^{-t} t^{-3} \mathcal{K}\left(e\hat{B} t\right)dt\geq 0\,.\label{eq:rel_non_M}
\end{align}
Using a formula provided in Eq.~\eqref{eq:zcotz}, we find
\begin{align}
    \frac{1}{2}\int_0^{\infty} e^{-t}t^{-3} \mathcal{K}\left(e\hat{B} t\right) dt
    =
    \sum_{p=1}^\infty \int_0^\infty e^{-t}t^{-3}\frac{\left(e\hat{B}t/p \pi\right)^4}{1+\left(e\hat{B}t/p \pi\right)^2}dt \geq 0\,.\label{eq:pos_M}
\end{align}
Therefore, the non-negativity of the nonlinear EFT contribution, $ \mathcal{L}_{\rm F}\left((e\hat{B})^2\right)$, follows from the non-negativity of the relative entropy.
In particular, the contribution from the first pole ($p=1$) is consistent with the resummed result in Eq.~\eqref{eq:resum_pert_B}.

    \item {\bf Electric model} --- 
The nonlinear EFT effect $\mathcal{L}_{\rm F}$ in the electric model can be derived from Eq.~\eqref{eq:W_nonlin_ferm} (see also Appendix~\ref{sec:Electric EH}) as follows:
\begin{align}
    S\left(\rho_{\rm R}\| \rho_{\rm T}\right)
    =
    \int d^4x_{\rm E}\, \mathcal{L}_{\rm F}\left((ie\hat{E})^2\right)
    =
    \frac{m^4}{8\pi^2}\int d^4x_{\rm E}\int_0^{i\infty} e^{-t} t^{-3} \mathcal{K}\left(ie\hat{E} t\right)dt\,,\label{eq:rel_non_E}
\end{align}
where $\mathcal{K}\left(ix\right)
    =
    -x\cot x +1 -\tfrac{x^2}{3}$.
We now deform the integration contour of the Schwinger proper time, $s=t/m^2$, in Eq.~\eqref{eq:rel_non_E}; see also Fig.~\ref{fig:contour}.
To this end, we first enumerate the singularities of $e^{-t}t^{-3}\mathcal{K}\left(ie\hat{E} t\right)$.
This function satisfies the following relations:
\begin{align}
    &\lim_{t\to 0}\, e^{-t}t^{-3}\,\mathcal{K}\left(ie\hat{E}t\right)=0\,,
    \quad
    \lim_{t\to t_p^\mp}\, e^{-t}t^{-3}\,\mathcal{K}\left(ie\hat{E}t\right)=\pm \infty\,,
    \\
    &{\rm Res}\,\left[e^{-t}t^{-3}\,\mathcal{K}\left(ie\hat{E}t\right),\, t_p\right]\equiv\lim_{t\to t_p}\, \left(t-t_p\right)e^{-t}t^{-3}\,\mathcal{K}\left(ie\hat{E}t\right)=-\frac{e^{-t_p}}{t^2_p}\,,\label{eq:poles}
\end{align}
where $t_p=\pi p/|e\hat{E}|$ for $p = 1,2,3,\ldots$.
Thus, along the real axis of the Schwinger proper time $s=t/m^2$, the function $e^{-t}t^{-3}\mathcal{K}\left(ie\hat{E} t\right)$ exhibits singularities.
In particular, the first pole, $t_1=\pi/|e\hat{E}|$, corresponds to the Borel singularity discussed in Section~\ref{sec:Borel_fer}.
By avoiding the poles, we can perform the integral in Eq.~\eqref{eq:rel_non_E} as follows:
\begin{align}
   \mathcal{L}_{\rm F}\left((ie\hat{E})^2\right)
    =
    \frac{m^4}{8\pi^2}\,\left[\mathcal{P}\,\int_0^\infty e^{-t} t^{-3} \mathcal{K}\left(ie\hat{E} t\right)dt
    +
    \sum_{p=1}^\infty
    \lim_{\epsilon\to 0^+}\int_{C_p}
    e^{-t} t^{-3} \mathcal{K}\left(ie\hat{E} t\right)dt
    \right]
    \,.\label{eq:tlL_full}
\end{align}
The principal-value integral is defined as
\begin{align}
&\mathcal{P}\,\int_0^{\infty}\, e^{-t}t^{-3} \mathcal{K}\left(ie\hat{E}t\right) dt\notag
\\
&\qquad\equiv \lim_{\epsilon\to 0^+}
\Bigg(
\int_0^{t_1-\epsilon} \, e^{-t}t^{-3} \mathcal{K}\left(ie\hat{E}t\right) dt
+ \sum_{p=1}^\infty
\int_{t_p+\epsilon}^{t_{p+1}-\epsilon}
\, e^{-t}t^{-3} \mathcal{K}\left(ie\hat{E}t\right) dt
\Bigg)\,,
\end{align}
and the second term on the right-hand side of Eq.~\eqref{eq:tlL_full} evaluates to
\begin{align}
   \lim_{\epsilon\to 0^+}\int_{C_p} \, e^{-t}t^{-3} \mathcal{K}\left(ie\hat{E}t\right) dt
   &= i\int_{\pi}^{0} \,
   \lim_{\epsilon\to 0^+}(t-t_p)\,e^{-t}t^{-3} \mathcal{K}\left(ie\hat{E}t\right)d\theta\notag
   \\
   &= -\,i\pi\, \mathrm{Res}\,\left[e^{-t}t^{-3} \mathcal{K}\left(ie\hat{E}t\right),\,t_p\right]=
   i\pi\,\frac{e^{-t_p}}{t_p^2}
   \,,\label{eq:Res_int}
\end{align}
where for $p\ge 1$ we parametrize the contour as
$C_p:\ t(\theta)=t_p+\epsilon e^{i\theta}$ with $\theta\in[\pi,0]$,
and we have used $dt=i\epsilon e^{i\theta} d\theta = i(t-t_p)d\theta$.
See Appendix~\ref{sec:Electric EH} for further details.
From Eqs.~\eqref{eq:poles}, \eqref{eq:tlL_full}, and \eqref{eq:Res_int}, we obtain
\begin{align}
    \mathcal{L}_{\rm F}\left((ie\hat{E})^2\right)
    =
    \frac{m^4}{8\pi^2}\,
    \left[
    \mathcal{P}\,\int_0^\infty e^{-t} t^{-3} \mathcal{K}\left(ie\hat{E} t\right)dt
    +
    i\pi\,\sum_{p=1}^\infty
    \frac{e^{-t_p}}{t_p^2}
    \right]\,,
\end{align}
where the principal value integral is real, while the last term represents the imaginary part associated with the Schwinger effect.
Thus, the  non-perturbatively calculated relative entropy (nonlinear EFT effect) for $e\hat{E}$ is given by
\begin{align}
    S\left(\rho_{\rm R}\|\rho_{\rm T}\right)&=\int d^4x_{\rm E}\, \mathcal{L}_{\rm F}\left((ie\hat{E})^2\right)\notag
    \\
    &= \frac{m^4}{8\pi^2}\int d^4x_{\rm E}\,\left[
    \mathcal{P}\,\int_0^\infty e^{-t} t^{-3} \mathcal{K}\left(ie\hat{E} t\right)dt
    +
    i\pi\,\sum_{p=1}^\infty
    \frac{e^{-t_p}}{t_p^2}
    \right]\,.\label{eq:non_per_E}
\end{align}
The above relative entropy becomes complex due to non-perturbative effects (instability of system), and its non-negativity is manifestly violated.
The imaginary part in Eq.~\eqref{eq:non_per_E}, arising from the first pole $t_1=\pi/|e\hat{E}|$, corresponds to the leading resurgent contribution associated with the Borel singularity discussed in Section~\ref{sec:Borel_fer}.
While the Schwinger effect appears as an explicit instability through the imaginary part, the real part of Eq.~\eqref{eq:non_per_E} captures the resummed perturbative contribution.
We now evaluate this resummed perturbative contribution to the relative entropy, corresponding to the principal value integral in Eq.~\eqref{eq:non_per_E}.
By definition, it is given by the real part of Eq.~\eqref{eq:non_per_E}:
\begin{align}
    {\rm Re}\,S\left(\rho_{\rm R}\|\rho_{\rm T}\right)&=\frac{m^4}{8\pi^2}\int d^4x_{\rm E}\,
    \mathcal{P}\,\int_0^\infty e^{-t} t^{-3} \mathcal{K}\left(ie\hat{E} t\right)dt\,. 
\end{align}
As shown in Appendix~\ref{sec:Electric EH}, the above resummed perturbative contribution can be rewritten as
\begin{align}
    {\rm Re}\,S\left(\rho_{\rm R}\|\rho_{\rm T}\right)&
    =\frac{m^4}{4\pi^2}\sum_{p=1}^\infty \int d^4x_{\rm E}\, 
    \mathcal{P}
    \int_0^\infty e^{-t} t^{-3}
    \frac{(e\hat{E}t/p\pi)^4}{1-(e\hat{E} t/p\pi)^2}dt
    >0~~ {\rm for}~~\frac{\pi}{|e\hat{E}|}\geq 1\,.\label{eq:non_resum}
    \end{align}
The expression arising from the first pole ($p=1$) is in full agreement with the resummed result in Eq.~\eqref{eq:rel_resurgent}.
In other words, the Borel transformation~\eqref{eq:pole_Btau2}, inferred from the formal EFT expansion, captures the contribution from the nearest pole.
As analytically shown in Appendix~\ref{app:pos_func}, the resummed relative entropy in Eq.~\eqref{eq:non_resum} remains positive in the perturbative regime ({\it i.e.}, in the limit far from the pole), where ${\pi}/{|e\hat{E}|} \geq 1$.
That is, in this regime, the nonlinear EFT contribution $\mathcal{L}_{\rm F}$ arising from the resummed perturbative expansion remains non-negative, as guaranteed by the non-negativity of the relative entropy.
In the strongly coupled regime ({\it i.e.}, in the limit approaching a pole), however, the non-negativity of the resummed relative entropy in Eq.~\eqref{eq:non_resum} is violated even after including an arbitrary finite number of pole contributions, as shown analytically in Appendix~\ref{app:pos_func}.
Therefore, the system's instability associated with the Schwinger effect also enters the real part of the resummed relative entropy, leading to a violation of the positivity bound on the nonlinear sector of Maxwell theory, $ \mathcal{L}^{(\infty)}_{\rm F}$.

\end{itemize}

\section{Scalar QED}\label{sec:scalar qed}
As a second application, we study a charged scalar field in electromagnetic backgrounds.
While most computational strategies familiar from fermionic QED continue to apply, defining the noninteracting theory requires particular care in gauge theories.
We therefore discuss this issue in detail, with a particular focus on scalar QED.

\subsection{Ambiguity of the noninteracting theory}
\label{sec:amb}
In gauge theories, the choice of the noninteracting reference theory associated with the density operator $\rho_{\rm R}$ is ambiguous due to gauge symmetry.
Throughout this work, we consider the minimal gauge coupling appearing in the covariant derivative of the bilinear kinetic term for the heavy charged fields; see also Ref.~\cite{Conzinu:2026cuf}.
A straightforward definition of the noninteracting reference theory is obtained by setting the interaction part of the Lagrangian to zero, $\mathcal{L}_{\rm I}=0$, {\it i.e.} taking $\mathcal{L}_0$, as discussed in Section~\ref{sec:qed}.
However, by gauge symmetry, pure-gauge configurations of the form $A_\mu=\partial_\mu\alpha$ do not produce physical effects in the low-energy theory obtained after integrating out the heavy fields $\Phi$.
Indeed, under the field redefinition of the heavy field $\Phi\to \Phi'=U(\alpha)\Phi$ associated with the gauge transformation $U(\alpha)$, the interaction term $\mathcal{L}_{\rm I}(\partial_\mu\alpha,\Phi)$ can be absorbed into the noninteracting term, such that
\begin{align}
\mathcal{L}_0(A_\mu,\Phi')=
\mathcal{L}_0(A_\mu,\Phi) + \mathcal{L}_{\rm I}(\partial_\mu\alpha,\Phi)\,,
\end{align}
where the kinetic term of the heavy charged field in $\mathcal{L}_0$ is bilinear in the fields.
Thus, the theory defined by $\mathcal{L}_0(A_\mu,\Phi) + \mathcal{L}_{\rm I}(\partial_\mu\alpha,\Phi)$ can equally well be regarded as a noninteracting reference theory.
We introduce this generalized reference theory and, following the notation of Section~\ref{sec:relative entropy}, define
\begin{align}
    \bar{\mathcal{L}}_\lambda(A_\mu,\Phi)\equiv \mathcal{L}_0(A_\mu,\Phi) + \mathcal{L}_{\rm I} \left(\partial_\mu\alpha,\Phi\right)
    +
    \lambda\, \left(\mathcal{L}_{\rm I} \left(A_\mu,\Phi\right)-\mathcal{L}_{\rm I} \left(\partial_\mu\alpha,\Phi\right)\right)\,.\label{eq:barL}
\end{align}
In what follows, we consider scalar QED as an example and provide further details.

\subsection{Setup and conventions}
Let us consider scalar QED, {\it i.e.}, a theory of charged spin-zero particles interacting with photons.
The Lagrangian in Minkowski space reads
\beq
\mathcal{L} =- \frac{1}{4} F^{\mu\nu} F_{\mu\nu} + (D^{\mu} \varphi)^{\dagger} D_{\mu} \varphi - m^{2} \varphi^{\dagger} \varphi \,,\label{eq:sQED}
\eeq
where $D_\mu=\partial_\mu +ie A_\mu$.
Unlike fermionic QED, this Lagrangian contains not only the usual first-order term in the coupling $e$, but also second-order contributions arising from the covariant derivative.
If one adopts a naive definition of the noninteracting theory $\mathcal{L}_0$, these second-order contributions can generate a non-vanishing $(dW_\lambda/d\lambda)_{\lambda=0}$ in Eq.~\eqref{eq:quanrel}. 
Consequently, the relative entropy is no longer sensitive exclusively to the nonlinear effects in Maxwell theory.
However, by exploiting the ambiguity of the reference theory discussed in Section~\ref{sec:amb}, $(dW_\lambda/d\lambda)_{\lambda=0}$ can vanish, and the relative entropy can reproduce nonlinear EFT effects similar to those in fermionic QED.

From Eq.~\eqref{eq:sQED}, we obtain the following terms in the Lagrangian:
\begin{align}
    \mathcal{L}_0(A_\mu,\varphi)&=-\frac{1}{4}F^{\mu\nu}F_{\mu\nu}
    + \left(\partial^\mu \varphi\right)^\dagger\partial_\mu \varphi
    -
    m^2 \varphi^\dagger \varphi\,,
    \\
    \mathcal{L}_{\rm I}(A_\mu,\varphi)&=i e A_\mu\left(
    (\partial^\mu \varphi)^\dagger\varphi
    -\varphi^\dagger (\partial^\mu \varphi)
    \right)
    + e^2 A^\mu A_\mu \varphi^\dagger \varphi\,.
\end{align}
Based on Eq.~\eqref{eq:barL}, we take the following Lagrangian as the noninteracting reference theory:
\begin{align}
    \bar{\mathcal{L}}_0(A_\mu,\varphi)
    &=\mathcal{L}_0(A_\mu,\varphi)
    +
    \mathcal{L}_{\rm I}(\partial_\mu \alpha,\varphi)\notag
    \\
    &=-\frac{1}{4}F^{\mu\nu}F_{\mu\nu}
    + \left(\partial^\mu \varphi\right)^\dagger\partial_\mu \varphi
    -
    m^2 \varphi^\dagger \varphi\notag
    \\
    &\qquad\qquad\qquad+i e (\partial_\mu \alpha) \left(
     (\partial^\mu \varphi)^\dagger \varphi-\varphi^\dagger
    (\partial^\mu \varphi)
    \right)
    + e^2 (\partial_\mu \alpha)^2 \varphi^\dagger \varphi\,
    .\label{eq:zero_gauge}
\end{align}
Under the field redefinition $\varphi' = e^{ie \alpha}\varphi$, one finds
\begin{align}
\bar{\mathcal{L}}_0(A_\mu,\varphi)=\mathcal{L}_0(A_\mu,\varphi')\,,
\end{align}
which can be regarded as a noninteracting theory.
It follows that, after integrating out $\varphi$, the theory~\eqref{eq:zero_gauge} is equivalent to the noninteracting theory, since $\partial_\mu \alpha$ does not contribute to the field strength $F_{\mu\nu}$.
The interaction terms can then be read off from Eq.~\eqref{eq:barL} as
\begin{align}
    \bar{\mathcal{L}}_{\rm I}&\equiv \mathcal{L}_{\rm I} \left(A_\mu,\varphi\right)-\mathcal{L}_{\rm I} \left(\partial_\mu\alpha,\varphi\right)\notag
    \\
    &=i e (A_\mu-\partial_\mu\alpha) \left(
    (\partial^\mu \varphi)^\dagger\varphi
    -\varphi^\dagger (\partial^\mu \varphi)
    \right)
    + e^2 \left(A^\mu A_\mu-(\partial_\mu\alpha)^2\right) \varphi^\dagger \varphi\,.\label{eq:int_zerofield}
\end{align}
Note that the last term in Eq.~\eqref{eq:int_zerofield} is invariant under the field redefinition $\varphi' = e^{ie\alpha}\varphi$ by virtue of its bilinear structure.
Based on this setup, in Appendix~\ref{app:scl_non}, we demonstrate that the relative entropy reproduces the nonlinear EFT corrections in scalar QED for static electric and magnetic fields, by choosing a suitable $\alpha$, as in the fermionic case.

\subsection{Summary of scalar QED results}
As explained in the previous sections and Appendix~\ref{app:scl_non}, the relative entropy can reproduce the nonlinear EFT effects of scalar QED.
The computational procedures for these effects in scalar QED are essentially the same as in the fermionic case.
We therefore summarize the results below.

For a constant electromagnetic field, the Euclidean one-loop effective action can be obtained using the Schwinger proper-time method as
\begin{align}
W_{\lambda=1}^{\rm E}[A^{\rm E,cl}]&=\int d^4x_{\rm E}\,\left[\mathcal{F}-\frac{e^2}{8\pi^2}
    \int_0^{i\infty}ds\, \frac{e^{-sm^2}}{s} \frac{  \mathcal{G}}{\cosh( es\,x_+)-\cosh( es\,x_-)}\right]\,,\label{eq:ScalarQEDWAe}
\end{align}
where $x_\pm=\sqrt{2(\mathcal{F}\pm \mathcal{G})}$ with $\mathcal{F}=(F_{IJ})^2/4$, and $\mathcal{G}=-F_{IJ}\widetilde{F}^{IJ}/4$.
$\mathcal{F}$ and $\mathcal{G}$ can be expressed in terms of the electric and magnetic fields $\vec{E}$ and $\vec{B}$ as follows:
\begin{align}
    \mathcal{F}=\frac{1}{2}\left(\vec{B}^2-\vec{E}^2\right)\, \left(1+e^2\,\delta_{\rm S}\right)^{-1}\,,\qquad
    \mathcal{G}=-i\left(\vec{E}\cdot \vec{B}\right)\, \left(1+e^2\,\delta_{\rm S}\right)^{-1}\,,\label{eq:WFR_scl}
\end{align}
where $\delta_{\rm S}=\tfrac{1}{48\pi^2}\int_0^\infty ds\, \tfrac{1}{s}e^{-sm^2}$ is the WFR effect.
By performing the EFT expansion, we obtain
\begin{align}
    W_{\lambda=1}[A^{\rm cl}]=\int d^4x_{\rm E}\,\left[
    \epsilon_{\rm S}
    +
    \frac{1}{2}\left(\vec{B}^2-\vec{E}^2\right)
    -
    \mathcal{L}_{\rm S}
    \right]\,,
\end{align}
where $\epsilon_{\rm S}=-\tfrac{1}{16\pi^2}\int_0^\infty ds\,\tfrac{1}{s^3}e^{-s m^2}$ is the vacuum energy, and $\mathcal{L}_{\rm S}$ is the nonlinear EFT correction defined by
\begin{align}
    \mathcal{L}_{\rm S}
    =\frac{e^4}{1440\pi^2m^4}
    \left(
    \frac{7}{4}\left(\vec{B}^2-\vec{E}^2\right)^2
    +\left(\vec{E}\cdot \vec{B}\right)^2
    \right)
    +\cdots\,.
\end{align}
The explicit form of this nonlinear EFT correction is given below for the magnetic case ($\vec{E}= 0$ and $\vec{B}\neq 0$) and the electric case ($\vec{E}\neq 0$ and $\vec{B}=0$):
\begin{align}
    \mathcal{L}_{\rm S}
    \left(o^2\right)
    =
    \frac{m^4}{16\pi^2}
    \begin{cases}
    \int_0^\infty \, e^{-t} t^{-3}\,\mathcal{J}\left(e\hat{B} t\right)\,dt\,, & \text{magnetic case}~o=e\hat{B}\,,
    \\
    \int_0^{i\infty} \, e^{-t} t^{-3}\,\mathcal{J}\left(ie\hat{E} t\right)\,dt\,, & \text{electric case}~o=ie\hat{E}\,,
\end{cases}\label{eq:scl_nonlin_L}
\end{align}
where
 \begin{align}
     \mathcal{J}(x)\equiv\frac{ix}{\sin (ix)}-1-\frac{1}{6}(ix)^2\,, \label{eq:scl_func_f}
    \end{align}
    with dimensionless parameters $\hat{E}= |\vec{E}|/m^2$, $\hat{B}= |\vec{B}|/m^2$, and $t=m^2 s$.
    As in the fermionic case, the relative entropy reproduces the nonlinear EFT correction:
    \begin{align}
    S\left(\rho_{\rm R}\|\rho_{\rm T}\right)
    =
    \int d^4x_{\rm E}\,\mathcal{L}_{\rm S}\,.
    \end{align}
    In what follows, we summarize the results of the perturbative analysis, Borel--Laplace resummation, and non-perturbative analysis.

\subsubsection{Perturbative analysis}
\label{sec:pert_scl_qed}
    The function defined in Eq.~\eqref{eq:scl_func_f}, truncated at order $N$, is given by
    \begin{align}
    \mathcal{J}(x)\simeq 2\sum_{n=2}^N  \zeta (2n)\left(1-\frac{1}{2^{2n-1}}\right)\, \frac{1}{(i\pi)^{2n}}\, x^{2n}\,.
    \end{align}
    Substituting this into Eq.~\eqref{eq:scl_nonlin_L}, we obtain
    \begin{align}
     \mathcal{L}^{(N)}_{\rm S}\left(o^2\right)
    =
    m^4\,\sum_{n=2}^N b_n 
    \begin{cases}
    \left(e\hat{B}\right)^{2n}\,,& \text{magnetic case}~o=e\hat{B}\,,
    \\
    \left(ie\hat{E}\right)^{2n}\,,& \text{electric case}~ o=ie\hat{E}\,,
    \end{cases}
    \end{align}
    where
    \begin{align}
    b_n=\frac{1}{8\pi^2}\zeta(2n)\left(1-\frac{1}{2^{2n-1}}\right)\frac{1}{(i\pi)^{2n}}\, (2n-3)!\,.
    \end{align}
    Therefore, as in fermionic QED, the relative entropy in the electric case remains non-negative at each order in the perturbative expansion ({\it i.e.}, $ b_n\,i^{2n}>0$).
    In contrast, in the magnetic case the perturbative coefficients $b_n$ alternate in sign, so finite-order truncations may violate non-negativity outside the regime of validity of the perturbative expansion.

\subsubsection{Borel--Laplace resummation}
Next, we apply Borel--Laplace resummation to the perturbative results presented in Section~\ref{sec:pert_scl_qed}.
We define the Borel transform of the scalar QED perturbative expansion as:
    \begin{align}
    \mathcal{B}_{\rm S}(\tau)\equiv \sum_{n=2}^\infty b_n \frac{\tau^{2n-3}}{(2n-3)!}
    =
    \sum_{n=2}^\infty \bar{b}_n \tau^{2n-3}\,,
    \end{align}
    where $b_n= \bar{b}_n (2n-3)!$, and, for $n\to\infty$, $\bar{b}_n$ is asymptotically given by
\begin{align}
    \bar{b}_n=\frac{1}{8\pi^2}\zeta(2n)\left(1-\frac{1}{2^{2n-1}}\right)(i\pi)^{-2n}\to \frac{1}{8\pi^2}(i\pi)^{-2n}\,,
    \end{align}
    where we have used $\lim_{n\to \infty}\zeta(2n)=1$.
   This result coincides with that of fermionic QED given in Eq.~\eqref{eq:an}, up to an overall factor.
    In fermionic QED, the nonlinear contribution is proportional to $m^4/4\pi^2$, whereas in scalar QED it is proportional to  $m^4/8\pi^2$.
    Therefore, apart from the overall normalization, the resummed relative entropy in scalar QED is identical to that in fermionic QED discussed in Section~\ref{sec:Borel_fer}.
    As will be shown in the next section, the one-loop non-perturbative analysis based on the Schwinger proper-time method reproduces the resummed results up to the contribution from the nearest pole.

\subsubsection{Non-perturbative analysis}
Finally, we present the non-perturbative results obtained using the Schwinger proper-time method.
    From Eq.~\eqref{eq:scl_nonlin_L}, we obtain
    {\fontsize{3mm}{3mm}\selectfont
\begin{align}
    \mathcal{L}_{\rm S}\left(o^2\right)
    =
    \frac{m^4}{8\pi^2}
    \sum_{p=1}^\infty
    \begin{cases}
    \displaystyle
    \int_0^\infty dt\, e^{-t} t^{-3}
    \left[
    \frac{\left(\frac{e\hat{B}t}{p\,i\pi}\right)^4}
    {1-\left(\frac{e\hat{B}t}{p\,i\pi}\right)^2}
    -
    \frac{
    \left(\frac{e\hat{B}t}{p\,i\pi}\right)^4
    \left(1+(-1)^p\right)}
    {1-\left(\frac{e\hat{B}t}{p\,i\pi}\right)^2}
    \right],
    & o=e\hat{B}\,,
    \\[1.5em]
    \displaystyle
    \mathcal{P}\,\int_0^\infty dt\, e^{-t} t^{-3}
    \left[
    \frac{\left(\frac{e\hat{E}t}{p\pi}\right)^4}
    {1-\left(\frac{e\hat{E}t}{p\pi}\right)^2}
    -
    \frac{
    \left(\frac{e\hat{E}t}{p\pi}\right)^4
    \left(1+(-1)^p\right)}
    {1-\left(\frac{e\hat{E}t}{p\pi}\right)^2}
    \right]
    +
    i\frac{\pi}{2}(-1)^{p+1}
    \frac{e^{-t_p}}{t_p^2},
    & o=ie\hat{E}\,.
    \end{cases}
    \label{eq:nonlin_scl_non}
\end{align}
}
    where $t_p= \pi\, p/|e\hat{E}|$ for $p\in \mathbb{N}$.
    For odd values of $p$, the second term in the real part of the integral in Eq.~\eqref{eq:nonlin_scl_non} vanishes, and the resulting contributions in Eq.~\eqref{eq:nonlin_scl_non} are consistent with those in fermionic QED discussed in Section~\ref{sec:nonpe_fer}, up to an overall factor.
    In particular, the nearest pole contribution ($p=1$) reproduces the corresponding fermionic QED result. 
    Consequently, the resummed relative entropy in scalar QED coincides with that in fermionic QED up to an overall factor.

    As analytically shown in Appendix~\ref{app:pos_func}, the resummed relative entropy in Eq.~\eqref{eq:nonlin_scl_non} remains positive in the perturbative regime for the electric case, where ${\pi}/{|e\hat{E}|} \geq 1$.
    In the strongly coupled regime for the electric case, however, the non-negativity of the resummed relative entropy in Eq.~\eqref{eq:nonlin_scl_non} is violated even after including an arbitrary finite number of pole contributions.

\section{Dirac-Born-Infeld model}\label{sec:DBI}
In this section, we study the implications of the sign of the relative entropy in the Dirac-Born-Infeld (DBI) model~\cite{Born:1934gh,Dirac:1962iy}.
As discussed in Sections~\ref{sec:qed} and \ref{sec:scalar qed}, these implications were derived under the assumption of specific interactions between heavy and light fields.
In the following, we relax this assumption and consider cases where such interactions are not explicit, but can emerge through an equivalent description involving auxiliary heavy degrees of freedom.
As discussed below, the bounds on nonlinear electrodynamics EFTs extracted from the relative entropy are insensitive to the detailed structure of the interactions between heavy and light fields.
Rather, they merely require the existence of heavy degrees of freedom from which the EFT can be derived, {\it i.e.}, that the theory potentially admits a UV completion.
As will be discussed in Section~\ref{sec:gen}, we derive the bounds on the nonlinear EFT under minimal assumptions, thereby further highlighting these points.

\subsection{Setup and conventions}
The DBI model~\cite{Born:1934gh,Dirac:1962iy} provides a low-energy EFT, which provides an illustrative example for the nonlinear electrodynamics with a well-motivated UV origin.
In four dimensions, the explicit form of the Lagrangian in Minkowski spacetime is given by
\begin{align}
    \mathcal{L}&=\Lambda^4\,\left(
    1-
    \sqrt{1+\frac{1}{2g_e^2 \Lambda^4}F_{\mu\nu}F^{\mu\nu}-\frac{1}{g_e^4 \Lambda^8}\left(F_{\mu\nu}\widetilde{F}^{\mu\nu}\right)^2}
    \right)\,,\label{eq:DBI_Lag}
\end{align}
where $\Lambda^4$ represents the brane tension, while $\Lambda$ sets the scale of the nonlinear interactions.
By introducing two real auxiliary fields $X$ and $Y$, the Lagrangian~\eqref{eq:DBI_Lag} can be rewritten as follows:
\begin{align}
    \mathcal{L}_{XY}=&-\frac{1}{4g_e^2}F_{\mu\nu}F^{\mu\nu}-\frac{M_X^2}{2}X^2 + \frac{M_X }{g_e^2\Lambda^2} F_{\mu\nu}\widetilde{F}^{\mu\nu} X\notag
    \\
    &-
    \frac{M_Y^2}{2} Y^2 +M_Y \Lambda^2  \left(1-\sqrt{1+\frac{1}{2g_e^2 \Lambda^4}F_{\mu\nu}F^{\mu\nu}-\frac{1}{g_e^4\Lambda^8}\left(F_{\mu\nu}\widetilde{F}^{\mu\nu}\right)^2}\right)Y\,,\label{eq:DBI_XY}
\end{align}
where $M_X$ and $M_Y$ are positive parameters with mass dimension, and
the auxiliary fields $X$ and $Y$ are non-propagating and can be integrated out algebraically, after which Eq.~\eqref{eq:DBI_XY} reduces to the original DBI Lagrangian~\eqref{eq:DBI_Lag}.
In other words, the DBI Lagrangian~\eqref{eq:DBI_Lag} can be interpreted as arising from interactions involving auxiliary fields.
From Eq.~\eqref{eq:DBI_XY}, we obtain the corresponding Hamiltonian as 
\begin{align}
    H=H_0+H_{\rm I}=-\int d^3x\,\mathcal{L}_{XY}\,,\label{eq:Ham_DBI}
\end{align}
where
\begin{align}
    H_{0}&\equiv\int d^3 x\, \left[\frac{1}{4g_e^2}F_{\mu\nu}F^{\mu\nu}
    +
    \frac{M^2_X}{2}X^2
    +
    \frac{M^2_Y}{2}Y^2\right]\label{eq:H0_DBI}
    \,,
    \\
    H_{\rm I}&\equiv\int d^3x\, \left[-\frac{M_X}{g^2_e \Lambda^2}F_{\mu\nu}\widetilde{F}^{\mu\nu} X
    -
    M_Y \Lambda^2
    \left(
    1- \sqrt{1+\frac{1}{2 g_e^2 \Lambda^4}F_{\mu\nu}F^{\mu\nu}-\frac{1}{g^4_e \Lambda^8}\left(F_{\mu\nu}\widetilde{F}^{\mu\nu}\right)^2}
    \right)Y\right]\label{eq:HI_DBI}\,
    .
\end{align}
For simplicity in the subsequent analysis, we treat the electromagnetic fields as constant  background fields and do not include them as dynamical degrees of freedom ({\it i.e.}, we do not perform the path integral over the electromagnetic fields).
In what follows, we evaluate the relative entropy in three different ways, using the above conventions.

\subsection{Perturbative analysis}
\label{sec:perp_DBI}
We now examine the DBI model~\eqref{eq:DBI_Lag} in the perturbative regime of $g_e^{-2}$.
In this regime, the interacting Hamiltonian~\eqref{eq:HI_DBI} is expanded up to order $N$ as
\begin{align}
    H_{\rm I}=& \int d^3x \left[g_e^{-2}\, \left(J_{X}^{(1)}\, X
    +
    J_{Y}^{(1)}\, Y
    \right)
    +
    \sum_{n=2}^N (g_e^{-2})^n \, J_{Y}^{(n)}\, Y\right]\,,\label{eq:int_H_DBI_pert}
\end{align}
where, for instance, up to the third order of $g_e^{-2}$, we obtain
\begin{align}
    J_{X}^{(1)}&=-\frac{M_X }{\Lambda^2} \,F_{\mu\nu}\widetilde{F}^{\mu\nu}\,,\qquad J_{Y}^{(1)}=\frac{M_Y}{4\Lambda^2}\, F_{\mu\nu}F^{\mu\nu}\,,  \\
    J_{Y}^{(2)}&=-\frac{M_Y}{32\Lambda^6}\, 
    \left[
    \left(F_{\mu\nu}F^{\mu\nu}\right)^2
    +16 \left(F_{\mu\nu}\widetilde{F}^{\mu\nu}\right)^2
    \right]\,,
    \\
    J_{Y}^{(3)}&=\frac{M_Y}{128\Lambda^{10}}\,\left[
    \left(F_{\mu\nu}F^{\mu\nu}\right)^3
    +16 \left(F_{\mu\nu}F^{\mu\nu}\right)\left(F_{\rho\sigma}\widetilde{F}^{\rho\sigma}\right)^2
    \right]\,.
\end{align}
It should be emphasized that the interacting Hamiltonian~\eqref{eq:int_H_DBI_pert}, defined perturbatively in $g_e^{-2}$, is Hermitian for arbitrary values of the background field $F_{\mu\nu}$.
Therefore, starting from the Hamiltonians~\eqref{eq:H0_DBI} and \eqref{eq:int_H_DBI_pert}, the density operators appearing in the relative entropy~\eqref{eq:quanrel} are Hermitian, and consequently the relative entropy is guaranteed to be non-negative.
By introducing the parameter $\lambda$ for convenience, and $H_\lambda= H_0+\lambda H_{\rm I}$, the effective action is given as 
\begin{align}
    W_\lambda [A^{\rm cl}]=-\ln Z_\lambda [A^{\rm cl}]\,,\quad Z_\lambda [A^{\rm cl}]=\lim_{\beta\to \infty}{\rm Tr}\left[e^{-\beta H_\lambda}\right]=\int \mathcal{D}X\mathcal{D}Y\, e^{-I_{\lambda}}\,,
\end{align}
with $I_{\lambda}=\int d \tau\, H_\lambda$, where $\tau$ is defined on the interval $\tau \in [0,\,\infty]$.
By explicit calculations, we obtain
\begin{align}
     W_\lambda^{(N)} [A^{\rm cl}]=\int d^4x_{\rm E}\,\left[
     \frac{1}{4g_e^2 }F_{\mu\nu}F^{\mu\nu} 
     -
     \mathcal{L}_{\rm DBI}^{(N)}
     \right]\,,
\end{align}
where the nonlinear effects $ \mathcal{L}_{\rm DBI}^{(N)}$ on the Maxwell theory are given as
\begin{align}
     \mathcal{L}_{\rm DBI}^{(N)}=
     \lambda^2 \frac{1}{2M^2_X} \left(g_e^{-2}\,J^{(1)}_X\right)^2+
     \lambda^2 
     \frac{1}{2M^2_Y}
     \left(
     g_e^{-2}\,
      J^{(1)}_Y
      +
      \sum_{n=2}^N (g_e^{-2})^n \, J^{(n)}_Y
     \right)^2\,.\label{eq:DBI_Wlambda}
\end{align}
This result is perturbative in $g_e^{-2}$ and is truncated at the $N$th order.
Substituting Eq.~\eqref{eq:DBI_Wlambda} into Eq.~\eqref{eq:quanrel}, we obtain
\begin{align}
    S^{(N)}\left(\rho_{\rm R}\|\rho_{\rm T}\right)
    &=
    \int d^4 x_{\rm E}\, \mathcal{L}_{\rm DBI}^{(N)}\notag
    \\
    &=\lambda^2 \int d^4 x_{\rm E}\, 
    \left[ \frac{1}{2M^2_X} \left(g_e^{-2}\,J^{(1)}_X\right)^2+ 
     \frac{1}{2M^2_Y}
     \left(
     g_e^{-2}\,
      J^{(1)}_Y
      +
      \sum_{n=2}^N (g_e^{-2})^n \, J^{(n)}_Y
     \right)^2\right]\geq 0\,.\label{eq:DBI_per_rel}
\end{align}
From this expression, it is clear that the non-negativity of relative entropy holds and yields bounds on the nonlinear corrections to Maxwell theory encoded in $ \mathcal{L}_{\rm DBI}^{(N)}$.
That is, within the perturbative analysis above, the relative entropy yields bounds on the EFT without requiring the detailed form of the UV interactions, but only the existence of heavy degrees of freedom, such as auxiliary fields, from which the DBI Lagrangian can be generated.
%
In what follows, we further examine this bound by considering two scenarios: the magnetic model and the electric model.

\begin{itemize}
    \item {\bf Magnetic model} --- We first consider the magnetic model ($\vec{B}\neq 0$ and $\vec{E}=0$).
    Using $F^{i0}=0$ and $-\epsilon^{ijk}F_{jk}/2=B^i$ and setting $\lambda=1$, Eq.~\eqref{eq:DBI_per_rel} is rewritten as
     \begin{align}
    S^{(N)}\left(\rho_{\rm R}\|\rho_{\rm T}\right)
    =\int d^4x_{\rm E}\, \mathcal{L}^{(N)}_{\rm DBI}\left((g_e^{-1}\bar{B})^2\right)=
    \Lambda^4 \int d^4x_{\rm E} \sum_{n= 2}^N  d_n\, (g_e^{-1}\bar{B})^{2n}\,,\label{eq:rel_DBI_B_n}
    \end{align}
    where
    \begin{align}
    d_n\equiv  i^{2n}\frac{(2n)!}{(2n-1)(n!)^2 4^n}~{\rm for}~n\geq 2\,,\qquad \bar{B}\equiv |\vec{B}|/{\Lambda^2}\,.\label{eq:dn}
    \end{align}
    For instance, the coefficients $d_n$ up to sixth order in $(g_e^{-1}\bar{B})^2$ are given by
    \begin{align}
    d_2 =\frac{1}{8}\,,\quad d_3 = -\frac{1}{16}\,,\quad  d_4=\frac{5}{128}\,,\quad  d_5=-\frac{7}{256},\quad d_6 =\frac{21}{1024}\,.
    \end{align}
    As in the QED cases, the first-order contribution in $(g_e^{-1}\bar{B})^2$ is non-negative, and the leading term in the operator expansion respects the non-negativity of the relative entropy.
    At higher orders, however, subleading terms in $(g_e^{-1}\bar{B})^2$ can become negative, and the non-negativity of the relative entropy may be violated when the perturbative expansion is no longer valid.
     This apparent violation of non-negativity is understood as arising from the spurious non-Hermitian density operators explained in Section~\ref{sec:pert}, which are generated by expanding in $(g_e^{-1}\bar{B})^2$ instead of $\lambda$.
    That is, the non-negativity of the relative entropy --- an implicit indicator of unitarity --- can be used to identify the region of $(g_e^{-1}\bar{B})^2$ in which the perturbative analysis is valid.    
    See also Section~\ref{sec:pe_fer} for a related discussion.

    \item {\bf Electric model} --- Replacing $|\vec{B}|$ with $i|\vec{E}|$ in Eq.~\eqref{eq:rel_DBI_B_n}, we obtain the perturbative definition of the relative entropy for the electric model ($\vec{E}\neq0$ and $\vec{B}= 0$), as
    \begin{align}
    S^{(N)}\left(\rho_{\rm R}\|\rho_{\rm T}\right)
    =\int d^4x_{\rm E}\, \mathcal{L}^{(N)}_{\rm DBI}\left((i g_e^{-1}\bar{E})^2\right)=
    \Lambda^4 \int d^4x_{\rm E}\, \sum_{n= 2}^N d_n\, (ig_e^{-1}\bar{E})^{2n}\,,\label{eq:rel_DBI_n}
    \end{align}
    where $\bar{E}\equiv |\vec{E}|/{\Lambda^2}$.
    From Eq.~\eqref{eq:rel_DBI_n} and $ d_n\,i^{2n}>0$, we find, analogously to the cases of fermionic and scalar QED, that the relative entropy remains non-negative at each order in the perturbative expansion.
This perturbative analysis is valid provided that higher-order contributions are sufficiently suppressed compared to lower-order terms.
Within this regime, the positivity of the relative entropy implies the positivity of the sum of all higher-dimensional operators in the EFT, denoted by $ \mathcal{L}^{(N)}_{\rm DBI}$.
If higher-order effects become comparable to the leading contributions, one must take the limit $N \to \infty$ and resum the perturbative results.
In Section~\ref{sec:Borel_DBI}, we study this limit employing a resummation technique.

\end{itemize}

\subsection{Singularities and resummation of the relative entropy}
\label{sec:Borel_DBI}
Using the results of Section~\ref{sec:perp_DBI}, we analyze the asymptotic behavior of the operator expansion of the relative entropy.
Combining the behavior of the relative entropy near its singularities with its general properties, we then construct the resummed relative entropy.

\begin{itemize}
    \item {\bf Magnetic model} --- Let us first consider the purely magnetic case ($\vec{B}\neq 0$ and $\vec{E}=0$). %
    Using Eq.~\eqref{eq:dn}, the coefficient $d_n$ can be rewritten as
     \begin{align}
    d_n =i^{2n} \frac{(2n)!}{(2n-1)(n!)^2 4^n}=-i^{2n}\frac{\Gamma(n-1/2)}{\Gamma(-1/2)\Gamma(n+1)}\,,
    \end{align}
    where we have used $\Gamma(-1/2)=-2\sqrt{\pi}$, $\Gamma(n+1)=n!$, and $\Gamma(n-1/2)=\tfrac{(2n-2)!}{4^{n-1}(n-1)!}\sqrt{\pi}$.
    In the limit $n\to \infty$, the asymptotic behavior of the coefficient $d_n$ is given by
    \begin{align}
    d_n\to i^{2n}\frac{n^{-3/2}}{2\sqrt{\pi}}  \,,\label{eq:dn_asym0}
    \end{align}
    where we have used the Stirling expansion $\Gamma(z)= \sqrt{2\pi}z^{z-1/2}e^{-z}\left(1+\mathcal{O}(z^{-1})\right)$ in the limit $z\to \infty$.
     Using Eq.~\eqref{eq:dn_asym0} and the ratio test, we find that the radius of convergence of the corresponding series is given by $\lim_{n\to \infty}|d_n/d_{n+1}|=1$, implying the existence of a singularity in the resummed series.
    Therefore, the parameter region $|g_e^{-1}\bar{B}|^2 \leq 1$ defines the domain of validity of perturbation theory. 
    In other words, the series~\eqref{eq:rel_DBI_B_n} has a finite radius of convergence, so Borel--Laplace resummation is not required within its domain of convergence.
    As discussed below, even in this case the relative entropy can be resummed in the vicinity of a singularity, based solely on its general properties.

    Using Eq.~\eqref{eq:rel_DBI_B_n}, we obtain the nonlinear contribution $ \mathcal{L}^{(\infty)}_{\rm DBI}\left((g_e^{-1}\bar{B})^2\right)$ as
    \begin{align}
 \mathcal{L}^{(\infty)}_{\rm DBI}\left((g_e^{-1}\bar{B})^2\right)=\Lambda^4\, \mathcal{I}_{\rm DBI}\left((g_e^{-1}\bar{B})^2\right)\,,\quad \mathcal{I}_{\rm DBI}\left(x\right)\equiv \sum_{n=2}^\infty d_n x^{n}\,.
    \end{align}
    By the Cauchy integral formula, the coefficient can be written as
    \begin{align}
    d_n=\frac{1}{2\pi i}\oint_{\mathcal{C}_0}\frac{\mathcal{I}_{\rm DBI}\left(z\right)}{z^{n+1}}dz\underset{n\to \infty}{=}  i^{2n}\frac{n^{-3/2}}{2\sqrt{\pi}}\,,\label{eq:asym_tayl}
    \end{align}
    where $\mathcal{C}_0$ denotes a closed integration contour encircling $z=0$, and the function $\mathcal{I}_{\rm DBI}\left(z\right)$ is assumed to possess a singularity at $z=R$. 
    Given the large-order asymptotics of the Taylor coefficients~\eqref{eq:asym_tayl}, one can infer the location $R$ of the nearest singularity, the associated algebraic exponent $\alpha$, and an overall normalization constant $c$, under the assumption that the singularity is of algebraic type, $\mathcal{I}_{\rm DBI}(z)|_{\rm sing}= c \left(R-z\right)^{-\alpha}$.
    Since the large-radius contour avoiding the singularities does not contribute in the limit $n\to\infty$, only the Hankel contour survives.
    By introducing the variable transformation $z=R (1+t/n)$, we find $z^{-(n+1)}=R^{-(n+1)}(1+t/n)^{-(n+1)}\underset{n\to \infty}{=} R^{-(n+1)} e^{-t}$, and $R-z=-Rt/n$.
    Substituting these expressions into Eq.~\eqref{eq:asym_tayl}, we obtain
    \begin{align}
    d_n &\underset{n\to \infty}{=} \frac{1}{2\pi i} \int_{-\mathcal{C}_H}\frac{c\, (-Rt/n)^{-\alpha}}{R^{n+1}e^{t}}\frac{R}{n}dt
    =-e^{i\pi \alpha}\frac{c}{2\pi i\, R^{\alpha+n}}\, n^{\alpha-1}\int_{\mathcal{C}_H} e^{-t} t^{-\alpha}dt\,,\label{eq:dn_asym}
    \end{align}
    where $\mathcal{C}_H$ denotes the Hankel contour (around the origin counter clockwise).
     Thus, using a well-known formula,
    \begin{align}
   \int_{\mathcal{C}_H} e^{-t} t^{-\alpha}dt&=-i2e^{i\alpha \pi} \sin (\pi\alpha)\Gamma \left(1-\alpha\right)\,,\label{eq:CH_int}
    \end{align}
    the asymptotic behavior of the coefficient $d_n$ is given by
    \begin{align}
        d_n&\underset{n\to \infty}{=} 
        \frac{c}{\pi  R^{\alpha+n}}\, n^{\alpha-1} \,  \sin (\pi\alpha)\, \Gamma(1-\alpha)=\frac{c}{\Gamma(\alpha)  R^{\alpha+n}}\, n^{\alpha-1}\,,\label{eq:dn_asym_2}
    \end{align}
    where we have used $\Gamma(z)\Gamma(1-z)=\tfrac{\pi}{\sin\pi z}$.
    By comparing Eq.~\eqref{eq:asym_tayl} with Eq.~\eqref{eq:dn_asym_2}, we find
    \begin{align}
    \alpha=-\frac{1}{2}\,,\quad R=-1\,,\quad c=i\,,
    \end{align}
    where $\Gamma\left(-1/2\right)=-2\sqrt{\pi}$ is used.
    That is, the singular part of $\mathcal{I}_{\rm DBI}\left(z\right)$ near the singularity at $z=-1$ behaves as
    \begin{align}
    \mathcal{I}_{\rm DBI}\left(z\right)\bigg|_{\rm sing}= -\sqrt{1+z}\,.
    \end{align}
    From this expression, we obtain the singular part of the relative entropy as
    \begin{align}
    S^{(\infty)}\left(\rho_{\rm R}\|\rho_{\rm T}\right)\bigg|_{\rm sing}
    =\int d^4x_{\rm E}\, \mathcal{L}^{(\infty)}_{\rm DBI}\left((g_e^{-1}\bar{B})^2\right)\bigg|_{\rm sing}=
    -\Lambda^4 \int d^4x_{\rm E}\, \sqrt{1+(g_e^{-1}\bar{B})^2}\,.\label{eq:sing_rel_DBI}
    \end{align}
    Since Eq.~\eqref{eq:sing_rel_DBI} represents only the singular part, the regular part governing the low-order Taylor coefficients has not been properly captured.
    However, the leading two Taylor coefficients can be inferred from general properties of the relative entropy.
    In particular, the relative entropy vanishes for $g_e^{-1}\bar{B}=0$, where the interaction terms disappear, and it is sensitive only to nonlinear corrections rather than renormalizable terms.
    Consequently, we obtain the resummed relative entropy as
    \begin{align}
    S^{(\infty)}\left(\rho_{\rm R}\|\rho_{\rm T}\right)
    &=\int d^4x_{\rm E}\, \mathcal{L}^{(\infty)}_{\rm DBI}\left((g_e^{-1}\bar{B})^2\right)\notag
    \\
    &\simeq
    \Lambda^4 \int d^4x_{\rm E} \left[\left(1-\sqrt{1+(g_e^{-1}\bar{B})^2}\right)+\frac{1}{2}(g_e^{-1}\bar{B})^2\right]\geq 0\,.\label{eq:DBI_B_resum}
    \end{align}
    This resummed relative entropy captures the asymptotic behavior of the Taylor coefficients, corresponding to the singular part, as well as the leading two Taylor coefficients, which encode part of the regular contribution.
    As will be shown below, the resummed relative entropy above is in full agreement with the exact results presented in Section~\ref{sec:non_DBI}.
    This resummed relative entropy is always non-negative.
    Namely, a proper reconstruction of the relative entropy removes the spurious violation of unitarity observed in the perturbative analysis of Section~\ref{sec:perp_DBI}.

    \item {\bf Electric model} --- We next consider the purely electric case ($\vec{E}\neq 0$ and $\vec{B}=0$).
    The analysis proceeds in parallel with the magnetic case. 
    By replacing $|\vec{B}|$ with $i|\vec{E}|$ in Eq.~\eqref{eq:DBI_B_resum}, we obtain the resummed relative entropy for the electric case as
    \begin{align}
    S^{(\infty)}\left(\rho_{\rm R}\|\rho_{\rm T}\right)
    &=\int d^4x_{\rm E}\, \mathcal{L}^{(\infty)}_{\rm DBI}\left((ig_e^{-1}\bar{E})^2\right)\notag
    \\
    &\simeq
    \Lambda^4 \int d^4x_{\rm E} \left[\left(1-\sqrt{1-(g_e^{-1}\bar{E})^2}\right)-\frac{1}{2}(g_e^{-1}\bar{E})^2\right]\,.\label{eq:DBI_E_resum}
    \end{align}
    This expression implies that the resummed relative entropy is non-negative in the perturbative regime $(g_e^{-1}\bar{E})^2\leq 1$, which lies within the radius of convergence of the operator expansion.
    In contrast, in the non-perturbative regime $(g_e^{-1}\bar{E})^2>1$, the resummed relative entropy becomes complex, and its non-negativity is violated.
    Consequently, even in the DBI model, a violation of the non-negativity of the relative entropy signals the presence of non-perturbative effects.

\end{itemize}
    
\subsection{Non-perturbative analysis}
\label{sec:non_DBI}
Finally, we evaluate the relative entropy non-perturbatively, incorporating operator effects to all orders, and verify the results in Section~\ref{sec:Borel_DBI} within a more robust framework.
Assuming constant electric and magnetic fields $\vec{E}$ and $\vec{B}$, the relative entropy without performing the operator expansion is given by
    \begin{align}
    S\left(\rho_{\rm R}\|\rho_{\rm T}\right)
    &=\int d^4x_{\rm E}\, \mathcal{L}_{\rm DBI}\notag
    \\
    &=
   \int d^4x_{\rm E}\,
   \left[\Lambda^4\left(
    1-
    \sqrt{1+\frac{1}{g_e^2 \Lambda^4}(-E^2+B^2)-\frac{16}{g_e^4 \Lambda^8}\left(\vec{E}\cdot \vec{B}\right)^2}
    \right)
    +
    \frac{1}{2g_e^2} (-E^2+B^2)
    \right]\,.\label{eq:non_DBI_rel}
    \end{align}
In what follows, we present the details for two cases: the magnetic and electric models.
\begin{itemize}
    
    \item {\bf Magnetic model} --- From Eq.~\eqref{eq:non_DBI_rel}, the relative entropy for the magnetic case ($\vec{E}= 0$ and $\vec{B}\neq 0$), is obtained as
    \begin{align}
    S\left(\rho_{\rm R}\|\rho_{\rm T}\right)
    &=\int d^4x_{\rm E}\, \mathcal{L}_{\rm DBI}\left((g_e^{-1}\bar{B})^2\right)\notag
   \\ 
    &=
   \Lambda^4\,\int d^4x_{\rm E}\,
   \left[\left(
    1-
    \sqrt{1+(g_e^{-1}\bar{B})^2}
    \right)
    +
    \frac{1}{2}  (g_e^{-1}\bar{B})^2
    \right]\,.
    \end{align}
    This result is consistent with the resummed relative entropy derived in Eq.~\eqref{eq:DBI_B_resum}.

     \item {\bf Electric model} --- Similarly, from Eq.~\eqref{eq:non_DBI_rel}, the relative entropy for the electric case ($\vec{E}\neq 0$ and $\vec{B}= 0$), is obtained as
      \begin{align}
    S\left(\rho_{\rm R}\|\rho_{\rm T}\right)
    &=\int d^4x_{\rm E}\,\mathcal{L}_{\rm DBI}\left((ig_e^{-1}\bar{E})^2\right)\notag
    \\
    &=
   \Lambda^4\,\int d^4x_{\rm E}\,
   \left[\left(
    1-
    \sqrt{1-(g_e^{-1}\bar{E})^2}
    \right)
    -
    \frac{1}{2}  (g_e^{-1}\bar{E})^2
    \right]\,.
    \end{align}
    This result is in agreement with the resummed relative entropy derived in Eq.~\eqref{eq:DBI_E_resum}.
    As discussed in Section~\ref{sec:Borel_DBI}, in the non-perturbative regime $(g_e^{-1}\bar{E})^2>1$, the non-negativity of the relative entropy is violated. 
    This violation is understood as arising from a non-Hermitian Hamiltonian.
    Indeed, the Hamiltonian~\eqref{eq:HI_DBI} for the electric case is given by
    \begin{align}
    H_{\rm I}=\int d^3x\, \left[
    -
    M_Y \Lambda^2
    \left(
    1- \sqrt{1-(g_e^{-1}\bar{E})^2}
    \right)Y\right]\,.
    \end{align}
    It is evident that this interaction term becomes non-Hermitian in the non-perturbative regime $(g_e^{-1}\bar{E})^2>1$.

\end{itemize}

\section{Relative entropy in nonlinear electrodynamics}
\label{sec:gen}
So far, we have considered UV theories underlying the EFT of nonlinear electrodynamics and examined the resulting positivity bounds.
In this section, we derive these bounds from minimal assumptions, focusing on typical nonlinear EFTs classified by the growth behavior of their operator expansions.
Specifically, we assume that a given EFT is generated by interactions between heavy and light degrees of freedom, {\it i.e.}, that the theory admits a UV completion (as discussed in Section~\ref{sec:DBI}, including cases in which the EFT can be generated by auxiliary fields).
We consider the following EFT of electromagnetism in Minkowski spacetime:
\begin{align}
    \mathcal{L}_{\rm EFT}
    =
    -\frac{1}{4}F_{\mu\nu}F^{\mu\nu}+\mathcal{L}_{\rm nonlin}\,,
\end{align}
where the nonlinear correction $ \mathcal{L}_{\rm nonlin}$ is assumed to be generated by interactions between heavy and light degrees of freedom. 
In particular, following the previous sections, we consider two cases: magnetic and electric models.
Under these assumptions, the relative entropy is given by
\begin{align}
    S\left(\rho_{\rm R}\|\rho_{\rm T}\right)
    =\int d^4x_{\rm E}\, \mathcal{L}_{\rm nonlin}\,.
\end{align}
Since, for constant electromagnetic fields, the independent Lorentz and gauge invariants are $F_{\mu\nu}F^{\mu\nu}$ and $F_{\mu\nu}\widetilde{F}^{\mu\nu}$, the nonlinear corrections $\mathcal{L}_{\rm nonlin}$ depend only on even powers of $|\vec{B}|$ ($|\vec{E}|$) in the magnetic (electric) case.
The nonlinear corrections to the EFT are formally expressed as
\begin{align}
     \mathcal{L}_{\rm nonlin}\left(o^2\right)
     =
     \Lambda^4\, \sum_{n=2}^\infty c_n\, o^{2n}\,, \label{eq:non_form_exp}
\end{align}
with $o^2= \bar{B}^2$ with $\bar{B}=|\vec{B}|/\Lambda^2$ for the magnetic case, $o^2= \left(i\bar{E}\right)^2$ with $\bar{E}=|\vec{E}|/\Lambda^2$ for the electric case, and $\Lambda$ a characteristic UV mass scale.
In what follows, we consider two representative classes of nonlinear EFTs, distinguished by the growth behavior\footnote{More complicated behaviors can also be considered. 
However, the following method can be straightforwardly generalized to such cases.} of the operator expansion coefficients: factorial growth, $c_n = C\, (\tau_{\rm pole})^{-2n}\, (2n-\ell)!$, and power-law growth, $c_n = C\, (\tau_{\rm pole})^{-2n}\, n^\gamma$, where $C$, $\tau_{\rm pole}$, $\ell$, and $\gamma$ are constants characterizing the large-order behavior.

\subsection{EFT with factorial growth}
Let us consider the EFT in Eq.~\eqref{eq:non_form_exp} with factorially growing coefficients\footnote{The analysis below can be straightforwardly generalized to the case $c_n= C\, (\tau_{\rm pole})^{-2n}\, \Gamma(2n-\ell+1)$.}, $c_n= C\, (\tau_{\rm pole})^{-2n}\, (2n-\ell)!$.
The Borel transform is then given by
\begin{align}
    \mathcal{B}(\tau)= \sum_{n=2}^\infty \frac{c_n }{(2n-\ell)!}\,\tau^{2n-\ell}\,,\qquad \frac{c_n}{(2n-\ell)!}= C \,(\tau_{\rm pole})^{-2n}\,.\label{eq:Borel_fact_E}
\end{align}
Accordingly, the singular part of the Borel transformation near the singularity is given by 
\begin{align}
    \tau^\ell\, \mathcal{B}_{\rm sing}(\tau)= \frac{C}{1-(\tau/\tau_{\rm pole})^2}\,.
\end{align}
The leading two coefficients in the Borel transformation are fixed by the properties of the relative entropy: it vanishes in the zero-field limit and is sensitive only to nonlinear effects beyond renormalizable terms.
We therefore obtain
\begin{align}
    \tau^\ell\, \mathcal{B}(\tau)&\simeq  C\frac{(\tau/\tau_{\rm pole})^4}{1-(\tau/\tau_{\rm pole})^2}\,.\label{eq:Borel_rel_E}
\end{align}
When the EFT coefficients are fully determined by $c_n= C\, (\tau_{\rm pole})^{-2n}\, (2n-\ell)!$, the Borel transform is uniquely fixed, and the resulting expression becomes exact.
For example, when multiple poles are present, this provides an approximation that retains only the nearest-pole contribution, and $c_n= C\, (\tau_{\rm pole})^{-2n}\, (2n-\ell)!$ should be interpreted as the asymptotic behavior of the coefficients.
Then, the Borel--Laplace resummation of the relative entropy is given by (see Eq.~\eqref{eq:Borel_just})
\begin{align}
    S^{(\infty)}\left(\widetilde{\rho}_{\rm R}\|\widetilde{\rho}_{\rm T}\right)
    &=
    \int d^4x_{\rm E}\, \mathcal{L}_{\rm nonlin}\left(o^2\right)\notag
    \\
    &\simeq   C\,\Lambda^4\,\int d^4 x_{\rm E}\, 
    \begin{cases}
        \int_0^{\infty}e^{-t} t^{-\ell}\frac{ (o t/\tau_{\rm pole})^{4}}{1-(o t/\tau_{\rm pole})^2}dt\,, & (o/\tau_{\rm pole})^2 <0\,,
        \\
        \mathcal{P}\,\int_0^{\infty}e^{-t} t^{-\ell}\frac{ (o t/\tau_{\rm pole})^{4}}{1-(o t/\tau_{\rm pole})^2}dt\pm i \frac{\pi}{2}\frac{e^{-|\tau_{\rm pole}/o|}}{(\tau_{\rm pole}/o)^{\ell-1}}\,, & (o/\tau_{\rm pole})^2 >0\,,
    \end{cases}\label{eq:gen_resum_rel}
\end{align}
In the following, we consider the two cases $(o/\tau_{\rm pole})^2 <0$ and $(o/\tau_{\rm pole})^2 >0$.
In the examples discussed above, these two cases correspond to the magnetic and electric models, respectively.
As will be seen, this correspondence can be inferred from the non-negativity of the relative entropy and the presence of system instability, even without detailed knowledge of the underlying UV theory.

\begin{itemize}
    \item {\bf $(o/\tau_{\rm pole})^2 <0$} --- In this case, the Borel integral does not encounter the nearest singularity along the integration contour.
    Then, the non-negativity of the relative entropy implies 
    \begin{align}
    S^{(\infty)}\left(\widetilde{\rho}_{\rm R}\|\widetilde{\rho}_{\rm T}\right)
    =
     C\,\underbrace{\Lambda^4\,\int d^4 x_{\rm E}\,\int_0^{\infty}e^{-t} t^{-\ell}\frac{ (o t/\tau_{\rm pole})^{4}}{1-(o t/\tau_{\rm pole})^2}dt}_{>0}>0\quad \Rightarrow \quad C>0\,,
    \end{align}
    where we have used the fact that the integrand $e^{-t}t^{-\ell} (o t/\tau_{\rm pole})^{4}/ (1-(o t/\tau_{\rm pole})^2)$ is non-negative when $(o/\tau_{\rm pole})^2 <0$.
    Therefore, from the non-negativity of the resummed perturbative relative entropy, we obtain a bound on the asymptotic behavior of the EFT operator expansion.
    For $C>0$, the above non-negativity condition holds throughout the entire parameter region of $(o/\tau_{\rm pole})^2<0$.
    As we will see, this contrasts with the case $(o/\tau_{\rm pole})^2>0$, where singularities lie on the integration contour and can signal an instability.
    In other words, the non-negativity of the relative entropy  for this case is insensitive to the analytic continuation from Euclidean to Minkowski spacetime.
    Therefore, from a purely IR perspective, this case is naturally identified with the magnetic background, $o=\bar{B}$.
    In fact, the case $(o/\tau_{\rm pole})^2>0$ can be obtained via the replacement $o \to io$ (equivalently, $|\vec{B}| \to i|\vec{E}|$, corresponding to the analytic continuation from Euclidean to Minkowski spacetime), as we will see below.

    \item {\bf $(o/\tau_{\rm pole})^2 >0$} --- The nearest singularity lies on the integration contour.
The resulting imaginary part exhibits a sign ambiguity, depending on whether the pole is avoided from above or below in the choice of contour.
    It therefore reflects the contour dependence of the resummation.
    Since perturbation theory alone does not generally fix this choice, an additional prescription is required to resolve the resulting non-perturbative ambiguity.
    If an imaginary part remains after resolving this ambiguity, it signals a violation of the non-negativity of the relative entropy and indicates an instability, reflecting an effective non-Hermiticity of the density operator.
    As shown in fermionic and scalar QED, this instability is realized as the Schwinger effect, which arises from the analytic continuation of the electric field.
    Moreover, the real part of the resummed perturbative result also encodes this instability, as it manifests through the violation of the non-negativity of the relative entropy.

    The perturbative expansion in $o^2$ does not capture the imaginary part associated with the Schwinger effect, since it arises from a genuinely non-perturbative contribution.
Consequently, the resummed perturbative relative entropy corresponds to the real part of Eq.~\eqref{eq:gen_resum_rel}:
\begin{align}
    {\rm Re}\, S^{(\infty)}\left(\widetilde{\rho}_{\rm R}\|\widetilde{\rho}_{\rm T}\right)\simeq   C\,\underbrace{\Lambda^4\,\int d^4 x_{\rm E}\, \mathcal{P}\,\int_0^{\infty}e^{-t} t^{-\ell}\frac{ (o t/\tau_{\rm pole})^{4}}{1-(o t/\tau_{\rm pole})^2}dt }_{\geq 0~{\rm for}~ |\tau_{\rm pole}/\kappa o|\geq 1}\,.\label{eq:resum_pert}
\end{align}
As shown in Ref.~\cite{Conzinu:2026cuf}, for positive $C$ the resummed perturbative relative entropy is positive in the weak-coupling regime, $|\tau_{\rm pole}/\kappa o|\geq 1$, where $\kappa\equiv 4-\ell$.
Therefore, the non-negativity of the resummed perturbative relative entropy in the weak-coupling regime yields a positivity bound on the asymptotic behavior of the EFT operator expansion:
\begin{align}
    {\rm Re}\, S^{(\infty)}\left(\widetilde{\rho}_{\rm R}\|\widetilde{\rho}_{\rm T}\right)>0\quad \Rightarrow \quad C>0\,.
\end{align}
In perturbative computations, no instability arises, and the non-negativity in the weak-coupling regime follows from perturbative unitarity and stability.
For consistency across all parameter regimes, the positivity condition on the coefficient $C$, derived in the weak-coupling regime, should remain valid even in the strong-coupling regime within our framework.

By contrast, the resummed relative entropy itself can become negative in the strong-coupling regime.
As shown in Ref.~\cite{Conzinu:2026cuf}, for $\kappa\leq 0$, the resummed relative entropy is positive even for the strong-coupling regime.
However, for $\kappa>0$, the resummed relative entropy becomes negative in the strong-coupling regime, signaling the non-perturbative effects such as instabilities of the system.
Therefore, from a purely IR perspective, this case is naturally identified with the electric model, $o=i\bar{E}$.
Even for $\kappa\leq 0$, where the real part remains non-negative, the same identification may follow from an analytic continuation in $\ell$.

\end{itemize}

\subsection{EFT with power-law growth}
We next consider an EFT with power-law growth, focusing on both the magnetic and electric cases.
For the power-law growth, the nonlinear contribution is given as
    \begin{align}
    \mathcal{L}_{\rm nonlin}\left(o^2\right)=\Lambda^4\, \mathcal{I}\left(o^2\right)\,,\quad \mathcal{I}\left(z\right)\equiv\sum_{n=2}^\infty c_n z^{n},\quad c_n= C\, (\tau_{\rm pole})^{-2n}\, n^\gamma\,,
    \end{align}
    with a radius of convergence defined by $\lim_{n\to\infty}|c_n/c_{n+1}|=|\tau_{\rm pole}|^2$.
    By the Cauchy integral formula, the coefficient can be written as
    \begin{align}
    c_n=\frac{1}{2\pi i}\oint_{\mathcal{C}_0}\frac{\mathcal{I}(z)}{z^{n+1}}dz\underset{n\to \infty}{=}  C\, (\tau_{\rm pole})^{-2n}\, n^\gamma\,.\label{eq:cn_gen}
    \end{align}
    For an algebraic singularity of the form $\mathcal{I}_{\rm sing}(z) = c\, (R-z)^{-\alpha}$, we obtain from Eq.~\eqref{eq:dn_asym_2},
    \begin{align}
        c_n&\underset{n\to \infty}{=} 
       \frac{c}{\Gamma(\alpha)  R^{\alpha+n}}\, n^{\alpha-1}\,.\label{eq:asy_gen}
    \end{align}
    It should be noted that $\Gamma(\alpha)$ diverges for $\alpha \in {0,-1,-2,\dots}$, so the asymptotic formula~\eqref{eq:asy_gen} applies only for other values of $\alpha$.
    By comparing Eqs.~\eqref{eq:cn_gen} and \eqref{eq:asy_gen}, we obtain
    \begin{align}
    \alpha=\gamma+1\,,\quad R=(\tau_{\rm pole})^2\,,\quad c= C \,(\tau_{\rm pole})^{2(\gamma+1)}\,  \Gamma (\gamma+1)\,.
    \end{align}
    That is, the singular part of $\mathcal{I}(z)$ near $z=(\tau_{\rm pole})^2$ behaves as
    \begin{align}
    \mathcal{I}_{\rm sing}(z)=\frac{C \,   \Gamma (\gamma+1)}{ \left(1-z/(\tau_{\rm pole})^2\right)^{\gamma+1}}\,.
    \end{align}
    The leading coefficients in $\mathcal{I}(z)$ are constrained by general properties of the relative entropy, namely its vanishing in the zero-field limit and its sensitivity only to nonlinear effects beyond renormalizable terms:
    \begin{align}
    \mathcal{L}_{\rm nonlin}\left(o^2\right)&\simeq 
    C\,  \Lambda^4 \,\Gamma (\gamma+1)
    \left[
    \frac{1}{\left(1-\left(o/\tau_{\rm pole}\right)^2\right)^{\gamma+1}}
    -
    \left(1+(\gamma+1) \left(o/\tau_{\rm pole}\right)^2\right)
    \right]\,,\label{eq:Lele_non_power}
    \end{align}
    where, in particular, $\Gamma(\gamma+1) > 0$ for $\gamma+1 > 0$.
    This resummation captures the leading two Taylor coefficients and the asymptotic behavior of the higher-order coefficients, but it remains ambiguous up to additional polynomial contributions.
    As a result, this resummation should be regarded as an approximation.
    However, for $\gamma+1>0$, the singular contribution dominates near the singularity, rendering this expression quantitatively reliable.
    From Eq.~\eqref{eq:Lele_non_power}, the resummed perturbative relative entropy is given by
    \begin{align} S^{(\infty)}\left(\widetilde{\rho}_{\rm R}\|\widetilde{\rho}_{\rm T}\right)
    &=\int d^4 x_{\rm E}\, \mathcal{L}_{\rm nonlin}\left(o^2\right)\notag
    \\
    &\simeq   C\times\underbrace{\Lambda^4 \, \Gamma (\gamma+1)\,\int d^4 x_{\rm E} \left[
    \frac{1}{\left(1-\left(o/\tau_{\rm pole}\right)^2\right)^{\gamma+1}}
    -
    \left(1+(\gamma+1) \left(o/\tau_{\rm pole}\right)^2\right)
    \right]}_{\geq 0~{\rm for}~\gamma+1> 0\,,~{\rm and}~(o/\tau_{\rm pole})^2<1}\,.\label{eq:rel_pow}
\end{align}
When $\gamma+1 > 0$, the factor $C$ must be positive, as a consequence of the non-negativity of the relative entropy\footnote{To see this, consider the function $f(x)=\left(1-x\right)^{-(\gamma+1)}-
    \left(1+(\gamma+1) x\right)$.
    The relative entropy for $(o/\tau_{\rm pole})^2>0$ is described by this function.
For $\gamma+1>0$, its derivative is given by
\begin{align}
\frac{df(x)}{dx}
=(1+\gamma)\left[
\frac{1}{(1-x)^{\gamma+2}}-1
\right] \,,
\end{align}
which is non-negative for $0\le x<1$.
Together with $f(0)=0$, this implies that $f(x)$ is monotonically increasing on $0\le x<1$.
Similarly, the function $f(-x)$ corresponding to $(o/\tau_{\rm pole})^2<0$ is also non-negative for $x\ge0$.
    }.
    Consequently, we obtain the positivity bound on the asymptotic behavior of the EFT operator expansion:
\begin{align} S^{(\infty)}\left(\widetilde{\rho}_{\rm R}\|\widetilde{\rho}_{\rm T}\right)>0\quad \Rightarrow \quad C>0\quad {\rm for}~\gamma+1>0\,,~{\rm and}~(o/\tau_{\rm pole})^2<1\,.
\end{align}
In the following, we further consider the two cases $(o/\tau_{\rm pole})^2<0$ and $(o/\tau_{\rm pole})^2>0$ by focusing on $\gamma+1>0$.
As we will see, the non-negativity of the relative entropy implies that these correspond to the magnetic and electric cases, respectively.

\begin{itemize}

\item {$(o/\tau_{\rm pole})^2<0$} --- In this case, the relative entropy does not encounter the nearest singularity along the real axis of the normalized operator $o^2$. 
Moreover, the relative entropy remains non-negative throughout the entire parameter region for $C>0$.
Thus, the non-negativity of the relative entropy is consistent with the stability of the system throughout the entire parameter region.
As we will see below, in the case $(o/\tau_{\rm pole})^2>0$, the non-negativity of the relative entropy can be violated, signaling the appearance of an instability.
Since the case $(o/\tau_{\rm pole})^2>0$ is obtained through the analytic continuation $o\to i o$ (equivalently, $|\vec{B}|\to i |\vec{E}|$), once this case is identified with the electric model, the present case $(o/\tau_{\rm pole})^2<0$ is naturally identified with the magnetic model.
That is, the non-negativity of the relative entropy in the present case is insensitive to the analytic continuation from Euclidean to Minkowski spacetime, implying that this case corresponds to the magnetic model with $o=\bar{B}$.

\item {$(o/\tau_{\rm pole})^2>0$} --- The relative entropy encounters the nearest singularity along the real axis of the normalized operator $o^2$. 
In the weakly coupled regime, $(o/\tau_{\rm pole})^2<1$, the relative entropy remains non-negative.
However, in the strong-coupling regime beyond the nearest singularity, the standard non-negativity property of the relative entropy can fail, signaling an instability associated with the analytic continuation from Euclidean to Minkowski spacetime.
Therefore, this case corresponds to the electric model with $o=i\bar{E}$. 

\end{itemize}

\section{Summary}\label{sec:summary}
In this paper, we have investigated nonlinear electrodynamic EFTs in which the relative entropy provides a natural probe of an infinite tower of higher-dimensional operators.
The central idea is that, when evaluated in suitable background electromagnetic fields, the relative entropy is sensitive to the full structure of the operator expansion.
In this sense, the relative entropy provides a framework for organizing information about the Wilson coefficients of nonlinear electrodynamic EFTs.
We illustrated this idea through several representative examples: fermionic QED, scalar QED, and the DBI model.
These examples exhibit qualitatively different types of UV behavior and therefore provide useful laboratories for understanding how relative entropy constrains nonlinear EFT effects.
We analyzed the relative entropy using three complementary approaches: a perturbative EFT analysis, resummation techniques such as Borel--Laplace resummation, and non-perturbative approaches including the Schwinger proper-time method.
In the perturbative analysis, we showed that the non-negativity of the relative entropy imposes sign constraints on a finite truncation of the tower of higher-dimensional operators in the weak-coupling regime, where the perturbative expansion remains valid.
These constraints generalize the familiar positivity properties of leading EFT coefficients to a broader class of nonlinear electrodynamic interactions.
When the perturbatively analyzed relative entropy shows a violation of non-negativity, this does not necessarily indicate a fundamental inconsistency of the underlying theory. 
Rather, it signals the breakdown of the truncated perturbative expansion. 
In this way, the apparent violation of non-negativity provides a quantitative diagnostic of the regime of validity of the perturbative expansion, reflecting perturbative unitarity inherent in the framework.
The situation is different once resummed or genuinely non-perturbative descriptions are considered. 
In these cases, the relative entropy can capture information that is invisible at any finite order in the EFT expansion.
We found that violations of non-negativity in the resummed or purely non-perturbative relative entropy are associated with physical instabilities of the system. 
Thus, while perturbative violations diagnose the failure of the perturbative expansion, non-perturbative violations diagnose an instability of the underlying background or vacuum. 
This distinction is one of the main lessons of our analysis.
We then extended the discussion beyond the explicit examples of QED and DBI theory to more general classes of UV completions.
In particular, we classified nonlinear electrodynamic EFTs according to the growth behavior of their operator expansions, including cases with factorial growth and cases with power-law growth. 
This classification allowed us to formulate general bounds on the nonlinear EFT effects.
We show that the non-negativity of the resummed relative entropy constrains the sign of the growth of the EFT coefficients, and that violations of the non-negativity of the resummed relative entropy themselves quantify the system’s
non-perturbative effects such as instabilities.
Our results therefore suggest that relative entropy furnishes a powerful and universal diagnostic for nonlinear EFTs. 
It simultaneously probes perturbative consistency and non-perturbative effects.
In particular, relative entropy offers a framework in which perturbative bounds on Wilson coefficients and non-perturbative stability criteria can be understood as different manifestations of the same underlying principle.
Several directions remain open for future work.
One natural extension is to study the consequences of these bounds for the entropy of charged black holes and their implications for the weak gravity conjecture.
We leave a detailed study of these questions for future work.


\begin{acknowledgments}
We thank the CERN Theory Department for financial support and for providing the environment in which this work was initiated.
DU is supported by grants from the ISF (No. 1002/23 and 597/24) and the BSF (No. 2021800).
PC acknowledges the support of INFN under the program ``QGSKY:Quantum Gravity in the SKY''.
\end{acknowledgments}

\appendix

\section{QED in Euclidean Space}
\label{sec:QED in E-space}
We consider the following Lagrangian in Minkowski spacetime:
\begin{align}
    \mathcal{L}=-\frac{1}{4}F_{\mu\nu}F^{\mu\nu} +\overline{\psi} \left(i\slashed{D}-m\right)\psi\,, \label{eq:qed}
\end{align}
where $\slashed{D} \equiv \gamma^{\mu} D_{\mu}$, $D_{\mu} \equiv \partial_{\mu} + i e A_{\mu}$, and $\overline{\psi}\equiv \psi^\dagger \gamma^0$.
Throughout this paper, we adopt the conventions 
$g_{\mu\nu} = (+,-,-,-)$, 
Greek (Latin) indices run over $0,1,2,3$ ($1,2,3$), 
and $\{\gamma^{\mu}, \gamma^{\nu}\} = 2 g^{\mu\nu}$.
We use the Weyl representation
\begin{align}
    \gamma^0=\begin{pmatrix}
    0 & \mathbb{I}_2
    \\
    \mathbb{I}_2 & 0
    \end{pmatrix}\,,\qquad \gamma^i = \begin{pmatrix}
    0 & \sigma_i
    \\
    -\sigma_i & 0
    \end{pmatrix}\,,
\end{align}
where $\sigma_i$ denotes the Pauli matrices.

\subsection{Fermionic path integral in QED}
\label{sec:C3}
The partition function for a given background field $A_{\mu}$ in Minkowski spacetime is
\begin{align}
    Z_{\rm M}[A]=\int \mathcal{D}\overline{\psi}\mathcal{D}\psi \, e^{i \int d^4x_{\rm M}\, \mathcal{L}}\,.\label{eq:part_Mink}
\end{align}
We evaluate this path integral in Euclidean spacetime by performing the Wick rotation $x^0 \to -i x^4$, with $x^4 \in \mathbb{R}$, 
under which $d^4x_{\rm M} = -i \, d^4x_{\rm E}$.
Hence,
\begin{align}
    {\rm exp}\left({i \int d^4x_{\rm M}\,\mathcal{L}}\right)\to {\rm exp}\left({i(-i) \int d^4x_{\rm E}\,\mathcal{L}^{\text{(analytic)}}}\right)\,,
\end{align}
where $\mathcal{L}^{\text{(analytic)}}$ denotes the analytically continued Lagrangian.
The Euclidean action is then defined by
\begin{align}
    I_{\rm E}\equiv -\int d^4x_{\rm E}\, \mathcal{L}^{\text{(analytic)}}\,.
\end{align}

We define the Euclidean gamma matrices by 
$\gamma^4_{\rm E} \equiv \gamma^0$, 
$\gamma^i_{\rm E} \equiv -i \gamma^i$ for $i = 1,2,3$.
These matrices satisfy $\{\gamma^I_{\rm E}, \gamma^J_{\rm E}\} = 2 \delta^{IJ}$ and 
$(\gamma_{\rm E}^I)^{\dagger} = \gamma_{\rm E}^I$, 
for $I,J = 1,\cdots,4$.
Thus, the Euclidean gamma matrices are Hermitian and satisfy the SO(4) Clifford algebra.
In what follows, indices $I,J$ are raised and lowered with $\delta^{IJ}$, and we therefore do not distinguish between upper and lower indices.
Under the Wick rotation, 
$\partial_0 = \partial/\partial x^0 \to i \partial_4 = i \partial/\partial x^4$, the Euclidean action becomes
\begin{align}
    I_{\rm E}&=-\int d^4 x_{\rm E }\, \left[-\frac{1}{4}(F_{IJ})^2+\overline{\psi}\left(i\gamma^{\mu}D_{\mu}-m\right)_{x^0\to -ix^4,\, A_0\to iA_4^{\rm E}}\psi\right]\notag
    \\
    &= \int d^4 x_{\rm E}\,  \left[\frac{1}{4}(F_{IJ})^2+\overline{\psi}\left(\gamma^{I}_{\rm E}D_{I}+m\right)\psi\right]\,.\label{eq:EucAction}
\end{align}
For the gauge field $A_{\mu}$, we adopt the conventions $A^{\rm M}_i\to A^{\rm E}_i$, and $A^{\rm M}_0\to i A_4^{\rm E}$.
The covariant derivative then takes the form $D_I=\partial_I +ie A_I^{\rm E}$.
The Euclidean field-strength tensor $F_{IJ} \equiv \partial_I A_J^{\rm E} - \partial_J A_I^{\rm E}$ is related to the physical electric and magnetic fields, $\vec{E}$ and $\vec{B}$, in Minkowski spacetime via the inverse analytic continuation ($\partial_4 \to -i\partial_0$ and $A_4^{\rm E} \to -i A_0^{\rm M}$) as follows:
\begin{align}
    F_{IJ}\to \begin{pmatrix}
    0 & -iF_{01} & -iF_{02} & -iF_{03}
    \\
    -iF_{10} & 0 & F_{12} & F_{13}
    \\
    -iF_{20} & F_{21} & 0 & F_{23}
    \\
    -iF_{30} & F_{31} & F_{32} & 0
    \end{pmatrix}
    =
    \begin{pmatrix}
    0 & -i E^1 & -i E^2 & -i E^3
    \\
    i E^1 & 0 & -B^3 & B^2
    \\
    i E^2 & B^3 & 0 & -B^1
    \\
    i E^3 & -B^2 & B^1 & 0
    \end{pmatrix}\,,\label{eq:BG_F}
\end{align}
where the electric and magnetic fields are defined by $E^i=F^{i0}$ and $B^i=-\epsilon^{ijk}F_{jk}/2$.
For the purpose of relative entropy calculations, we introduce a bookkeeping parameter $\lambda$ that tracks interaction terms and their perturbative contributions:
\begin{align}
    I_{{\rm E},\,\lambda}\equiv \int d^4x_{\rm E}\, \left[
    \frac{1}{4}(F_{IJ})^2+\overline{\psi}\left(\gamma^{I}_{\rm E}\partial_{I}+m\right)\psi
    +i\lambda e\, \overline{\psi}\gamma^I_{\rm E} A_I^{\rm E} \psi
    \right]\,.\label{eq:Ilambda_E}
\end{align}
The free and interacting theories correspond to $\lambda=0$ and $\lambda=1$, respectively.

\subsection{Free theory}
\label{sec:free_qed}
In the free limit of QED~\eqref{eq:qed}, the partition function is given by
\begin{align}
    Z_{0}^{\rm E}[A]\equiv \int \mathcal{D}\overline{\psi}\mathcal{D}\psi \,e^{-\int d^4x_{\rm E}\,(F_{IJ})^2/4}\,  {\rm exp}\left(-\int d^4x_{\rm E}\, \overline{\psi}\, {\rm M}_0 \psi \right)=e^{-\int d^4x_{\rm E}\,(F_{IJ})^2/4}\,{\rm det}\, {\rm M}_0\,  ,\label{eq:free_qed_part}
\end{align}
where ${\rm M}_0\equiv \gamma^I_{\rm E} \partial_I +m$.
The corresponding effective action is then given by
\begin{align}
    W_0^{\rm E}[A]&\equiv-\ln Z_0^{\rm E}[A]\notag
    \\
    &=\int d^4x_{\rm E}\,\left[\frac{1}{4}(F_{IJ})^2\right]-\ln {\rm det}\, {\rm M}_0=\int d^4x_{\rm E}\,\left[\frac{1}{4}(F_{IJ})^2\right]-{\rm Tr}\,\left[\ln \left(\slashed{\partial}_{\rm E}+m\right)\right]\,,\label{eq:W0}
\end{align}
where ${\rm Tr}$ denotes the trace, including both the spacetime integration $\int d^4x_{\rm E}$ and the trace over Dirac indices, while ${\rm tr_D}$ denotes the trace over Dirac indices only, and $\slashed{\partial}_{\rm E}\equiv \gamma_{\rm E}^I \partial_I$.
Here we have used the identity $\ln \det\, {\rm M}_0 = {\rm Tr}\,\left[\ln \, {\rm M}_0\right]$.
From Eq.~\eqref{eq:W0}, we obtain
\begin{align}
    \frac{d }{d m^2}W_0^{\rm E}[A]& =-\frac{1}{2}\int d^4x_{\rm E}\, {\rm tr_D}\left[{\langle x|}\frac{1}{-(\slashed{\partial}_{\rm E})^2+m^2}{|x\rangle}\right]\notag
    \\
    &=-\frac{1}{2} \int_0^{\infty} ds\, e^{-s m^2}\int d^4x_{\rm E}\, {\rm tr_D}\left[{\langle x|}e^{s (\slashed{\partial}_{\rm E})^2}{|x\rangle}\right]\,,\label{eq:dW0_free}
\end{align}
using the representation $1/\mathcal{A}=\int_0^\infty ds\,e^{-s \mathcal{A}}$ for real $\mathcal{A}>0$.
This is possible because $-(\slashed{\partial}_{\rm E})^2+m^2$ is a positive-definite operator in the free massive theory on Euclidean spacetime.
Integrating Eq.~\eqref{eq:dW0_free}, we obtain the proper-time representation of the effective action,
\begin{align}
W_0^{\rm E}[A]=\int d^4x_{\rm E}\,\left[\frac{1}{4}(F_{IJ})^2+\frac{1}{2} \int_0^{\infty} d s\, {1\over s}\, e^{-s m^2}\, {\rm tr_D} \left[ \langle x | e^{-\hat{H}_0 s } | x \rangle \right]\right]+\text{const}\,.\label{eq:W02}
\end{align}
where $\hat{H}_0 \equiv -(\partial_I)^2 = (\hat{p}_I)^2$ with ${\langle x|}\hat{p}_I = -i\tfrac{\partial}{\partial x^I}{\langle x|}$.
The Hamiltonian is manifestly Hermitian, $\hat{H}_0^{\dagger}=\hat{H}_0$.
We now evaluate the transition amplitude ${\langle y;0|x;s\rangle}$ by inserting a complete set of momentum eigenstates:
\begin{align}
    {\langle y;0|x;s\rangle}&= {\langle y|e^{-\hat{H}_0 s}|x\rangle}=\int \frac{d^4p_{\rm E}}{(2\pi)^4}\, {\langle y|p\rangle} {\langle p|x\rangle}e^{-p_I^2 s}\notag
    \\
    &=\int \frac{d^4p_{\rm E}}{(2\pi)^4}\, e^{i p_I(y-x)^I }e^{-p_I^2 s}=\frac{1}{16\pi^2 s^2}e^{-\frac{(y-x)_I^2}{4s}},\label{eq:free_xy}
\end{align}
where we have used ${|y\rangle}={|y; 0\rangle}$, ${|x;s\rangle}\equiv e^{-\hat{H}_0 s}{|x\rangle}$, $\hat{1}=\int \frac{d^4p_E}{(2\pi)^4}\,{|p\rangle}{\langle p|}$ and ${\langle p|x\rangle}=e^{-ip_I x^I}$.
Combining Eqs.~\eqref{eq:W02} and \eqref{eq:free_xy}, we obtain
\begin{align}
    W_0^{\rm E}[A]= \int d^4x_{\rm E}\,\left[\frac{1}{4}(F_{IJ})^2+\frac{1}{8\pi^2}\int_0^{\infty}ds\, \frac{1}{s^3}e^{-s m^2}\right]\,.
    \label{eq:finalW0QED}
\end{align}

\subsection{Interacting theory}\label{sec:interacting theory}
We now turn to the interacting theory, namely QED.
From Eq.~\eqref{eq:Ilambda_E}, the partition function in Euclidean spacetime is given by
\begin{align}
    Z_{\lambda}^{\rm E}[A]&\equiv \int \mathcal{D}\overline{\psi} \mathcal{D}\psi\, e^{-\int d^4x_{\rm E}\,(F_{IJ})^2/4}\,{\rm exp}\left[-\int d^4x_{\rm E}\, \overline{\psi}\,  {\rm M}_\lambda \psi \right]=e^{-\int d^4x_{\rm E}\,(F_{IJ})^2/4}\,{\rm det}\,{\rm M}_\lambda\,,\label{eq:ZA}
\end{align}
where ${\rm M}_\lambda\equiv \gamma^I_{\rm E} \left(\partial_I+i\lambda e A_I^{\rm E}\right) +m$.
From Eq.~\eqref{eq:ZA}, the effective action is given by
\begin{align}
    W_{\lambda}^{\rm E}[A]&\equiv -\ln Z_{\lambda}^{\rm E}[A]\notag
    \\
    &=\int d^4x_{\rm E}\,\left[\frac{1}{4}(F_{IJ})^2\right]-\ln {\rm det}\,{\rm M}_\lambda=\int d^4x_{\rm E}\,\left[\frac{1}{4}(F_{IJ})^2\right]-{\rm Tr}\ln \left(\slashed{D}_{\rm E}+m\right)\,,\label{eq:WA}
\end{align}
where $\slashed{D}_{\rm E}\equiv \gamma^I_{\rm E}D_I$ with $D_I\equiv \partial_I+i \lambda e A_I^{\rm E}$.
Using Eq.~\eqref{eq:WA}, we obtain
\begin{align}
\frac{d}{dm^2}W_{\lambda}^{\rm E}[A] & =-\frac{1}{2}\int d^4 x_{\rm E}\, {\rm tr_D} \left[ \langle x | \frac{ 1}{-(\slashed{D}_{\rm E})^2 + m^2} | x \rangle \right]
 \notag
 \\
&=-\frac{1}{2} \int d^4 x_{\rm E}\int_0^{\infty} d s \, e^{-s m^2}\, {\rm tr_D} \left[ \langle x | e^{-\hat{H}_\lambda s } | x \rangle \right]\,,\label{eq:dW/dm2}
\end{align}
where $1/\mathcal{A}=\int_0^\infty ds\,e^{-s \mathcal{A}}$ for real $\mathcal{A}>0$.
This is possible because $-(\slashed{D}_{\rm E})^2+m^2$ is a positive-definite operator in Euclidean spacetime.
However, after the inverse analytic continuation in Eq.~\eqref{eq:BG_F}, the Hamiltonian ceases to be Hermitian, and the integration contour for $s$ must be deformed to avoid the poles.
As discussed below, the standard $i\epsilon$ prescription defining the forward-time evolution fixes the sign of the imaginary part of the Hamiltonian $-(\slashed{D}_{\rm E})^2+m^2$.
This sign ensures that $-1/\left(\hat{H}_\lambda+m^2\right)=-i \int_0^\infty ds\, e^{-is \left(\hat{H}_\lambda+m^2\right)}$ is valid, provided that $\lim_{s\to \infty}\,e^{-is\hat{H}_\lambda}=0$.
Consequently, the integration contour for $s$ is uniquely determined.
The Hamiltonian $\hat{H}_\lambda$ is
\begin{align}
    \hat{H}_\lambda\equiv -(\slashed{D}_{\rm E})^2=(\hat{p}_I + \lambda e A^{\rm E}_I)^2 - \tfrac{\lambda e}{2} F_{IJ} \sigma^{IJ}_{\rm E}\,.\label{eq:Ham_qed}
\end{align}
This Hamiltonian is Hermitian in Euclidean spacetime, whereas it becomes non-Hermitian in Minkowski spacetime under the inverse analytic continuation in Eq.~\eqref{eq:BG_F}.
Here, we have used the relations
\begin{align}
D_I &= \partial_I + i \lambda e A_I^{\rm E}=i\left(-i\partial_I + \lambda e A_I^{\rm E}\right)=i\left( \hat{p}_I+\lambda e A_I^{\rm E}\right)~,
\\
\notag
\\
(\slashed{D}_{\rm E})^2
&=
\gamma^I_{\rm E}\gamma^J_{\rm E}D_I D_J
=
\left(
\frac{1}{2}\{\gamma^I_{\rm E}\,,\,\gamma^J_{\rm E}\}
+
\frac{1}{2}[\gamma^I_{\rm E}\,,\,\gamma^J_{\rm E}]
\right)D_I D_J
\notag
\\
&=
\delta^{IJ}D_I D_J
+
\frac{1}{2}[\gamma^I_{\rm E}\,,\,\gamma^J_{\rm E}]D_I D_J
=
D_I D_I
+
\frac{1}{4}[\gamma^I_{\rm E}\,,\,\gamma^J_{\rm E}][D_I\,,\,D_J]
\notag
\\
&=
D_I D_I
+
\frac{\lambda e}{2}F_{IJ}\sigma_{\rm E}^{IJ}\,,
\end{align}
where
\begin{align}
 \sigma^{IJ}_{\rm E}\equiv \frac{i}{2}[\gamma^I_{\rm E}\,,\,\gamma^J_{\rm E}]\,, \quad \{\gamma^I_{\rm E}\,,\,\gamma^J_{\rm E}\}= 2 \delta^{IJ}\,, \quad  [D_I\,,\, D_J]=i\lambda e\, F_{IJ}\,.
\end{align}
By integrating Eq.~\eqref{eq:dW/dm2}, we obtain
\begin{align}
W_{\lambda}^{\rm E}[A]=\int d^4x_{\rm E}\,\left[\frac{1}{4}(F_{IJ})^2+\frac{1}{2} \int_0^{\infty} d s {1\over s}\, e^{-s m^2}\, {\rm tr_D} \left[ \langle x | e^{-\hat{H}_\lambda s } | x \rangle \right]\right]+\text{const}\,.\label{eq:WA2}
\end{align}
To evaluate Eq.~\eqref{eq:WA2}, we need to calculate $\langle x | e^{- \hat{H}_\lambda s } | x \rangle$. 
To this end, we consider the following differential equation:
\begin{align}
\langle y,0|x,s\rangle =& \langle y|e^{-\hat{H}_\lambda s}|x\rangle \quad
\Rightarrow \quad  -\partial_s \langle y,0|x,s\rangle = \langle y|e^{-\hat{H}_\lambda s}\hat{H}_\lambda|x\rangle\,,\label{eq:diff_qed}
\end{align}
where
\[\hat{x}^I|x\rangle = x^I|x\rangle~, \qquad |x,s\rangle = e^{-\hat{H}_\lambda s}|x\rangle~, \qquad  \hat{x}^I(s) = e^{\hat{H}_\lambda s} \hat{x}^I e^{-\hat{H}_\lambda s}\]
\[[\hat{x}^I,\, \hat{p}_J] = i \delta^I_{J}, \qquad [\hat{x}^I(s),\, \hat{p}_J(s)] = i \delta^I_{J}, \qquad \hat{\Pi}_J\equiv -i D_J\,\]
\[[\hat{x}^I(s),\, \hat{\Pi}_J(s)] =[\hat{x}^I(s),\, \hat{p}_J(s) + \lambda e A^{\rm E}_J(\hat{x}(s))] = i\delta^I_{J}~,\qquad  [\hat{\Pi}_I(s),\, \hat{\Pi}_J(s)] = -i\lambda e \,F_{IJ}~.\]
To solve this equation, we first rewrite the Hamiltonian~\eqref{eq:Ham_qed}.  
We begin by evaluating $\hat{\Pi}_I(s)$ appearing in $\hat{H}_\lambda$. 
From the Heisenberg equation, we obtain
\begin{align}
    \frac{d \hat{\Pi}_I(s)}{ds} = [\hat{H}_\lambda, \hat{\Pi}_I(s)] = i2\lambda e\,F_{IJ}\hat{\Pi}_J(s) \quad \Rightarrow \quad\hat{\Pi}_I(s) = e^{i2\lambda esF_{IJ}}\hat{\Pi}_J(0)\,.
\label{eq:momentum}
\end{align}
To further rewrite $\hat{\Pi}_J(0)$ appearing in Eq.~\eqref{eq:momentum}, we make use of the following differential equation:
\begin{align}
     \frac{d\hat{x}^I(s)}{ds} = [\hat{H}_\lambda, \hat{x}^I(s)] = -i2\,\hat{\Pi}^I(s)\quad  \Rightarrow\quad \hat{x}_I(s) &= \hat{x}_I(0) -i2\int_0^s dt\, \hat{\Pi}_I(t)\,.\label{eq:difx}
\end{align}
Substituting Eq.~\eqref{eq:momentum} into Eq.~\eqref{eq:difx}, we obtain
\begin{align}
    \hat{x}_I(s) &=\hat{x}_I(0) -i 2\int_0^s dt\, e^{i2\lambda etF_{IJ}} \hat{\Pi}_J(0)=\hat{x}_I(0) +
    \left(J^{-1}\right)_{IJ}\, \hat{\Pi}_J(0)
    \,,\label{eq:x_sol}
\end{align}
where $\boldsymbol{J}^{-1}\equiv -i2s e^{i\lambda es \boldsymbol{F}} \frac{\sin (\lambda es\boldsymbol{F})}{s\lambda e \boldsymbol{F}}$, with $\boldsymbol{F}$ being a $4\times4$ matrix whose components are $F_{IJ}$. 
From Eq.~\eqref{eq:x_sol}, $\hat{\Pi}_J(0)$ is given by
\begin{align}
    \hat{\Pi}_J(0) &
=J_{JI}\, \left[
\hat{x}_I(s)-\hat{x}_I(0)
\right]\,,\label{eq:Pi0_re}
\end{align}
where $\boldsymbol{J}= i\tfrac{1}{2s} e^{-i\lambda es \boldsymbol{F}} \frac{s\lambda e \boldsymbol{F}}{\sin (\lambda es\boldsymbol{F})}$.
Substituting Eq.~\eqref{eq:Pi0_re} back into Eq.~\eqref{eq:momentum}, we obtain, in terms of $\hat{x}_I$,
\begin{align}
\hat{\Pi}_I(s) &= e^{i2\lambda esF_{IJ}}\hat{\Pi}_J(0)
={L}_{IJ} \left[ \hat{x}_J(s)-\hat{x}_J(0)\right]\,,\label{eq:mom_re}
\end{align}
where $\boldsymbol{L}\equiv i\tfrac{1}{2s}e^{i\lambda es \boldsymbol{F}}\frac{s\lambda e \boldsymbol{F}}{\sin (\lambda es\boldsymbol{F})}$.
With these preparations, the Hamiltonian can be rewritten in terms of the position operators.
Substituting Eq.~\eqref{eq:mom_re} into Eq.~\eqref{eq:Ham_qed}, we obtain
\begin{align}
    \hat{H}_\lambda (s)&=\hat{\Pi}_I(s)^2-\frac{\lambda e}{2}F_{IJ} \sigma^{IJ}_{\rm E}\notag
    \\
    &=\hat{x}_I(s) K_{IJ} \hat{x}_J(s)  + \hat{x}_I(0) K_{IJ} \hat{x}_J(0)
    -2\hat{x}_I(s) K_{IJ} \hat{x}_J(0) \notag
    \\
    &\qquad\qquad\qquad -i
    \frac{1}{2} \left(-\lambda  e \boldsymbol{F}+ i\lambda  e \boldsymbol{F} \cot(\lambda  e \boldsymbol{F}s)\right)_{II} 
    -\frac{\lambda  e}{2}F_{IJ} \sigma^{IJ}_{\rm E}\,,\label{eq:Hs_re}
\end{align}
where we have used the following several relations:
\begin{align}
K_{IJ}\equiv (\boldsymbol{L}^T \boldsymbol{L})_{IJ}= -\frac{1}{4s^2} \left(\frac{s\lambda e \boldsymbol{F}}{\sin (\lambda  es\boldsymbol{F})}\right)^2_{IJ}\,,\qquad K_{IJ}=K_{JI}\,,
\end{align}
\begin{align}
    [\hat{x}_I(s), \hat{x}_J(0)]&=[\hat{x}_I(0)+\left(J^{-1}\right)_{IK}\hat{\Pi}_K(0), \hat{x}_J(0)]=-i\left(J^{-1}\right)_{IJ}\quad\Rightarrow \quad[\hat{x}_I(0), \hat{x}_J(s)]=i\left(J^{-1}\right)_{JI}\,,
\end{align}
and
\begin{align}
    K_{IJ} [\hat{x}_I(0), \hat{x}_J(s)]
    &=i\frac{\lambda e}{2} \left( \boldsymbol{F}\left(-1+i\cot (s\lambda e \boldsymbol{F})\right)
    \right)_{II}\,.
\end{align}
Substituting Eq.~\eqref{eq:Hs_re} into Eq.~\eqref{eq:diff_qed}, the differential equation can be rewritten as follows:
\begin{align}
    -\partial_s {\langle y;0|x;s\rangle}& = {\langle y|e^{-\hat{H}_\lambda s}\hat{H}_\lambda(s)|x\rangle}\notag
    \\
    &=\bigg[
     -(y-x)_I \left(\frac{(\lambda e)^2 \boldsymbol{F}^2}{4} \frac{1}{\sin^2 (\lambda es \boldsymbol{F})}\right)_{IJ}(y-x)_J \notag
     \\
    &\qquad\qquad\qquad+
    \frac{1}{2}\left(\lambda e \boldsymbol{F} \cot (\lambda es \boldsymbol{F})\right)_{II}
    -
    \frac{\lambda e}{2} F_{IJ}\sigma^{IJ}_{\rm E}
    \bigg]
    {\langle y;0|x;s\rangle}\,,
    \label{eq:diffeq}
\end{align}
where we have used \[
\begin{aligned}
&\hat{x}_I(s) = e^{\hat{H}_\lambda s}\hat{x}_I e^{-\hat{H}_\lambda s},  \qquad
|x;s\rangle = e^{-\hat{H}_\lambda s}|x;0\rangle, \\
&\langle y;0|e^{-\hat{H}_\lambda s}\hat{x}_I(s)\hat{x}_J(s)|x;0\rangle
= y_I y_J \langle y;0|x;s\rangle, \\
&\langle y;0|e^{-\hat{H}_\lambda s}\hat{x}_I(0)\hat{x}_J(0)|x;0\rangle
= x_I x_J \langle y;0|x;s\rangle, \\
&\langle y;0|e^{-\hat{H}_\lambda s}\hat{x}_I(s)\hat{x}_J(0)|x;0\rangle
= y_I x_J \langle y;0|x;s\rangle.
\end{aligned}
\]
The solution of Eq.~\eqref{eq:diffeq} is given by
\begin{align}
    {\langle y;0|x;s\rangle}
    &=C(x,y)\, {\rm exp}\bigg[
    -\left(y-x\right)_I \left(\frac{\lambda e\boldsymbol{F}}{4} \cot \left(\lambda es\boldsymbol{F}\right)\right)_{IJ} \left(y-x\right)_J\notag
    \\
    &\qquad\qquad\qquad\qquad\qquad\qquad\qquad-\frac{1}{2} \left[\ln \left(\frac{\sin\left(\lambda es\boldsymbol{F}\right)}{\lambda e\boldsymbol{F}}\right)\right]_{II}
    + \frac{\lambda es}{2}F_{IJ}\sigma^{IJ}_{\rm E}
    \bigg]\,
    .\label{eq:sol0}
\end{align}

Next, we determine $C(x,y)$ in Eq.~\eqref{eq:sol0}.  
To this end, we present several formulas as follows:
\begin{align}
    {\langle y;0|\hat{\Pi}_I(0)e^{- \hat{H}_\lambda s}|x;0\rangle}
    &= {\langle y;0|\left(\hat{p}_I +\lambda eA^{\rm E}_I(\hat{x})\right)e^{- \hat{H}_\lambda  s}|x;0\rangle}=\left(-i\frac{\partial}{\partial y^I}+\lambda e A^{\rm E}_I(y)\right) {\langle y;0|x;s\rangle}\,.\label{eq:C_eq}
\end{align}
Also, from Eq.~\eqref{eq:mom_re}, we obtain
\begin{align}
    {\langle y;0|\hat{\Pi}_I(0)e^{-\hat{H}_\lambda s}|x;0\rangle}&=-i\left(e^{i\lambda e s \boldsymbol{F}}\frac{\lambda e \boldsymbol{F}}{2\sin (\lambda e s \boldsymbol{F})}\right)_{IJ} {\langle y;0|\left[\hat{x}_J(-s)-\hat{x}_J(0)\right]e^{-\hat{H}_\lambda s}|x;0\rangle}\notag
    \\
    &=-i\left(e^{i\lambda e s \boldsymbol{F}}\frac{\lambda e \boldsymbol{F}}{2\sin (\lambda e s \boldsymbol{F})}\right)_{IJ} \left[x_J-y_J\right]{\langle y;0|x;s\rangle}\,.\label{eq:C_eq2}
\end{align}
Combining Eqs.~\eqref{eq:C_eq} and \eqref{eq:C_eq2}, we obtain
\begin{align}
    \left(i\frac{\partial}{\partial y_I}-\lambda e A^{\rm E}_I(y)\right) {\langle y;0|x;s\rangle}=-i\left(e^{i\lambda es{\bf F}} \frac{\lambda e{\bf F}}{2 \sin \left(\lambda es{\bf F}\right)}\right)_{IJ} \left[y_J-x_J\right] {\langle y;0|x;s\rangle}\,.\label{eq:diffad}
\end{align}
Upon substituting Eq.~\eqref{eq:sol0} into Eq.~\eqref{eq:diffad}, we obtain
\begin{align}
    \left[i \frac{\partial}{\partial y_I}-\lambda e A^{\rm E}_I(y) +\frac{\lambda e}{2}F_{IJ}(x-y)^J\right]C(x,y)=0\,.\label{eq:diffad2}
\end{align}
The solution of Eq.~\eqref{eq:diffad2} is given by
\begin{align}
    C(x,y)&=C\, {\rm exp}\left[i\int_y^x dz_I \left(\lambda e A^{\rm E}_I(z)-\frac{\lambda e}{2}F_{IJ}(x-z)^J\right) \right]\,.\label{eq:Cxy}
\end{align}
Combining Eqs.~\eqref{eq:sol0} and \eqref{eq:Cxy} yields
\begin{align}
    {\langle y;0|x;s\rangle} &= C\, {\rm exp}\left[i\int_y^x dz_I \left(\lambda e A^{\rm E}_I(z)-\frac{\lambda e}{2}F_{IJ}(x-z)^J\right) \right]\notag
    \\
    &\times {\rm exp} \left[
    - \left(\vec{y}-\vec{x}\right)^{\rm T}\frac{\lambda e\boldsymbol{F}}{4} \cot \left(\lambda es\boldsymbol{F}\right) \left(\vec{y}-\vec{x}\right)
    -\frac{1}{2} \left[\ln \left(\frac{\sin\left(\lambda es\boldsymbol{F}\right)}{\lambda e\boldsymbol{F}}\right)\right]_{II}
    +\frac{\lambda es}{2}F_{IJ}\sigma^{IJ}_{\rm E}
    \right]\,,\label{eq:sol2}
\end{align}
where the prefactor $C$ can be determined as $C=\tfrac{1}{16\pi^2 }$ from the free theory discussed in Appendix~\ref{sec:free_qed} (see Eq.~\eqref{eq:free_xy}).

We next evaluate Eq.~\eqref{eq:WA2} based on Eq.~\eqref{eq:sol2}.
For this purpose, let us evaluate the following quantity:
\begin{align}
    (F_{IJ}\sigma^{IJ}_{\rm E})^2 &= 2 (F_{IJ})^2 -\gamma_{5,{\rm E}} \epsilon^{IJKL}F_{IJ}F_{KL}= 8 \left(\mathcal{F}+\gamma_{5,{\rm E}}\, \mathcal{G}\right)=\begin{pmatrix}
    8 \left(\mathcal{F}+\mathcal{G}\right)\mathbb{I}_2& 0
    \\
    0 & 8 \left(\mathcal{F}-\mathcal{G}\right)\mathbb{I}_2
    \end{pmatrix}\,,\label{eq:Fsigma}
\end{align}
where we have used a formula $\tfrac{1}{2}\left\{\sigma_{\rm E}^{IJ},\sigma_{\rm E}^{KL}\right\}=-\gamma_{5,{\rm E}}\,\epsilon^{IJKL} +\delta^{IK}\delta^{JL} -\delta^{JK}\delta^{IL}$ with $\epsilon^{1234}=+1$, $\gamma_{5,{\rm E}}\equiv \gamma^1_{\rm E}\,\gamma^2_{\rm E}\,\gamma^3_{\rm E}\, \gamma^4_{\rm E}$, $\gamma^4_{\rm E}=\gamma^0$, and $\gamma^i_{\rm E}= -i\gamma^i$, {\it i.e.}, $\gamma_{5,{\rm E}}=\begin{pmatrix} \mathbb{I}_2 & 0
\\
0 & -\mathbb{I}_2
\end{pmatrix}$. 
For convenience, we define the following quantities:
$\mathcal{F}\equiv (F_{IJ})^2/4$, $\mathcal{G}\equiv -F_{IJ}\widetilde{F}^{IJ}/4$, and $\widetilde{F}^{IJ}\equiv \epsilon^{IJKL}F_{KL}/2$.
From Eq.~\eqref{eq:Fsigma}, the four eigenvalues of $F_{IJ}\sigma^{IJ}_{\rm E}$ are given by
\begin{align}
    \left(2\,x_+,\,-2\,x_+,\,2\,x_-,\,-2\,x_-\right)\,,\label{eq:eign_Fsigma}
\end{align}
with $(x_+)^2=2\left(\mathcal{F}+ \mathcal{G}\right)$, and $(x_-)^2=2\left(\mathcal{F}- \mathcal{G}\right)$.
Note that $F_{IJ}\sigma^{IJ}_{\rm E}$ is Hermitian and therefore has real eigenvalues.
Equation~\eqref{eq:eign_Fsigma} then yields
\begin{align}
    {\rm tr_D}\,\left[
    e^{\frac{\lambda es}{2}(F_{IJ}\sigma_{\rm E}^{IJ})}
    \right]&=2\left(\cosh\left(\lambda es\, x_+\right)+\cosh\left(\lambda es \,x_-\right)\right)\,.\label{eq:Tr1}
\end{align}
Also, the four eigenvalues of $\boldsymbol{F}$ are given by
\begin{align}
    \left(\lambda^F_+,\, -\lambda^F_+,\,\lambda^F_-,\,-\lambda^F_-\right)\,, \label{eq:Fmat}
\end{align}
with
\begin{align}
(\lambda^F_+)^2=-\frac{1}{4}\left(x_+ +x_-\right)^2\,,\qquad (\lambda^F_-)^2=-\frac{1}{4}\left(x_+ -x_-\right)^2\,.
\end{align}
From Eq.~\eqref{eq:Fmat}, we evaluate the trace appearing in Eq.~\eqref{eq:sol2} as follows:
\begin{align}
    \frac{1}{2}\left[\ln \left(\frac{\sin (\lambda es\boldsymbol{F})}{\lambda e\boldsymbol{F}}\right)\right]_{II}&= \frac{1}{2}\,{\rm tr}_{4\times 4}\,\left[\ln \left(\frac{\sin (\lambda es\boldsymbol{F})}{\lambda e\boldsymbol{F}}\right)\right]\,,\label{eq:traLog}
\end{align}
where ${\rm tr}_{4\times4}$ denotes the trace over the Euclidean spacetime indices $I,J=1,\dots,4$.
We then obtain
\begin{align}
\frac{1}{2}\left[\ln \left(\frac{\sin (\lambda es\boldsymbol{F})}{\lambda e\boldsymbol{F}}\right)\right]_{II}
&=\ln \left(\frac{\sin (\lambda es\lambda^F_+)\,\sin (\lambda es\lambda^F_-)}{(\lambda e)^2\lambda^F_+\lambda^F_-}\right)=\ln \left(\frac{\cosh\left(\lambda es \,x_+\right)-\cosh\left(\lambda es\,x_-\right)}{2(\lambda e)^2\, \mathcal{G}}\right)\,.\label{eq:Tr2}
\end{align}
Up to this point, the calculation of the building blocks of the effective action has been completed.
We now assemble the effective action from these components.  
By combining Eqs.~\eqref{eq:WA2}, \eqref{eq:sol2}, \eqref{eq:Tr1}, and \eqref{eq:Tr2}, the effective action is obtained as follows:
\begin{align}
    W_{\lambda}^{\rm E}[A]&=\int d^4x_{\rm E}\,\left[\mathcal{F} +\frac{(\lambda e)^2}{8\pi^2}
    \int_0^{\infty}ds\, \frac{e^{-sm^2}}{s}
    \,\frac{\cosh\left(\lambda es\, x_+\right)+\cosh\left(\lambda es \,x_-\right)}{\cosh\left(\lambda es \,x_+\right)-\cosh\left(\lambda es\,x_-\right)}\,
    \mathcal{G}
    \right]\,,\label{eq:WAe}
\end{align}
where $x_\pm=\sqrt{2\left(\mathcal{F}\pm\mathcal{G}\right)}$.
Performing a series expansion of Eq.~\eqref{eq:WAe} with respect to $\lambda e$, we obtain
\begin{align}
    W_{\lambda}^{\rm E}[A]&=\int d^4x_{\rm E}\,\left[
   \epsilon_{\rm F}
    +\left(1+(\lambda e)^2\,\delta_{\rm F}\right)\mathcal{F}-\frac{(\lambda e)^4}{32\pi^2} \int_0^{\infty}ds\,\frac{e^{-s m^2}}{s}
    \left(
    \frac{4}{45}s^2 \left(4\mathcal{F}^2 -7 \mathcal{G}^2\right)
    +\cdots
    \right)
    \right]\,,
\end{align}
where $\epsilon_{\rm F} \equiv \tfrac{1}{8\pi^2}\int_0^{\infty} ds\,\tfrac{1}{s^3} e^{-s m^2}$ is defined as the vacuum energy, and $\delta_{\rm F} \equiv \tfrac{1}{12\pi^2}\int_0^{\infty} ds\tfrac{1}{s} e^{-s m^2}$ represents the wavefunction renormalization.
Using the inverse analytic continuation in Eq.~\eqref{eq:BG_F}, $\mathcal{F}$ and $\mathcal{G}$ can be expressed in terms of the physical electromagnetic fields as
\begin{align}
    \mathcal{F}=\frac{1}{2}\left(\vec{B}^2-\vec{E}^2\right)\, \left(1+(\lambda e)^2\,\delta_{\rm F}\right)^{-1}\,,\qquad \mathcal{G}=-i \left(\vec{E}\cdot \vec{B}\right)\, \left(1+(\lambda e)^2\,\delta_{\rm F}\right)^{-1}\,,\label{eq:FG_EB}
\end{align}
where the one-loop wavefunction renormalization factor $(1+(\lambda e)^2\,\delta_{\rm F})^{-1}$ is absorbed into the classical solutions $\mathcal{F}$ and $\mathcal{G}$.
Consequently, at the one-loop level, we obtain
\begin{align}
    W_{\lambda}[A^{\rm cl}]=\int d^4x_{\rm E}\,\left[
    \epsilon_{\rm F}
    +
    \frac{1}{2}\left(\vec{B}^2-\vec{E}^2\right)
    -
    \frac{(\lambda e)^4}{360\pi^2 m^4}
    \left\{
    \left(\vec{B}^2-\vec{E}^2\right)^2
    +
    7\left(\vec{E}\cdot \vec{B}\right)^2
    \right\}
    +\cdots
    \right]\,.
\end{align}
That is,
\begin{align}
    W_{\lambda=1}[A^{\rm cl}]=\int d^4x_{\rm E}\,\left[
    \epsilon_{\rm F}
    +
    \frac{1}{2}\left(\vec{B}^2-\vec{E}^2\right)
    -
    \frac{e^4}{360\pi^2 m^4}
    \left\{
    \left(\vec{B}^2-\vec{E}^2\right)^2
    +
    7\left(\vec{E}\cdot \vec{B}\right)^2
    \right\}
    +\cdots
    \right]\,.\label{eq:EH_lam1}
\end{align}
Also, we obtain
\begin{align}
    \left(\frac{d W_{\lambda}[A^{\rm cl}]}{d \lambda}\right)_{\lambda=0}=0\,,\qquad W_{\lambda=0}[A^{\rm cl}]=\int d^4x_{\rm E}\,\left[
    \epsilon_{\rm F}
    +
    \frac{1}{2}\left(\vec{B}^2-\vec{E}^2\right)
    \right]\,. 
\end{align}
From these results, the relative entropy can be evaluated  perturbatively as 
\begin{align}
    S\left(\rho_{\rm R}\|\rho_{\rm T}\right)&= W_{\lambda=0}[A^{\rm cl}]-W_{\lambda=1}[A^{\rm cl}]+\left(\frac{d W_{\lambda}[A^{\rm cl}]}{d \lambda}\right)_{\lambda=0}\notag
    \\
    &=\frac{e^4}{360\pi^2 m^4} \int d^4x_{\rm E}\,\left[    \left(\vec{B}^2-\vec{E}^2\right)^2
    +
    7\left(\vec{E}\cdot \vec{B}\right)^2
    \right]+\cdots \geq 0\,.
\end{align}
Hence, the non-negativity of the perturbative relative entropy implies that the sum of all nonlinear higher-derivative terms in Eq.~\eqref{eq:EH_lam1} is non-negative.
In what follows, we focus on two scenarios, which we refer to as the magnetic model ($\vec{E}=0$) and the electric model ($\vec{B}=0$).

\subsection{Magnetic model}\label{sec:Magnetic EH}
For a purely magnetic background ($\vec{E}=0$), the system remains stable under the inverse analytic continuation~\eqref{eq:BG_F}.
Consequently, the Hamiltonian in Eq.~\eqref{eq:Ham_qed} remains Hermitian, and the Euclidean analysis continues to apply. 
In particular, the proper-time representation $1/\mathcal{A}=\int_0^\infty ds\,e^{-s \mathcal{A}}$ remains valid for real $\mathcal{A}>0$.
From Eq.~\eqref{eq:FG_EB}, we find
\begin{align}
    \mathcal{F}=\frac{1}{2}\vec{B}^2\, \left(1+(\lambda e)^2\,\delta_{\rm F}\right)^{-1}\,,\qquad \mathcal{G}=0\,.
\end{align}
From Eq.~\eqref{eq:WAe}, we obtain
\begin{align}
    W_\lambda[A^{\rm cl}]=\int d^4x_{\rm E}\,
    \left[
    \epsilon_{\rm F} 
    +
    \frac{1}{2} \vec{B}^2
    -
    \mathcal{L}_{\rm F}\left((\lambda e \hat{B})^2\right)
    \right]\,,
\end{align}
where $\hat{B}\equiv |\vec{B}|/m^2$, and the nonlinear EFT effect is given by
\begin{align}
    \mathcal{L}_{\rm F}\left((\lambda e \hat{B})^2\right)&=\frac{1}{8\pi^2}\int_0^{\infty} e^{-sm^2} s^{-3}
    \,\mathcal{K}\left(\lambda e \hat{B} m^2 s\right)\,ds=\frac{m^4}{8\pi^2}\int_0^{\infty} e^{-t} t^{-3}
    \,\mathcal{K}\left(\lambda e \hat{B} t\right)\,dt\,,
\end{align}
where $\mathcal{K}\left(x\right)\equiv -i x \cot \left(ix\right)+1- \tfrac{(ix)^2}{3}$, and $t\equiv m^2 s$.
Note that the function $\mathcal{K}\left(x\right)$ has no poles on the real $x$-axis.

\subsection{Electric model}\label{sec:Electric EH}
For a purely electric background ($\vec{B}=0$), the system becomes unstable under the inverse analytic continuation~\eqref{eq:BG_F}.
As a result, the Hamiltonian in Eq.~\eqref{eq:Ham_qed} is no longer Hermitian, and the Euclidean analysis requires a slight modification.
As we explain below, this modification is equivalent to deforming the integration contour of the Schwinger proper time $s$.

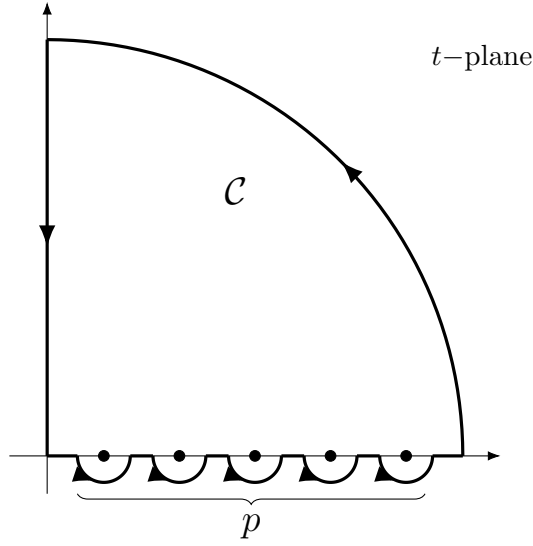
\begin{figure}
    \centering
\begin{tikzpicture}[scale=5, >=Latex]

\tikzset{
  arrowpos/.style 2 args={
    postaction={decorate},
    decoration={markings, mark=at position #1 with {\arrow{#2}}}
  }
}

\draw[->, thin] (-0.1,0) -- (1.2,0) node[right] {};
\draw[->, thin] (0,-0.1) -- (0,1.2) node[above] {};

\node[below left] at (0,0) { };

\draw[black, very thick, arrowpos={0.5}{>}] (0,1.1) -- (0,0); %
\draw[black, very thick, arrowpos={0.5}{>}] plot[domain=0:90,samples=50]
   ({1.1*cos(\x)}, {1.1*sin(\x)}); %

\foreach \x in {0.15,0.35,0.55,0.75,0.95}{
  \draw[black, very thick, arrowpos={0.4}{>}]
     plot[domain=180:360, samples=40, variable=\t]
     ({\x + 0.07*cos(\t)}, {0.07*sin(\t)});
  \filldraw[black] (\x,0) circle (0.4pt);
}

\draw[black, very thick]
  (0,0) -- (0.08,0)
  (0.22,0) -- (0.28,0)
  (0.42,0) -- (0.48,0)
  (0.62,0) -- (0.68,0)
  (0.82,0) -- (.88,0)
  (1.02,0) -- (1.1,0);

\draw[decorate, decoration={brace, amplitude=4pt, mirror}] 
  (0.08,-0.1) -- (1,-0.1)
  node[midway, below=4pt] {\Large $p$};

\node at (0.5,0.7) {\Large $\mathcal{C}$};

\node[above right] at (.99,.99) { $t-$plane};

\end{tikzpicture}
\caption{Integration contour $\mathcal{C}$ used in the evaluation of the proper-time integral.
The contour is defined by $\mathcal{C}\equiv
\lim_{\epsilon\to0^+\,,R\to\infty}
\left(
A\cup B_0
\cup
\bigcup_{p=1}^{\infty}(B_p\cup\mathcal{C}_p)
\cup D
\right)$, where $A$ is the line segment $t\in(+iR,0]$, $B_0$ is the interval $t\in[0,t_1-\epsilon]$, $B_p$ ($p\ge1$) is the interval $t\in[t_p+\epsilon,\,t_{p+1}-\epsilon]$, $\mathcal{C}_p$ ($p\ge1$) is the semicircle $t=t_p+\epsilon e^{i\theta}$ with $\theta\in[\pi,2\pi]$, and $D$ is the arc $t=R e^{i\theta}$ with $\theta\in[0,\pi/2]$.
}
\label{fig:contour}
\end{figure}

We now explain the prescription for deforming the integration contour.
From Eq.~\eqref{eq:sol2} with $s\to is$, we obtain
\begin{align}
    {\langle y|e^{-i\hat{H}_\lambda s}|x\rangle} &= {\rm exp}\left[i\int_y^x dz_I \left(\lambda e A^{\rm E}_I(z)-\frac{\lambda e}{2}F_{IJ}(x-z)^J\right) \right]\notag
    \\
    &\times {\rm exp} \left[
    i\left(\vec{y}-\vec{x}\right)^{\rm T}\frac{\lambda e\boldsymbol{F}}{4} \coth \left(\lambda es\boldsymbol{F}\right) \left(\vec{y}-\vec{x}\right)
    \right]\,{\langle x|e^{-i\hat{H}_\lambda s}|x\rangle}\,,
\end{align}
where
\begin{align}
    \frac{\lambda e\boldsymbol{F}}{4} \coth \left(\lambda es\boldsymbol{F}\right)
    =
    \frac{1}{4s}\,\boldsymbol{O}^T_{F}\,
    \begin{pmatrix}
    \lambda e s|\vec{E}| \coth \left(\lambda es|\vec{E}|\right) & 0 & 0 & 0
    \\
    0 & \lambda e s|\vec{E}| \coth \left(\lambda es|\vec{E}|\right) & 0 & 0
    \\
    0 & 0 & 1 & 0
    \\
    0 & 0 & 0 & 1
    \end{pmatrix}\, \boldsymbol{O}_{F}\,,
\end{align}
and
\begin{align}
    {\langle x|e^{-i\hat{H}_\lambda s}|x\rangle}
    &= -\frac{1}{16\pi^2}\,\frac{\left(\lambda e |\vec{E}|\right)^2}{\lambda e s|\vec{E}|\, \sinh\left(\lambda e s |\vec{E}|\right)}
    \,
    \boldsymbol{O}_{\rm D}^T\,\begin{pmatrix}
    e^{\lambda es |\vec{E}|} & 0 & 0 & 0
    \\
    0 & e^{-\lambda es |\vec{E}|} & 0 & 0
    \\
    0 & 0 & e^{\lambda es |\vec{E}|} & 0
    \\
    0 & 0 & 0 & e^{-\lambda es |\vec{E}|}
    \end{pmatrix}
    \boldsymbol{O}_{\rm D}
    \,,
\end{align}
with the orthogonal matrices $\boldsymbol{O}_{F}$ and $\boldsymbol{O}_{D}$.
Using these expressions, we find $ \lim_{s\to \pm \infty}{\langle y|e^{-i\hat{H}_\lambda s}|x\rangle}
    =
    0$ ({\it i.e.}, $\lim_{s\to \pm\infty}e^{-i\hat{H}_\lambda s}=0$).
    Therefore, both positive- and negative-time contour prescriptions are formally allowed.
To select the contour corresponding to forward-time evolution, we introduce the standard $i\epsilon$ prescription, $\hat{H}_\lambda\to \hat{H}_\lambda -i\epsilon^+$, with $\epsilon^+>0$.
We then obtain
\begin{align}
    \lim_{s\to +\infty} e^{-i(\hat{H}_\lambda-i\epsilon^+)s}=0\,,
\end{align}
whereas the evolution operator diverges as $s\to -\infty$.
Thus the forward-time contour is uniquely selected.
In this case, after the inverse analytic continuation~\eqref{eq:BG_F}, Eq.~\eqref{eq:dW/dm2} can be rewritten as
\begin{align}
\frac{d}{dm^2}W_{\lambda}^{\rm E}[A] & =-\frac{1}{2}\int d^4 x_{\rm E}\, {\rm tr_D} \left[ \langle x | \frac{ 1}{-(\slashed{D}_{\rm E})^2 + m^2-i\epsilon^+} | x \rangle \right]
 \notag
 \\
&=-\frac{i}{2} \int d^4 x_{\rm E}\int_0^{\infty} d s \, e^{-is m^2}\, {\rm tr_D} \left[ \langle x | e^{-i(\hat{H}_\lambda-i\epsilon^+) s } | x \rangle \right]\,,
\end{align}
where $-i/\mathcal{A}=\int_0^\infty ds\,e^{-i s \mathcal{A}}$ provided that $\lim_{s\to +\infty} e^{-is \mathcal{A}}=0$\footnote{
From $d e^{-i s\mathcal{A}}/ds=-i \mathcal{A}e^{-i s\mathcal{A}}$, we obtain
\begin{align}
    -i \mathcal{A}^{-1}= \int_0^\infty ds e^{-is \mathcal{A}}-i \mathcal{A}^{-1} \lim_{s\to +\infty} e^{-is \mathcal{A}}\,.
\end{align}
}.
That is, for the electric model, Eq.~\eqref{eq:WAe} remains valid upon the contour deformation $s\to is$. 
Consequently, from Eq.~\eqref{eq:WAe}, we obtain
\begin{align}
    W_\lambda[A^{\rm cl}]=\int d^4x_{\rm E}\,
    \left[
    \epsilon_{\rm F} 
    -
    \frac{1}{2} \vec{E}^2
    -
    \mathcal{L}_{\rm F}\left((i\lambda e \hat{E})^2\right)
    \right]\,,
\end{align}
where $\hat{E}\equiv |\vec{E}|/m^2$, and the nonlinear EFT effect is given by
\begin{align}
    \mathcal{L}_{\rm F}\left((i\lambda e \hat{E})^2\right)&=\frac{1}{8\pi^2}\int_0^{i\infty} e^{-sm^2} s^{-3}
    \,\mathcal{K}\left(i\lambda e \hat{E} m^2 s\right)\,ds=\frac{m^4}{8\pi^2}\int_0^{i\infty} e^{-t} t^{-3}
    \,\mathcal{K}\left(i\lambda e \hat{E} t\right)\,dt\,,\label{eq:E_LF_def}
\end{align}
where $\mathcal{K}\left(x\right)= -i x \cot \left(ix\right)+1- \tfrac{(ix)^2}{3}$, and $t= m^2 s$.
Note that the function $\mathcal{K}\left(ix\right)$ has poles on the real $x$-axis. 
We now evaluate the integral in Eq.~\eqref{eq:E_LF_def}.
Applying the residue theorem to the contour $\mathcal{C}$ shown in Fig.~\ref{fig:contour}, we obtain
\begin{align}
    \int_{\mathcal{C}}  \,e^{-t} t^{-3}
    \,\mathcal{K}\left(i\lambda e \hat{E} t\right)\,dt= 2\pi i\,\sum_{p=1}^\infty {\rm Res}\,\left[e^{-t} t^{-3}
    \,\mathcal{K}\left(i\lambda e \hat{E} t\right),\, t_p\right]\,.\label{eq:C}
\end{align}
The poles are located at $t_p=\pi p/\lambda e\hat{E}$, where $p=1,2,3,\cdots$.
Using the simple pole structures, we find
\begin{align}
    {\rm Res}\,\left[e^{-t} t^{-3}
    \,\mathcal{K}\left(i\lambda e \hat{E} t\right),\, t_p\right]
    =
    \lim_{t\to t_p}\, \left(t-t_p\right)\,e^{-t} t^{-3}
    \,\mathcal{K}\left(i\lambda e \hat{E} t\right)= -\frac{e^{-t_p}}{t_p^2}\,.
\end{align}
From these, we obtain
\begin{align}
     \int_{\mathcal{C}} \,e^{-t} t^{-3}
    \,\mathcal{K}\left(i\lambda e \hat{E} t\right)\,dt&= -\int_0^{i\infty}\,e^{-t} t^{-3}
    \,\mathcal{K}\left(i\lambda e \hat{E} t\right)\,dt\notag
    \\
    &+ \mathcal{P}\,\int_0^{\infty} \,e^{-t} t^{-3}
    \,\mathcal{K}\left(i\lambda e \hat{E} t\right)\,dt+ \sum_{p=1}^{\infty}\,  \lim_{\epsilon\to 0^+}\int_{\mathcal{C}_p}\,e^{-t} t^{-3}
    \,\mathcal{K}\left(i\lambda e \hat{E} t\right)\,dt,\label{eq:C2}
\end{align}
where the principal value integral is defined as
\begin{align}
&\mathcal{P}\,\int_0^{\infty}\,e^{-t} t^{-3}
    \,\mathcal{K}\left(i\lambda e \hat{E} t\right)\,dt\notag
    \\
    &\qquad \equiv  \lim_{\epsilon\to 0^+}\left(\int_0^{t_1-\epsilon}  \,e^{-t} t^{-3}
    \,\mathcal{K}\left(i\lambda e \hat{E} t\right)\,dt+ \sum_{p=1}^\infty\int_{t_p+\epsilon}^{t_{p+1}-\epsilon}\,e^{-t} t^{-3}
    \,\mathcal{K}\left(i\lambda e \hat{E} t\right)\,dt\right)\,.
\end{align}
Furthermore, we find
\begin{align}
   \lim_{\epsilon\to 0^+} \int_{\mathcal{C}_p} \,e^{-t} t^{-3}
    \,\mathcal{K}\left(i\lambda e \hat{E} t\right)\,dt&= i\int_{\pi}^{2\pi}d\theta \lim_{\epsilon\to 0^+}(t-t_p)\,e^{-t} t^{-3}
    \,\mathcal{K}\left(i\lambda e \hat{E} t\right)\notag
    \\
    &=i\pi\, {\rm Res}\,\left[e^{-t} t^{-3}
    \,\mathcal{K}\left(i\lambda e \hat{E} t\right), t_p\right]\,, \label{eq:Cn}
\end{align}
where $dt=i\epsilon e^{i\theta}d\theta=i(t-t_p)d\theta$ has been used, parameterizing the contour as $t=t_p+\epsilon e^{i\theta}$.
Combining Eqs.~\eqref{eq:C}, \eqref{eq:C2}, and \eqref{eq:Cn}, we arrive at
\begin{align}
    \mathcal{L}_{\rm F}\left((i\lambda e \hat{E})^2\right)&=\frac{m^4}{8\pi^2}\int_0^{i\infty}\, e^{-t} t^{-3}
    \,\mathcal{K}\left(i\lambda e \hat{E} t\right)\,dt\notag
    \\
    &=\frac{m^4}{8\pi^2}\,\left[\mathcal{P}\,\int_0^{\infty}\,  e^{-t} t^{-3}
    \,\mathcal{K}\left(i\lambda e \hat{E} t\right)\,dt +i\pi \sum_{p=1}^{\infty}\frac{e^{-t_p}}{t_p^2}\right]\,,\label{eq:E_non_F_App}
\end{align}
where the principal value integral is purely real, while the last term gives the imaginary contribution associated with the Schwinger effect.

To make the contribution from each pole explicit, we rewrite the function $\mathcal{K}$.
Using Euler's infinite product formula for the sine function,
\begin{align}
    \sin x =x \prod_{p=1}^\infty \left(1-\frac{x^2}{p^2\pi^2}\right)\,,
\end{align}
taking the logarithmic derivative $\tfrac{d}{dx}\ln \sin x =\cot x$ yields
\begin{align}
    \cot x =\frac{1}{x}+ \sum_{p=1}^\infty \frac{2x}{x^2 -p^2\pi^2}\quad\Rightarrow \quad x \cot x = 1+\sum_{p=1}^\infty \frac{2x^2}{x^2 -p^2\pi^2}\,.\label{eq:xcot}
\end{align}
Furthermore, we can rewrite
\begin{align}
     \sum_{p=1}^\infty \frac{2x^2}{x^2-p^2\pi^2}
    =
    2x^2 \sum_{p=1}^\infty \frac{x^2}{p^2\pi^2 (x^2-p^2\pi^2)}-2x^2 \sum_{p=1}^\infty \frac{1}{p^2\pi^2}\,\label{eq:2xp}.
\end{align}
Combining Eqs.~\eqref{eq:xcot} and \eqref{eq:2xp}, we obtain
\begin{align}
    x \cot x -1=2x^2 \sum_{p=1}^\infty \frac{x^2}{p^2\pi^2 (x^2-p^2\pi^2)}-2x^2 \sum_{p=1}^\infty \frac{1}{p^2\pi^2}\,.\label{eq:ap1}
\end{align}
Using the well-known Basel sum, we obtain
\begin{align}
    \sum_{p=1}^\infty \frac{1}{p^2}= \frac{\pi^2}{6}\quad \Rightarrow \quad \sum_{p=1}^\infty \frac{1}{p^2\pi^2}= \frac{1}{6}\,,\label{eq:ap2}
\end{align}
Combining Eqs.~\eqref{eq:ap1} and~\eqref{eq:ap2}, we obtain
\begin{align}
    \mathcal{K}\left(ix\right)=-x \cot x +1-\frac{1}{3}x^2 =-2x^4\, \sum_{p=1}^\infty \frac{1}{p^2\pi^2 (x^2-p^2\pi^2)}\,.\label{eq:zcotz}
\end{align}
Substituting Eq.~\eqref{eq:zcotz} into Eq.~\eqref{eq:E_non_F_App}, we obtain
\begin{align}
    {\rm Re}\,\mathcal{L}_{\rm F}\left((i\lambda e \hat{E})^2)\right)
    &=
    \frac{m^4}{8\pi^2}\,\mathcal{P}\,\int_0^\infty \, e^{-t} t^{-3} \mathcal{K}\left(i\lambda e \hat{E}t\right)\,dt\notag
    \\
    &=
    \frac{m^4}{4\pi^2}\,\sum_{p=1}^\infty\,\mathcal{P}\,\int_0^\infty \, e^{-t} t^{-3}\,\frac{\left(\lambda e\hat{E}t/p\pi\right)^4}{1-\left(\lambda e\hat{E}t/p\pi\right)^2}\,dt \,
    .\label{eq:tildeLgE_exp}
\end{align}

\section{Details of the relative entropy in scalar QED}
\label{app:scl_non}
We show that, in scalar QED, the relative entropy reproduces the nonlinear EFT corrections by choosing a suitable gauge parameter $\alpha$.
For later generalizations, we keep the expressions in a formal form.
As in the fermionic case, the partition functions in Euclidean spacetime are given by
\begin{align}
    Z^{\rm E}_\lambda[A^{\rm E,cl}]
    =
    \int \mathcal{D} A^{\rm E}\, z_\lambda^{\rm E}[A^{\rm E}]\,,\qquad z_\lambda^{\rm E}[A^{\rm E}]
    =
    \int \mathcal{D}\varphi^{\rm E}\,e^{-\int d^4x_{\rm E}\,\bar{\mathcal{L}}^{\rm E}_\lambda}\,,\label{eq:zlam_scl}
\end{align}
where $\bar{\mathcal{L}}^{\rm E}_\lambda$ is the Euclidean counterpart of the Lagrangian defined in Eq.~\eqref{eq:barL}, together with Eqs.~\eqref{eq:zero_gauge} and~\eqref{eq:int_zerofield}.
At the one-loop level, the path integral is dominated by the stationary configuration $A^{\rm E,cl}$, yielding
\begin{align}
Z_\lambda^{\rm E}[A^{\rm E,cl}] = z_\lambda^{\rm E}[{A}^{\rm E,cl}]\,, \label{eq:scl_pat}
\end{align}
where $A^{\rm E,cl}$ includes the effect of wave-function renormalization, as in fermionic QED.
From these partition functions, we obtain the one-loop effective actions in Euclidean spacetime as
\begin{align}
    W^{\rm E}_\lambda[A^{\rm E,cl}] = -\ln \, Z_\lambda^{\rm E}[A^{\rm E,cl}]= w^{\rm E}_\lambda[{A}^{\rm E,cl}] \,,
\end{align}
where $w^{\rm E}_\lambda [{A}^{\rm E,cl}]= -\ln \, z_\lambda^{\rm E}[{A}^{\rm E,cl}]$.
From Eq.~\eqref{eq:zlam_scl}, we obtain the following expression for a fixed background field $A^{\rm E}$:
\begin{align}
    \left.\frac{dw_\lambda^{\rm E}}{d\lambda}\right|_{\lambda=0}=
    \int \mathcal{D}\varphi^{\rm E}\, \bar{I}_{\rm I}^{\rm E}\, \frac{e^{-\int d^4x_{\rm E}\,\bar{\mathcal{L}}_{0}^{\rm E} }}{z_0^{\rm E}}
    =
    \int \mathcal{D}\varphi^{\rm E}\, \left(e\,\bar{I}_{\rm I}^{(1)}+e^2\,\bar{I}_{\rm I}^{(2)}\right)\, \frac{e^{-\int d^4x_{\rm E}\,\bar{\mathcal{L}}_{0}^{\rm E} }}{z_0^{\rm E}}
    \,,\label{eq:dw_lam_scl}
\end{align}
where the interaction part of the Euclidean action is defined as
\begin{align}
    \bar{I}_{\rm I}^{\rm E}\equiv \int d^4x_{\rm E}\,\bar{\mathcal{L}}_{\rm I}^{\rm E}
    =
    e\,\bar{I}_{\rm I}^{(1)}+e^2\,\bar{I}_{\rm I}^{(2)}\,,
\end{align}
with
\begin{align}
    \bar I_{\rm I}^{(1)}&\equiv \int d^4x_{\rm E}\, 
    i (A_I-\partial_I\alpha) \left[
    (\partial_I \varphi^\dagger)\varphi
    -\varphi^\dagger (\partial_I \varphi)
    \right]\,,
    \\
    \bar I_{\rm I}^{(2)}&\equiv \int d^4x_{\rm E} \left(|A_I|^2-|\partial_I\alpha|^2\right) \varphi^\dagger \varphi\,.
\end{align}

We first focus on the term linear in $e$ in Eq.~\eqref{eq:dw_lam_scl}.
After the field redefinition  $\varphi'=e^{i e\alpha}\varphi$, we obtain
\begin{align}
     \int \mathcal{D}\varphi^{\rm E} \,\bar{I}_{\rm I}^{(1)}[A, \varphi]\, \frac{e^{-\int d^4x_{\rm E}\,\bar{\mathcal{L}}_{0}^{\rm E}(A, \varphi) }}{z_0^{\rm E}}=\int \mathcal{D}{\varphi'}^{\rm E}\, \bar{I}_{\rm I}^{(1)}[A, e^{-ie\alpha}\varphi']\, \frac{e^{-\int d^4x_{\rm E}\,{\mathcal{L}}_{0}^{\rm E}(A, \varphi') }}{z_0^{\rm E}}\,,\label{eq:I1_exp}
\end{align}
where we have used the invariance of the path-integral measure, $\mathcal{D}\varphi^{\rm E}=\mathcal{D}{\varphi'}^{\rm E}$, and the identities $\bar{I}_{\rm I}^{(1)}[A, \varphi]=\bar{I}_{\rm I}^{(1)}[A, e^{-ie\alpha}\varphi']$, and $\bar{\mathcal{L}}_0^{\rm E}(A,\varphi)=\mathcal{L}_0^{\rm E}(A,\varphi')$. 
To linear order in $e$, we find
\begin{align}
    \bar{I}_{\rm I}^{(1)}[A, e^{-ie\alpha}\varphi']&=
    \int d^4x_{\rm E}\, 
    i\left(
    A_I-\partial_I\alpha
    \right)
    \left[
    (\partial_I {\varphi'}^\dagger)\varphi'
    -{\varphi'}^\dagger (\partial_I \varphi') 
    +
    2ie \partial_I \alpha {\varphi'}^\dagger \varphi'
    \right]\notag
    \\
    &=
    \int d^4x_{\rm E}\, 
    i\left(
    A_I-\partial_I\alpha
    \right)
    \left[
    (\partial_I {\varphi'}^\dagger)\varphi'
    -{\varphi'}^\dagger (\partial_I \varphi') \right]\notag
    \\
    &+\int d^4x_{\rm E}\, 
    2e\alpha\,\left[
    \partial_I \left(A_I-\partial_I\alpha\right) {\varphi'}^\dagger \varphi'
    +
    \left(
    A_I-\partial_I\alpha
    \right)
    \partial_I  \left({\varphi'}^\dagger \varphi'\right)
\right]\,,\label{eq:I1_gau}
\end{align}
where, in the second equality, we have integrated by parts, assuming that the heavy field $\varphi$ vanishes at the boundary.
Combining Eqs.~\eqref{eq:I1_exp} and \eqref{eq:I1_gau}, we obtain
\begin{align}
    &\int \mathcal{D}\varphi^{\rm E} \,\bar{I}_{\rm I}^{(1)}[A, \varphi]\, \frac{e^{-\int d^4x_{\rm E}\,\bar{\mathcal{L}}_{0}^{\rm E}(A, \varphi) }}{z_0^{\rm E}}\notag
    \\
    &
    \qquad=
    \int d^4x_{\rm E}\, \left[i\left(
    A_I-\partial_I\alpha
    \right)\,\left(
    \langle (\partial_I {\varphi}^\dagger)\varphi
    -{\varphi}^\dagger (\partial_I \varphi) \rangle_0
    -
    2ie \alpha
    \, \langle \partial_I(\varphi^\dagger \varphi) \rangle_0
    \right)
    +
    2e \alpha\,\partial_I 
    \left(A_I-\partial_I \alpha\right)\langle\varphi^\dagger \varphi\rangle_0\right]\,,\label{eq:I1_exp2}
\end{align}
where
\begin{align}
    {\langle \mathcal{O}\rangle}_0\equiv 
    \int \mathcal{D}\varphi^{\rm E}\, \mathcal{O}[\varphi]\,\frac{e^{-\int d^4x_{\rm E}\,{\mathcal{L}}_{0}^{\rm E}(A, \varphi) }}{z_0^{\rm E}}\,. 
\end{align}
Using the $O(4)$ symmetry of the Euclidean spacetime, Eq.~\eqref{eq:I1_exp2} becomes
\begin{align}
    &\int \mathcal{D}\varphi^{\rm E} \,\bar{I}_{\rm I}^{(1)}[A, \varphi]\, \frac{e^{-\int d^4x_{\rm E}\,\bar{\mathcal{L}}_{0}^{\rm E}(A, \varphi) }}{z_0^{\rm E}}= 
    2e \,\int d^4x_{\rm E}\,\alpha\,\partial_I\left(
    A_I-\partial_I\alpha
    \right) \langle \varphi^\dagger \varphi \rangle_0\,,\label{eq:I1_exp3}
\end{align}
where we have used $\langle (\partial_I {\varphi}^\dagger)\varphi\rangle_0=0$.
As we show below for both the magnetic and electric backgrounds, the right-hand side of Eq.~\eqref{eq:I1_exp3} can be made to vanish by an appropriate choice of the gauge parameter.

We next consider the term quadratic in $e$ in Eq.~\eqref{eq:dw_lam_scl}.
As in the linear term, after the field redefinition  $\varphi'=e^{i e\alpha}\varphi$, we obtain
\begin{align}
\int \mathcal{D}\varphi^{\rm E}\, \bar{I}_{\rm I}^{(2)}[A,\varphi]\, \frac{e^{-\int d^4x_{\rm E}\,\bar{\mathcal{L}}_{0}^{\rm E}(A,\varphi) }}{z_0^{\rm E}}
=
\int \mathcal{D}{\varphi'}^{\rm E}\, \bar{I}_{\rm I}^{(2)}[A,\varphi']\, \frac{e^{-\int d^4x_{\rm E}\,\mathcal{L}_{0}^{\rm E}(A,\varphi') }}{z_0^{\rm E}}\,,
\end{align}
where we have used $\mathcal{D}\varphi^{\rm E}=\mathcal{D}{\varphi'}^{\rm E}$, $\bar{I}_{\rm I}^{(2)}[A, \varphi]=\bar{I}_{\rm I}^{(2)}[A, \varphi']$, and $\bar{\mathcal{L}}_0^{\rm E}(A,\varphi)=\mathcal{L}_0^{\rm E}(A,\varphi')$. 
Note that the relation
$\bar{I}_{\rm I}^{(2)}[A, \varphi]=\bar{I}_{\rm I}^{(2)}[A, \varphi']$
holds because $\bar{I}_{\rm I}^{(2)}$ originates from the quadratic gauge interaction generated by the minimal coupling in the kinetic term of the charged heavy field.
Using the $O(4)$ symmetry of the Euclidean spacetime, we obtain
\begin{align}
    \int \mathcal{D}\varphi^{\rm E}\, \bar{I}_{\rm I}^{(2)}[A,\varphi]\, \frac{e^{-\int d^4x_{\rm E}\,\bar{\mathcal{L}}_{0}^{\rm E}(A,\varphi) }}{z_0^{\rm E}}
=
\int d^4x_{\rm E}
\left(
|A_I|^2
- 
|\partial_I\alpha|^2
\right)\,\langle \varphi^\dagger \varphi\rangle_0\,.\label{eq:I2_exp}
\end{align}
Combining Eqs.~\eqref{eq:dw_lam_scl}, \eqref{eq:I1_exp3}, and \eqref{eq:I2_exp}, we obtain
\begin{align}
    \left.\frac{dw_\lambda^{\rm E}}{d\lambda}\right|_{\lambda=0}=
    e^2 
    \,\int d^4x_{\rm E}\left[2 \alpha\, \partial_I\left(
    A_I-\partial_I\alpha
    \right)
    +
    \left(
|A_I|^2
- 
|\partial_I\alpha|^2
\right)\right]\,\langle \varphi^\dagger \varphi \rangle_0\,.\label{eq:I2exp}
\end{align}
Expanding the effective action $w_\lambda^{\rm E}$ around $\lambda=0$, we obtain
\begin{align}
    w_\lambda^{\rm E}[A^{\rm E}]=w_{\lambda=0}^{\rm E}[A^{\rm E}]+
    \lambda\,\left.\frac{dw_\lambda^{\rm E}[A^{\rm E}]}{d\lambda}\right|_{\lambda=0}
    +
    \mathcal{O}(\lambda^2)\,,\label{eq:w_lambda_E}
\end{align}
where $w^{\rm E}_{\lambda=0}[A^{\rm E}]=\int d^4x_{\rm E}\left(\epsilon_{\rm S}+(F_{IJ})^2/4\right)$, with $\epsilon_{\rm S}$ denoting the vacuum energy.
As shown below, the term linear in $\lambda$ vanishes.
The stationary configuration of $A^{\rm E}$ therefore receives corrections only at $\mathcal{O}(\lambda^2)$ ({\it i.e.}, ${A}^{\rm E,cl}_{\lambda}={A}^{\rm E,cl}_0+\mathcal{O}(\lambda^2)$).
Using $W^{\rm E}_{\lambda}[A^{\rm E,cl}]=w_{\lambda}^{\rm E}[{A}^{\rm E,cl}]$, it follows that
\begin{align}
    \left(\frac{d W^{\rm E}_\lambda[A^{\rm E,cl}_\lambda]}{d\lambda}\right)_{\lambda=0}&=\left(\frac{dw_\lambda^{\rm E}[{A}^{\rm E}]}{d\lambda}\right)_{\lambda=0,\, A^{\rm E}=A^{\rm E,cl}_\lambda}\notag
    \\
    &=e^2 
    \,\int d^4x_{\rm E}\left[2 \alpha\, \partial_I\left(
    A_I-\partial_I\alpha
    \right)
    +
    \left(
|A_I|^2
- 
|\partial_I\alpha|^2
\right)\right]\,\langle \varphi^\dagger \varphi \rangle_0\Bigg|_{A^{\rm E}=A^{\rm E,cl}_\lambda}\,.\label{eq:dW_lam_scl_man}
    \end{align}
Note that at the one-loop level considered here, the $\lambda$-dependence of the stationary configuration does not modify Eq.~\eqref{eq:dW_lam_scl_man}.
Corrections arising from gauge-field fluctuations enter only beyond the one-loop approximation.
    As we show below, the right-hand side of Eq.~\eqref{eq:dW_lam_scl_man} vanishes for a suitable choice of the gauge parameter $\alpha$.
Therefore, evaluating the effective action at the stationary configuration, we obtain
\begin{align}
    W^{\rm E}_{\lambda=1}[A^{\rm E,cl}]=w_{\lambda=1}^{\rm E}[{A}^{\rm E,cl}]
    =
    \int d^4x_{\rm E}\,\left[\epsilon_{\rm S}+\frac{1}{4}(F_{IJ}^{\rm cl})^2-\mathcal{L}_{\rm S}\right]\,,\label{eq:W_lambda_e_sc}
\end{align}
where $\mathcal{L}_{\rm S}$ denotes the one-loop nonlinear EFT correction arising at $\mathcal{O}\left(\lambda^2\right)$.
After wave-function renormalization, the kinetic term is canonically normalized and therefore independent of $\lambda$.
Applying the analytic continuation to Eq.~\eqref{eq:W_lambda_e_sc}, and using Eq.~\eqref{eq:quanrel}, we obtain
\begin{align}
    S\left(\rho_{\rm R}\| \rho_{\rm T}\right)
    =
    \int d^4x_{\rm E}\, \mathcal{L}_{\rm S}\,.
\end{align}
In what follows, we choose the gauge parameter $\alpha$ separately in the electric and magnetic cases so that the right-hand side of Eq.~\eqref{eq:I2exp} vanishes.
With this choice, the relative entropy directly captures the nonlinear EFT effects.

\begin{itemize}
    \item {\bf Magnetic model} --- For simplicity, we consider a constant magnetic field $\vec{B}=(0,\, 0,\, B)$.
     A convenient gauge choice is
    \begin{align}
    A_1 = \frac{B}{2} y\,,\quad A_2 =-\frac{B}{2}x\,,\quad A_3=0\,,\quad A_4=0\,.
    \end{align}
    This gauge choice yields the magnetic field:
    \begin{align}
    B^1 =F_{32}=0\,,\quad B^2 =F_{13}=0\,,\quad B^3=F_{21}=B\,,\quad E^k= F_{4k}=0\,.
    \end{align}
    Then, we find 
    \begin{align}
     |A_I|^2=\frac{B^2}{4}\left(x^2+y^2\right)=\frac{B^2}{4}r^2\,,
    \end{align}
    where $r^2=x^2+y^2$.
    We choose the following gauge parameter
    \begin{align}
    \alpha= \frac{B}{2}xy\,,\quad  \partial_1\alpha
    =
    \frac{B}{2}y\,,\quad  \partial_2\alpha =\frac{B}{2}x\,,\quad \partial_3\alpha= \partial_4 \alpha=0\,,\label{eq:alpha_gauge} 
    \end{align}
    where
    \begin{align}
    |\partial_I\alpha|^2=|A_I|^2=\frac{B^2}{4}r^2\,.\label{eq:sqralpha}
    \end{align}
    Furthermore, we find
    \begin{align}
    A_2-\partial_2\alpha= -B x\,,\quad A_1-\partial_1\alpha=A_3-\partial_3\alpha=A_4-\partial_4\alpha=0\,.
    \end{align}
    It follows that
    \begin{align}
    \partial_I \left(A_I-\partial_I\alpha\right)=0\,.\label{eq:partalpa}
    \end{align}
    Therefore, by choosing $\alpha$ as Eq.~\eqref{eq:alpha_gauge}, and substituting Eqs.~\eqref{eq:sqralpha} and \eqref{eq:partalpa} into Eq.~\eqref{eq:dW_lam_scl_man}, the right-hand side of Eq.~\eqref{eq:dW_lam_scl_man} vanishes.

    \item {\bf Electric model} --- In Euclidean spacetime, the choice of the gauge parameter is analogous to that in the magnetic case.
    Let us consider a constant Euclidean electric field along the $z$ direction, $\vec{E}=(0,\,0,\,E)$.
    In this case, the gauge field $A_I$ can be chosen as
    \begin{align}
    A_1=0\,,\quad A_2=0\,, \quad A_3=\frac{E}{2} \tau\,, \quad A_4 =-\frac{E}{2} z\,.
    \end{align}
    For this gauge choice, the electric field is given by
    \begin{align}
    B^1=B^2=B^3=E^1=E^2=0\,,\qquad E^3=F_{43}=E\,.
    \end{align}
    We then find
    \begin{align}
    |A_I|^2 = \frac{E^2}{4} \left(z^2+\tau^2\right)\,.
    \end{align}
    We choose the  gauge parameter
    \begin{align}
    \alpha= \frac{E}{2}\tau z\,, \quad \partial_1\alpha=\partial_2\alpha=0\,,\quad \partial_3\alpha=\frac{E}{2}\tau\,,\quad \partial_4 \alpha=\frac{E}{2}z\,.\label{eq:alpha_E}
    \end{align}
    For this choice,
    \begin{align}
    |\partial_I\alpha|^2=|A_I|^2=\frac{E^2}{4} \left(z^2+\tau^2\right)\,.\label{eq:ge1}
    \end{align}
   Moreover,
    \begin{align}
    A_1-\partial_1\alpha= A_2-\partial_2\alpha= 
    A_3-\partial_3\alpha=0\,,\quad
    A_4-\partial_4\alpha=-E z\,.
    \end{align}
    Therefore,
    \begin{align}
    \partial_I \left(A_I-\partial_I\alpha\right)=0\,.\label{eq:ge2}
    \end{align}
    Therefore, with the choice of $\alpha$ in Eq.~\eqref{eq:alpha_E}, substituting Eqs.~\eqref{eq:ge1} and~\eqref{eq:ge2} into Eq.~\eqref{eq:dW_lam_scl_man}, shows that the right-hand side of Eq.~\eqref{eq:dW_lam_scl_man} vanishes. 
    
\end{itemize}
As discussed in detail in Ref.~\cite{Conzinu:2026cuf}, the above results can be generalized
to the case of minimal gauge couplings in the kinetic terms of charged fields.

\section{Sign structure of the relative entropy in QED}
\label{app:pos_func}
In the Schwinger proper-time method, the real part of the relative entropy in the electric model is given, for both fermionic and scalar QED, by
\begin{align}
    {\rm Re}\,S\left(\rho_{\rm R}\|\rho_{\rm T}\right)
    &=\frac{m^4}{4\pi^2}\sum_{p=1}^{\infty}\, \mathcal{P}\, \int_0^\infty dt\,e^{-t} t^{-3}
    \begin{cases}
    \frac{\left(e\hat{E}t/p\pi\right)^4}{1-\left(e\hat{E}t/p\pi\right)^2}\,,&\text{(fermion QED)}
    \\
    \tfrac{1}{2}\left(1-\left(1+(-1)^p\right)\right)\frac{\left(e\hat{E}t/p\pi\right)^4}{1-\left(e\hat{E}t/p\pi\right)^2}\,,&\text{(scalar QED)}
    \end{cases}\label{eq:rel_non}
\end{align}
Below, we analyze the sign of the relative entropy for both fermionic and scalar QED.

\subsection{Fermionic QED}
From Eq.~\eqref{eq:rel_non}, we obtain
\begin{align}
    {\rm Re}\,S\left(\rho_{\rm R}\|\rho_{\rm T}\right)
    &=\frac{m^4}{4\pi^2}\sum_{p=1}^{\infty}\, \mathcal{P}\, \int_0^\infty dt\,e^{-t} t^{-3} \frac{\left(e\hat{E}t/p\pi\right)^4}{1-\left(e\hat{E}t/p\pi\right)^2}=\frac{m^4}{4\pi^2}\sum_{p=1}^{\infty} \left|e\hat{E}/p \pi\right|^2\, \mathcal{I}_f\left(p\pi/e\hat{E}\right)\,,\label{eq:sign_ferm_QED}
\end{align}
where 
\begin{align}
    \mathcal{I}_f\left(a\right)\equiv \mathcal{P}\,\int_0^\infty e^{-at}\frac{t}{1-t^2}dt
    =
    \lim_{\epsilon\to 0^+}\int_0^{1-\epsilon}\frac{\mathcal{N}\left(t,a\right)}{t(1-t^2)}dt\,,
    \quad
    \mathcal{N}\left(t,a\right)\equiv t^2e^{-at}-e^{-a/t}\,.\label{eq:Fa}
\end{align}
For $\mathcal{N}\left(t,a\right)>0$, it is evident that $\mathcal{I}_f\left(a\right)>0$.
As we show below,  $\mathcal{I}_f$ becomes negative in the unstable regime.
To illustrate this behavior, we consider two limiting regimes: the strong coupling regime ($a=p \pi/|e\hat{E}|\ll 1$) and the weak coupling regime ($1\ll a=p \pi/|e\hat{E}|$).

\begin{itemize}

\item {\it Strong coupling regime} --- In the limit $a=p \pi/|e\hat{E}|=0$, corresponding to the strong coupling regime, the integrand $\mathcal{N}\left(t,0\right)$ becomes negative for $0\leq t<1$, since $t^2-1<0$.
Consequently, $\mathcal{I}_f$ is negative in this limit; see also Ref.~\cite{Conzinu:2026cuf}.

\item {\it Weak coupling regime} --- In the weak coupling regime, $1\ll a=p\pi/|e\hat{E}|$, we show that $\mathcal{N}\left(t,a\right)>0$.
To this end, consider the function
\begin{align}
    n\left(t,a\right)\equiv \ln \left(t^2 e^{-at}\right)-\ln \left(
    e^{-a/t}
    \right)=a \left(
    \frac{1}{t}-t
    \right)+2\ln t\,.
\end{align}
If $\mathcal{N}\left(t,a\right)>0$, it follows that $n\left(t,a\right)>0$.
To show this, consider
\begin{align}
    m_+\left(t\right)= \frac{1}{t}-t +2\ln t\,.
\end{align}
For $0\leq t\leq 1$, we find
\begin{align}
    \frac{dm_+\left(t\right)}{dt}= - \frac{1}{t^2}-1 + \frac{2}{t} = - \frac{(t-1)^2}{t^2}\leq 0\,.
\end{align}
Thus, $m_+\left(t\right)$ is monotonically decreasing.
Since $m_+(1)=0$, it follows that
\begin{align}
    \left(\frac{1}{t}-t\right)+2\ln t =m_+\left(t\right)\geq 0~{\rm for}~0\leq t\leq 1\,.
\end{align}
In the weak coupling regime, $a=p \pi/|e\hat{E}|\geq 1$, we obtain
\begin{align}
    n\left(t,a\right)=a \left(\frac{1}{t}-t\right)+2 \ln t
    \geq
    \left(\frac{1}{t}-t\right)+2 \ln t
    =
    m_+\left(t\right)\geq 0~{\rm for}~0\leq t\leq 1\,.
\end{align}
Hence, $n\left(t,a\right)\geq 0$ for $0\leq t\leq 1$.
Therefore, $\mathcal{I}_f>0$ in the weak coupling regime.

\end{itemize}

From these analyses, we find that each pole contribution in Eq.~\eqref{eq:sign_ferm_QED} remains positive in the weak coupling regime but turns negative in the strong coupling regime.
Since $\lim_{e\hat{E}\to \infty}\mathcal{I}_f\left(p \pi/e\hat{E}\right)<0$, it follows that
\begin{align}
    \lim_{e\hat{E}\to \infty}\frac{m^4}{4\pi^2}\sum_{p=1}^{N} \left|e\hat{E}/p \pi\right|^2\, \mathcal{I}_f\left(p\pi/e\hat{E}\right)<0\,,\label{eq:fer_ine_N}
\end{align}
where the sum over pole contributions is truncated at a finite order $N$.
Consequently, from Eqs.~\eqref{eq:sign_ferm_QED} and \eqref{eq:fer_ine_N}, a finite-order truncation of the pole contributions can yield a negative relative entropy in the strong-coupling regime.

\subsection{Scalar QED}
Similarly, from Eq.~\eqref{eq:rel_non}, we obtain
\begin{align}
    {\rm Re}\,S\left(\rho_{\rm R}\|\rho_{\rm T}\right)
    &=\frac{m^4}{8\pi^2}\sum_{p=1}^{\infty}\, \mathcal{P}\, \int_0^\infty e^{-t} t^{-3} \left(1-\left(1+(-1)^p\right)\right)\frac{\left(e\hat{E}t/p\pi\right)^4}{1-\left(e\hat{E}t/p\pi\right)^2}dt\notag
    \\
    &=\frac{m^4}{8\pi^2}\, \mathcal{P}\, \int_0^\infty e^{-t} t^{-3} g\left(e\hat{E}t\right)dt\,,
\end{align}
where we have used
\begin{align}
    g\left(x\right)&\equiv \frac{x^4}{\pi^4} \sum_{p=1}^\infty \frac{1}{p^4}
    \left(
    \frac{1}{1-(x/p\pi)^2}
    -
    \frac{1}{8} \frac{1}{1-(x/2p\pi)^2}
    \right)\notag
    \\
    &=\sum_{p=1}^\infty 
    \left(
    \frac{(x/p\pi)^4}{1-(x/p\pi)^2}
    -
    2 \frac{(x/2p\pi)^4}{1-(x/2p\pi)^2}
    \right)\,.
\end{align}
Thus, introducing
\begin{align}
\mathcal{I}_s(a)\equiv
2\mathcal{I}_f(a)-\mathcal{I}_f(2a)\,,
\end{align}
we find
\begin{align}
\operatorname{Re}S\left(\rho_{\rm R}\|\rho_{\rm T}\right)
    =
    \frac{m^4}{16\pi^2}
\sum_{p=1}^\infty
\left|\frac{e\hat{E}}{p\pi}\right|^2
\mathcal{I}_s\left(p\pi/e\hat{E}\right).
\label{eq:scal_rel_sign}
\end{align}
From Eq.~\eqref{eq:Fa}, we obtain
\begin{align}
    \frac{d\mathcal{I}_f\left(a\right)}{da}=-\lim_{\epsilon\to 0^+} \int_0^{1-\epsilon} \frac{\mathcal{O}\left(t,a\right)}{t^2\left(1-t^2\right)}dt\,,\quad \mathcal{O}\left(t,a\right)\equiv t^4 e^{-at}-e^{-a/t}\,.
\end{align}
Using formulae provided in Ref.~\cite{Conzinu:2026cuf}, for $a>0$, we also obtain
\begin{align}
    \frac{d\mathcal{I}_f\left(a\right)}{da}
    =
    \begin{cases}
    >0\, & a\ll 1~(\text{strong coupling})
    \\
    <0\, & 1\ll a~(\text{weak coupling})
    \end{cases}\,.\label{eq:ine_Fa}
\end{align}
Thus, from Eqs.~\eqref{eq:scal_rel_sign} and \eqref{eq:ine_Fa}, in the weak coupling regime, we find that the non-negativity of the relative entropy holds:
\begin{align}
    {\rm Re}\,S\left(\rho_{\rm R}\|\rho_{\rm T}\right)
    &>\frac{m^4}{16\pi^2}\sum_{p=1}^\infty \left|{e \hat{E}}/{p \pi}\right|^2\, \mathcal{I}_f \left(p \pi/e\hat{E}\right)>0\,,~{\rm for}~ 1\ll\pi/e\hat{E}~(\text{weak coupling})\,.
\end{align}
In the strong-coupling regime, the relative entropy~\eqref{eq:scal_rel_sign} obtained by truncating the pole contributions at order $N$ satisfies
\begin{align}
    \lim_{e\hat{E}\to \infty}\frac{m^4}{16\pi^2}
\sum_{p=1}^N
\left|\frac{e\hat{E}}{p\pi}\right|^2
\mathcal{I}_s\left(p\pi/e\hat{E}\right)
<\lim_{e\hat{E}\to \infty} \frac{m^4}{16\pi^2}\sum_{p=1}^N \left|{e \hat{E}}/{p \pi}\right|^2\, \mathcal{I}_f \left(p \pi/e\hat{E}\right)<0\,.
\end{align}
Therefore, as in fermionic QED, a finite-order truncation of the pole contributions can yield a negative relative entropy in the strong-coupling regime.

%
%
\bibliographystyle{JHEP}
\bibliography{BIB_main_jhep}
\end{document}